\documentclass[fleqn,usenatbib,useAMS]{mnras}

\usepackage{graphicx}             
\usepackage{amsmath}              
\usepackage{upgreek}              
\usepackage{amssymb}              

\usepackage{bm}                   
\usepackage[mathcal]{euscript}    
\usepackage{CJKutf8}              
\usepackage[T1]{fontenc}          
\usepackage{comment}              
\usepackage{transparent}          
\usepackage{tabularx}             
\usepackage{threeparttable}       
\usepackage{booktabs}             
\usepackage{soul}                 
\usepackage{ulem}                 
\usepackage{multirow}             
\usepackage{makecell}             
\usepackage[dvipsnames]{xcolor}   
\usepackage{tipa}
\usepackage{microtype}            
\DisableLigatures[f]{encoding = *, family = *} 
\usepackage[figuresleft]{rotating}
\emergencystretch=\maxdimen       
\newcommand{\fracd}[2]{\frac{\textnormal{d}{#1}}{\textnormal{d}{#2}}}

\newcommand{\fracp}[2]{\frac{\partial{#1}}{\partial{#2}}}

\newcommand{\hatomega}{\hat{\text{\bf \textscomega}}}
\def\approxprop{\def\p{
    \setbox0=\vbox{\hbox{$\propto$}}
    \ht0=0.6ex \box0 }
  \def\s{\vbox{\hbox{$\sim$}}}
  \mathrel{\raisebox{0.7ex}{\mbox{$\underset{\s}{\p}$}}}
}
\newcommand{\rlr}[1]{#1} 
\newcommand{\biborder}[1]{} 

\defcitealias{Baruteau2011}{B11}
\defcitealias{LiYaPing2021}{L21}
\defcitealias{Li2022b}{Paper II}

\usepackage{ae,aecompl}
\usepackage{multicol}        

\usepackage{newtxtext,newtxmath}

\title[Binaries Embedded in Discs]{Hydrodynamical Evolution of Black-Hole Binaries Embedded in AGN Discs}

\author[R. Li and D. Lai]{
  Rixin Li$^{1}$
  \begin{CJK*}{UTF8}{gbsn}
    (李日新)
  \end{CJK*} \thanks{Contact e-mail: \href{mailto:rixin.li@cornell.edu}{rixin.li@cornell.edu}}
  and 
  Dong Lai$^{1}$
  \begin{CJK*}{UTF8}{gbsn}
    (赖东)
  \end{CJK*}
\\
$^{1}$Center for Astrophysics and Planetary Science, Department of Astronomy, Cornell University, Ithaca, NY 14853, USA \\
}

\date{Accepted 2022 September 05. Received 2022 July 27; in original form 2022 February 15}

\pubyear{2021}

\begin{document}
\label{firstpage}
\pagerange{\pageref{firstpage}--\pageref{lastpage}}
\maketitle

\abovedisplayshortskip=6pt plus 3pt minus 2pt 
\belowdisplayshortskip=6pt plus 3pt minus 2pt 
\abovedisplayskip=6pt plus 3pt minus 2pt 
\belowdisplayskip=6pt plus 3pt minus 2pt 

\begin{abstract}
Stellar-mass binary black holes (BBHs) embedded in active galactic nucleus (AGN) discs are possible progenitors of black-hole mergers detected in gravitational waves by LIGO/VIRGO.
To better understand the hydrodynamical evolution of BBHs interacting with the disc gas, we perform a suite of high-resolution 2D simulations of binaries in local disc (shearing-box) models, considering various binary mass ratios, eccentricities and background disc properties.
We use the $\gamma$-law equation of state and adopt a robust post-processing treatment to evaluate the mass accretion rate, torque and energy transfer rate on the binary to determine its long-term orbital evolution. 
We find that circular comparable-mass binaries contract, with an orbital decay rate of a few times the mass doubling rate.
Eccentric binaries always experience eccentricity damping. Prograde binaries with higher eccentricities or smaller mass ratios generally have slower orbital decay rates, with some extreme cases exhibiting orbital expansion.  The averaged binary mass accretion rate depends on the physical size of the accretor.  
The accretion flows are highly variable, and the dominant variability frequency is the apparent binary orbital frequency (in the rotating frame around the central massive BH) for circular binaries but gradually shifts to the radial epicyclic frequency as the binary eccentricity increases.  Our findings demonstrate that the dynamics of BBHs embedded in AGN discs is quite different from that of isolated binaries in their own circumbinary discs.
Furthermore, our results suggest that the hardening timescales of the binaries are much shorter than their migration timescales in the disc, for all reasonable binary and disc parameters.
\end{abstract}


\section{Introduction}
\label{sec:intro}

Since the first detection of gravitational waves from the merging black-hole binary GW 150904 \citep{Abbott2016PRL}, the LIGO/VIRGO collaboration has reported about 90 merger events in the first three observing runs \citep{Abbott2021arXiv}.  A number of formation channels to produce such mergers from stellar-mass black holes (BHs) have been studied over the years \citep[e.g.,][]{Mapelli2020}.  In the isolated binary evolution channel, massive binary stars evolve into BHs and experience mass transfer and common envelope processes, leading to BH binaries in close orbits \citep[e.g.,][]{Lipunov1997, Podsiadlowski2003, Belczynski2010, Belczynski2016}.  A variant of the isolated binary channel involves chemically homogeneous evolution of very close massive stellar binaries \citep{Mandel2016, Marchant2016}.  Alternatively, binary BHs can form through several flavors of dynamical channels that involve either strong gravitational scatterings in dense star clusters \citep[e.g.,][]{PortegiesZwart2000, Oleary2006, Miller2009,Banerjee2010, Downing2010, Ziosi2014, Samsing2014, Samsing2018b, Rodriguez2015, Kremer2019}, or more gentle ``tertiary-induced mergers'' (often via Lidov-Kozai mechanism) -- the latter can take place either in stellar triple/quadrupole systems \citep[e.g.,][]{Miller2002, Silsbee2017, LiuLai2018, LiuLai2019, LiuBin2019b, Fragione2019, Fragione2019b}, or in nuclear clusters dominated by a central supermassive BH \citep[e.g.,][]{Antonini2012, VanLandingham2016, Petrovich2017, Hamers2018, LiuBin2019a, LiuLai2020PRD, LiuLai2021}.

In recent years, another flavor of dynamical BBH formation channel based on binaries in the discs of active galactic nuclei (AGN) has attracted much attention.  \citet{McKernan2012, McKernan2014} suggested that intermediate-mass BHs may be formed efficiently in AGN discs via collisions or accretions of smaller bodies (stars and/or compact objects).  It has also been proposed that BBHs may be captured in the inner AGN discs ($\sim 0.01$~pc from the super massive black hole (SMBH)) from nuclear star clusters \citep{Bartos2017} or form in situ in the extended region ($\sim$ pc) of AGN discs \citep{Stone2017}; in either case, the binaries may harden via gas dynamical friction or binary-single interactions.  The orbital migration of BHs in AGN discs and the possible migration traps may facilitate the formation of BBHs \citep{Bellovary2016, Secunda2019, Secunda2020, Yang2019PRL}.  \citet{McKernan2018, McKernan2020} and \citet{Tagawa2020a} attempted to incorporate many of these physical ingredients into population synthesis studies of BH mergers in AGN discs.  Although the ``predicted'' merger rates are highly uncertain, BBH mergers AGN discs may exhibit some distinct properties. For example, such mergers may be heavy enough to be in the pair-instability mass gap or beyond if their progenitors are higher-generation BHs \citep[e.g., GW190521;][]{Yang2019PRL, Abbott2020PRL, Abbott2020APJ}.  Moreover, they might have associated, observable electromagnetic counterparts \citep{deMink2017, McKernan2019, Graham2020, Ashton2021, Palmese2021}.

In the AGN disc channel for BBH mergers, a major uncertainty concerns hydrodynamical interactions between the BBH and the gaseous AGN disc.  The orbital evolution of binaries through gas torques has only been studied numerically by a handful of previous works.  \citet[][hereafter \citetalias{Baruteau2011}]{Baruteau2011} carried out global disc simulations in 2D, with a limited resolution of the accretion flow around each binary component and a relatively large gravitational softening length (about $1/4$ binary separation).  They found that a massive (gap-opening) prograde, equal-mass binary is hardened by dynamical friction from the lagging spiral tails trailing each binary component inside the Hill radius.  They also found that the hardening timescale for the binary is shorter than its migration timescale in the disc.

Motivated by recent studies on circumbinary accretion that show resolved circum-single discs (CSDs) may result in orbital expansion instead of decay \citep[see also \citealt{Miranda2017}]{Munoz2019, Munoz2020, Moody2019, Duffell2020, Tiede2020}, \citet[][hereafter \citetalias{LiYaPing2021}]{LiYaPing2021} revisited the problem of BBHs in AGN discs.  With an improved numerical resolution, a smaller gravitational softening length (about $\rlr{0.08}$ binary separation), and a prescription for BH accretion, they found that adequately resolved CSD regions lead to expanding binaries, contradicting the findings in \citetalias{Baruteau2011}.  A recent work from the same group \citep{LiYaPing2022} found that an enhanced temperature of the CSD may lead to binary orbital decay, suggesting that the gas thermodynamics can play an important role.

Resolving CSDs around each binary component in global disc simulations is computational expensive, limiting the choices of physical parameters.  In particular, the BBH to SMBH mass ratios adopted by \citetalias{Baruteau2011} and \citetalias{LiYaPing2021} are several orders of magnitude larger than those expected in AGN discs (e.g., in \citetalias{LiYaPing2021}, the BBH mass is $0.002M_{\rm SMBH}$).  Moreover, both \citetalias{Baruteau2011} and \citetalias{LiYaPing2021} included only the gravitational torque from the gas (i.e. dynamical friction) on the binary, but ignored the torques caused by hydrodynamical forces due to accretion and pressure, which may be non-negligible \citep{Thun2016, Munoz2019}.

\citet{Kaaz2021} studied the accretion flows around BBHs embedded in a local, wind tunnel box with a prescribed velocity profile at the boundaries.  However, they put the binary in an inertial frame and neglected the centrifugal and coriolis forces.  Such a setup did not represent the realistic environment that BBHs experience in AGN discs, where the shear flow is much stronger when taking into account the non-inertial forces in a rotating frame.

Motivated by these previous works, in this paper we carry out a suite of 2D hydrodynamical simulations of binaries embedded in AGN discs using a co-rotating local disc (``shearing-box'') model.  We adopt realistic BBH to SMBH mass ratios, and consider various binary eccentricities, mass ratios and semi-major axes (relative to the Hill radius).  We use the $\gamma$-law equation of state \rlr{(EOS)} and survey different background disc parameters (characterized by the disc scale height and velocity shear). By using multi-level mesh refinements and an absorbing accretion prescription, we resolve the flow around each binary component with a negligible gravitational softening length.  We take account of the gravitational forces and hydrodynamical forces to compute the torque and the energy transfer rate to the binary.  Our goal is to determine the flow structure, variability, and most importantly, the long-term accretion rate and orbital evolution of BBHs embedded in AGN discs.

The paper is organized as follows.  In Section \ref{sec:methods}, we describe our numerical scheme and setup, including the important dimensionless parameters for the problem (Eqs. \ref{eq:q}--\ref{eq:lambda} and Eqs. \ref{eq:c_s_v_b}--\ref{eq:Delta_V_K_v_b}) and our method for evaluating the mass accretion and torques (Section \ref{subsec:acc_T}).  Section \ref{sec:results} presents our results, starting with prograde equal-mass binaries on circular orbits in Section \ref{subsec:result_fiducial}, followed by equal-mass binaries on eccentric orbits in Section \ref{subsec:result_ecc}, unequal-mass binaries on circular orbits in Section \ref{subsec:result_qb}, and retrograde equal-mass, circular binaries in Section \ref{subsec:result_retro}.  Section \ref{sec:comp2previous} compares our results with previous studies in details.  Section \ref{sec:summary} summarizes our findings and discusses possible caveats and astrophysical implications.

\section{Methods}
\label{sec:methods}

To study the hydrodynamical evolution of binaries embedded in accretion discs, we use the code \texttt{ATHENA} \citep{Stone2008, Stone2010} in a shearing box.  Section \ref{subsec:schemes} describes our simulation setup for modelling the flows around the binary.  In section \ref{subsec:acc_T}, we explain how we compute the long-term (secular) evolution of the binary by post-processing our simulations.  Section \ref{subsec:setups} summarizes the parameter choices for our simulations.

\begin{figure}
  \centering
  \includegraphics[width=\linewidth]{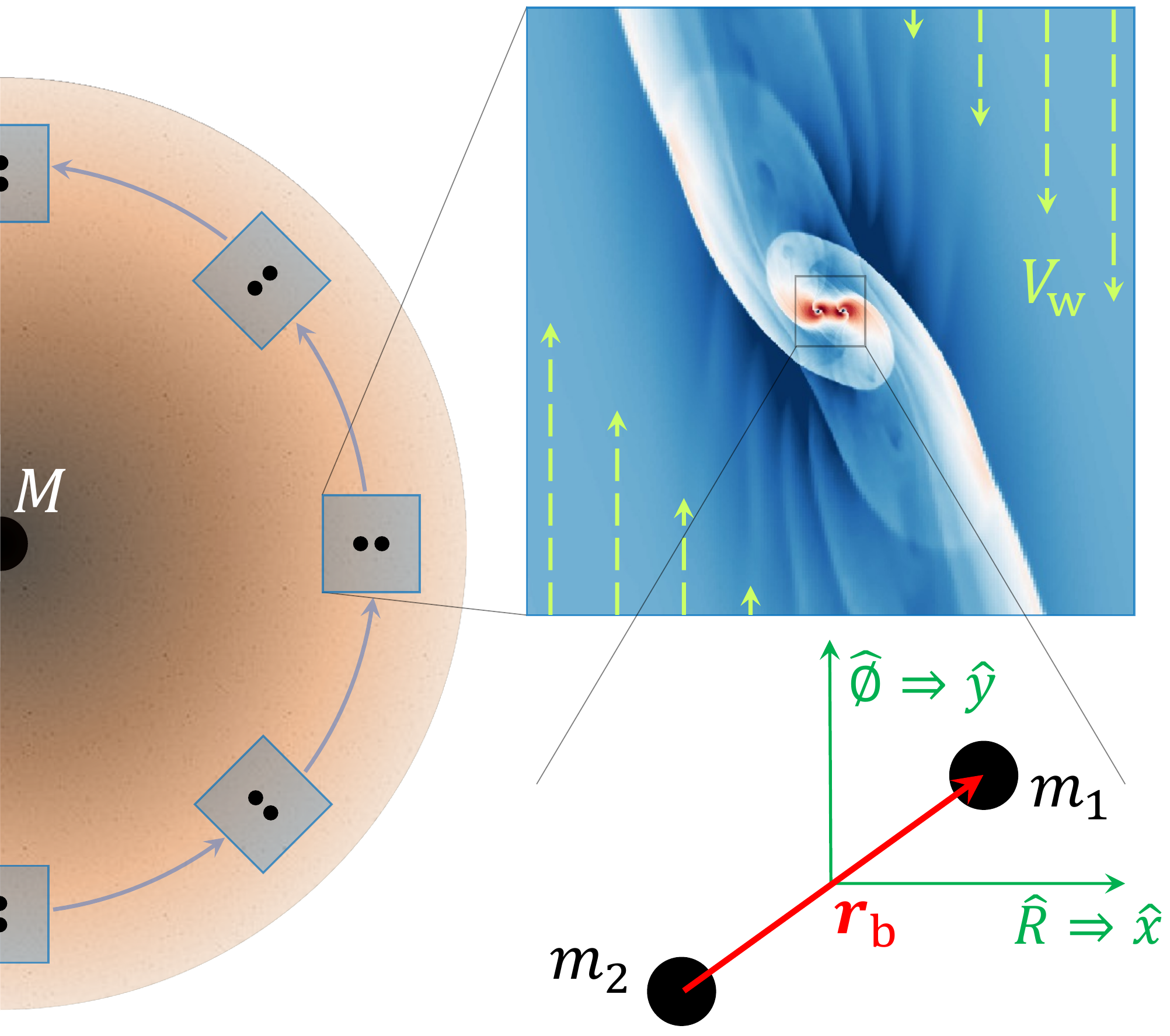}
  \caption{Cartoon illustration of our model to study the hydrodynamical evolution of a binary ($m_1$ and $m_2$) embedded in the disc around a SMBH ($M$).
  \label{fig:cartoon}}
\end{figure}

\subsection{Numerical Setup and Method}
\label{subsec:schemes}

We consider a binary (with component masses $m_1$ and $m_2$) centred in a small patch of an accretion disc around a massive object (e.g., a super massive black hole (SMBH) with mass $M$) using the local shearing box approximation \citep[see Fig. \ref{fig:cartoon};][]{Goldreich1965, Hawley1995, Stone2010}.  With this approximation, the global cylindrical geometry of the disc is mapped onto local Cartesian coordinates with unit vectors $\bm{\hat{x}}$ and $\bm{\hat{y}}$ in the radial and azimuthal directions, respectively.  The centre of mass (COM) of the binary, i.e., the centre of the computational domain --- $(x, y) = (0, 0)$ --- is located at a fiducial disc radius $R$ from the SMBH.  At this location, the Keplerian velocity is $V_{\rm K}=\sqrt{GM/R}$ and the Keplerian frequency is $\Omega_{\rm K}=V_{\rm K}/R$
\footnote{The COM of the binary in fact orbits around the central object at a frequency slightly larger than $\Omega_{\rm K}$ because of the mass quadrupole associated with the binary, with the correction of order $\Delta \Omega / \Omega_{\rm K} \sim (\mu_{\rm b}/m_{\rm b}) (a_{\rm b}/R)^2 \ll 1$ (where $\mu_{\rm b}$ and $m_{\rm b}$ are the reduced mass and total mass of the binary, $a_{\rm b}$ the binary separation).  This effect is negligible.}.
Our reference frame rotates at this frequency.

In the rotating frame, we simulate the dynamics of an inviscid compressible flow, with a gamma-law equation of state, by solving the following equations of gas dynamics in 2D:
\begin{align}
  \fracp{\Sigma_{\rm g}}{t} + \nabla \cdot \left(\Sigma_{\rm g} \bm{u}\right) &= 0, \label{eq:gascon} \\
  \begin{split}\label{eq:gasmom}
    \fracp{(\Sigma_{\rm g} \bm{u})}{t} + \nabla\cdot(\Sigma_{\rm g} \bm{u}\bm{u} + P\bm{I}) &=\\
    \Sigma_{\rm g} \biggl[ 2\bm{u}\times\bm{\Omega}_{\rm K} &+ 2 q_{\rm sh} {\Omega}_{\rm K}^2 \bm{x} - \nabla \phi_{\rm b} \biggr], 
  \end{split} \\
  \begin{split}\label{eq:gasE}
    \fracp{E}{t} + \nabla\cdot\left[ (E + P) \bm{u}\right] &= \Sigma_{\rm g} \bm{u} \cdot \left(2 q_{\rm sh}\Omega_{\rm K}^2 \bm{x} - \nabla\phi_{\rm b} \right) , \\
    E = \frac{P}{\gamma - 1} &+ \frac{1}{2} \Sigma_{\rm g} (\bm{u} \cdot \bm{u}),
  \end{split}
\end{align}
where $\Sigma_{\rm g}$, $\bm{u}$, $P$, $E$, and $\gamma$ are surface density, velocity, pressure, total energy surface density, and adiabatic index of gas, $\bm{I}$ is the identity matrix, $\bm{\Omega}_{\rm K}$ aligns with $\hat{\bm{z}}$, $q_{\rm sh} \equiv \mathbf{-} \textnormal{d}\ln \Omega_{\rm K}/\textnormal{d} \ln R$ is the background shear parameter and is $3/2$ for a Keplerian disc, $\phi_{\rm b}$ is the gravitational potential of the binary
\begin{equation}
    \phi_{\rm b}(\bm{r}_{k}) = -\frac{G m_1}{\sqrt{(\bm{r}_1 - \bm{r}_k)^2 + \xi_{\rm s}^2}} -\frac{G m_2}{\sqrt{(\bm{r}_2 - \bm{r}_k)^2 + \xi_{\rm s}^2}},
\end{equation}
where $\bm{r}_1$ and $\bm{r}_2$ denote the position vectors of the binary components, $\bm{r}_k$ is the centre position of the $k$-th cell in the computational domain, and $\xi_{\rm s}$ is the gravitational softening length.  Throughout this work, we adopt $\gamma = 1.6$ and neglect the self-gravity of the gas.  

To ensure the stability of our simulations, we adopt the van Leer integrator with first-order flux correction \citep{van_Leer2003, Stone2009}, a piecewise parabolic spatial reconstruction in the primitive variables, and the Roe's linearized Riemann solver with H-correction \citep{Stone2008, Xu2019}.  Furthermore, we use the static mesh refinement (SMR) to properly resolve the flow around the binary while the root domain is still large enough so that the flow far from the binary can be specified by the disc profile (see Section \ref{subsec:acc_T} and \ref{subsec:setups} for more details).

The binary in our models has total mass $m_{\rm b} = m_1 + m_2$ and orbits on a prescribed orbit with a semi-major axis of $a_{\rm b}$ and an eccentricity of $e_{\rm b}$.  The mass ratio between the binary components is defined as $q_{\rm b} \equiv m_2/m_1 \leqslant 1$.  The mean orbital frequency, orbital angular momentum, and energy in the inertial frame are thus
\begin{align}
  \bm{\Omega}_{\rm b} &= \sqrt{\frac{G m_{\rm b}}{a_{\rm b}^3}}\ \hatomega_{\rm b} = \frac{v_{\rm b}}{a_{\rm b}} \hatomega_{\rm b}, \qquad \text{with}~ v_{\rm b} \equiv \sqrt{\frac{G m_{\rm b}}{a_{\rm b}}}, \\
  \bm{L}_{\rm b} &= \mu_{\rm b} \bm{\ell}_{\rm b} = \mu_{\rm b} \bm{\Omega}_{\rm b} a_{\rm b}^2 \sqrt{1 - e_{\rm b}^2}, \\
  E_{\rm b} &= \mu_{\rm b} \mathcal{E}_{\rm b} = -\mu_{\rm b} \frac{G m_{\rm b}}{2 a_{\rm b}},
\end{align}
where $\hatomega_{\rm b}$ is binary normal unit vector, $\mu_{\rm b} = m_1 m_2 / m_{\rm b}$ is the reduced mass, and $\ell_{\rm b}$ and $\mathcal{E}_{\rm b}$ are the specific angular momentum and specific energy, respectively.  Throughout this paper, we consider co-planar binaries, but allow for both prograde ($\hatomega_{\rm b}\cdot \hat{z}=1$) and retrograde ($\hatomega_{\rm b}\cdot \hat{z}=-1$) orientations.

The code units of our simulations are set to the natural units of the binary, where the length unit and the time unit are $a_{\rm b}$ and $\Omega_{\rm b}^{-1}$, respectively.  The velocity unit is then $v_{\rm b}$.  The mass unit is $m_{\rm b}$ for the binary and $\Sigma_{\infty} a_{\rm b}^2$ for the background gas, where $\Sigma_{\infty}$ is the gas density far away from the binary.

To establish the background flow (``wind'') profile in the vicinity of the binary, we define the following three dimensionless parameters, namely the mass ratio of the binary to the SMBH at disc centre, the disc aspect ratio at $R$, and the ratio of binary Hill radius
\footnote{Note that this definition of $R_{\rm H}$ differs from the usual expression for the Hill radius, $R (m_{\rm b}/3 M)^{1/3}$.}
$R_{\rm H} \equiv R (m_{\rm b}/M)^{1/3}$ to $a_{\rm b}$,
\begin{align}
    q &= \frac{m_{\rm b}}{M}, \label{eq:q}\\
    h &= \frac{H_{\rm g}}{R} = \frac{c_{\rm s,\infty}}{V_{\rm K}}, \label{eq:h} \\
    \uplambda &= \frac{R_{\rm H}}{a_{\rm b}} = \frac{R}{a_{\rm b}} \left(\frac{m_{\rm b}}{M}\right)^{1/3}, \label{eq:lambda}
\end{align}
where $H_{\rm g}$ is gas scale height of the accretion disc at $R$, and $c_{\rm s,\infty} = \sqrt{\gamma P / \Sigma_{\infty}}$ is the sound speed.  The stability of the binary requires $\lambda \gtrsim$ a few.  The time-independent background wind profile (far away from the binary) in the shearing box can be then expressed as
\begin{equation}\label{eq:v_wind}
  \begin{aligned}
    \bm{V}_{\rm w} &= \bm{V}_{\rm sh} + \bm{\Delta V}_{\rm K} \\
    &= -q_{\rm sh} \Omega_{\rm K} x \bm{\hat{y}} - \beta h^2 V_{\rm K} \bm{\hat{y}},
  \end{aligned}
\end{equation}
where $\bm{V}_{\rm w}$ contains two parts: the Keplerian shear $\bm{V}_{\rm sh}(x)$ and the deviation from Keplerian velocity $\bm{\Delta V}_{\rm K}$.  The latter term accounts for the sub-Keplerian orbital velocity of the disc gas due to its own pressure support, where $\beta$ ($\simeq \textnormal{d}\ln P / \textnormal{d}\ln R $) is an order unity coefficient determined by (background) disc pressure profile.  Including this sub-Keplerian adjustment equivalently shifts the Keplerian shear slightly (i.e., $\Delta V_{\rm K} \ll c_{\rm s,\infty}$) towards smaller $R$.

From the perspective of the binary, the following three characteristic velocity ratios determine the flow dynamics:
\begin{align}
    \frac{c_{\rm s,\infty}}{v_{\rm b}} &= h\ q^{-1/3}\ \uplambda^{-1/2}, \label{eq:c_s_v_b} \\
    \frac{V_{\rm s}}{v_{\rm b}} = \frac{|V_{\rm sh}(x=a_{\rm b})|}{v_{\rm b}} &= q_{\rm sh} \frac{\Omega_{\rm K}}{\Omega_{\rm b}} = q_{\rm sh}\ \uplambda^{-3/2}, \label{eq:V_s_v_b} \\
    \frac{\Delta V_{\rm K}}{v_{\rm b}} &= \beta h^2\ q^{-1/3}\ \uplambda^{-1/2}, \label{eq:Delta_V_K_v_b}
\end{align}
where $V_{\rm s}$ is the magnitude of Keplerian shear across a radial length of $a_{\rm b}$.  Throughout the paper, we fix $q_{\rm sh}=3/2$ and $\beta=1$.

To model the gravitational potential $\phi_{\rm b}$, we prescribe the binary orbit in two steps.  (i) We prescribe the elliptical Kepler orbit in the inertial frame with two base vectors \citep{SSD2000}
\begin{align}
  \bm{r}_{\rm b}(t) &= g_1(t) \bm{r}_{\rm b0} + g_2(t) \bm{w}_{\rm b0}, \\
  g_1(t) &= \frac{a_{\rm b}}{r_{\rm b0}} (1 - \cos E ) - 1, \\
  g_2(t) &= \frac{1}{\Omega_{\rm b}} \left[E(t) - \sin E \right] - t,
\end{align}
where $\bm{r}_{\rm b} = \bm{r}_1 - \bm{r}_2$ and $\bm{w}_{\rm b} \equiv \dot{\bm{r}}_{\rm b} = \bm{v}_1 - \bm{v}_2$ are the relative position vector and velocity vector, $\bm{r}_{\rm b0}$ and $\bm{w}_{\rm b0}$ are their initial values at pericentre, with $r_{\rm b0} = a_{\rm b} (1 - e_{\rm b})$ and $w_{\rm b0} = v_{\rm b} \sqrt{(1 - e_{\rm b})/(1 + e_{\rm b})}$, $E(t)$ is the eccentric anomaly and is obtained by solving the corresponding Kepler's equation using the Newton–Raphson root-finding method.  (ii) We then take into account both the physical apsidal precession of the binary due to the tidal gravity of $M$ \citep[$\dot{\varpi}$; e.g.,][]{LiuBin2015} and the apparent precession due to the rotating frame ($\Omega_{\rm K}$) by rotating the base vectors, $\bm{r}_{\rm b0}$ and $\bm{w}_{\rm b0}$, on an angular frequency $\Omega_{\rm pre}$, where
\begin{equation}\label{eq:Omega_pre}
  \frac{\bm{\Omega}_{\rm pre}}{\Omega_{\rm b}} = \frac{\dot{\varpi}\hatomega_{\rm b} - \bm{\Omega}_{\rm K}}{\Omega_{\rm b}} = \frac{3}{4} \frac{\sqrt{1 - e_{\rm b}^2}}{\uplambda^{3}} \hatomega_{\rm b} - \uplambda^{-3/2} \hat{\bm{z}}.
\end{equation}
Note that the inclusion of the $\dot{\varpi}$ term implies that the binary orbit is not exactly Keplerian.  Since $\dot{\varpi}\ll \Omega_{\rm b}$, this correction is small, and we treat $\dot{\varpi}$ as a reference shift from the inertial frame for simplicity.  Appendix \ref{appsec:no_pomega} demonstrates that the $\dot{\varpi}$ term only has a moderate influence on the binary orbital evolution.

Our prescriptions revert back to a fixed circular orbit when $e_{\rm b}=0$.  In the rotating shearing box frame, the apparent orbital frequency, orbital velocity and period are
\begin{align}
    \bm{\Omega}_{\rm b}' &= \bm{\Omega}_{\rm b} + \bm{\Omega}_{\rm pre}, \label{eq:Omega_bprime} \\
    \bm{w}_{\rm b}' & = \bm{\Omega}'_{\rm b}\times \bm{r}_{\rm b}, \label{eq:w_bprime} \\
    P_{\rm b}' &= 2\pi / \Omega_{\rm b}', \label{eq:P_bprime} 
\end{align}
where prime denotes quantities in the rotating frame.

\begingroup 
\setlength{\medmuskip}{0mu} 
\setlength\tabcolsep{4pt} 
\setcellgapes{3pt} 
\begin{table*}
  \nomakegapedcells
  \caption{Simulation Setups and Results for \texttt{Run I} Series 
  ($q=1$e-$6$, $h=0.01$, $\lambda=2.5$)}\label{tab:run_I}
  \makegapedcells 
  \linespread{1.025} 
  \begin{tabular}{lcccccccc|rrrrrr}
    \hline
    \texttt{Run}
    & $q_{\rm b}$
    & $e_{\rm b}$
    & $\displaystyle \frac{r_{\rm s}}{a_{\rm b}}$
    & $\displaystyle \frac{r_{\rm e}}{a_{\rm b}}$
    & $L_X\times L_Y$
    & \footnotesize $N_{\rm SMR}$
    & $\displaystyle \frac{a_{\rm b}}{\delta_{\rm fl}}$
    & Remarks
    & \makecell[c]{$\langle\dot{m}_{\rm b}\rangle$}
    & \makecell[c]{$\langle\dot{\mathcal{E}}_{\rm b}\rangle$}
    & \makecell[c]{$\ell_0$}
    & \makecell[c]{$\displaystyle \frac{\langle\dot{a}_{\rm b}\rangle}{a_{\rm b}}$}
    & \makecell[c]{$\langle\dot{e_{\rm b}^2}\rangle$}
    & \makecell[c]{$\eta$} \\

    &
    &
    & $[\%]$
    & $[\%]$
    & $[a_{\rm b}^2]$
    &
    &
    &
    & \makecell[c]{\footnotesize $[\Sigma_{\infty} v_{\rm b} a_{\rm b}]$}
    & \makecell[c]{\footnotesize $\displaystyle \left[\frac{\Sigma_\infty v_{\rm b}^3 a_{\rm b}}{m_{\rm b}}\right]$}
    & \makecell[c]{$[v_{\rm b} a_{\rm b}]$}
    & \makecell[c]{\small $\displaystyle \left[\frac{\langle\dot{m}_{\rm b}\rangle}{m_{\rm b}}\right]$}
    & \makecell[c]{\small $\displaystyle \left[\frac{\langle\dot{m}_{\rm b}\rangle}{m_{\rm b}}\right]$}
    & \\
    (1)
    & (2)
    & (3)
    & (4)
    & (5)
    & (6)
    & (7)
    & (8)
    & (9)
    & \makecell[c]{(10)}
    & \makecell[c]{(11)}
    & \makecell[c]{(12)}
    & \makecell[c]{(13)}
    & \makecell[c]{(14)}
    & \makecell[c]{(15)} \\
    \hline\hline
    \texttt{I-FID}
    & $1.0$ & $0.0$ & $4$ & \makecell{$4.75$\\$5.5$\\$6.5$} & $25\times25$ & $6$ & $245.76$
    & \makecell{fiducial\\-\\-}
    & \makecell[r]{$0.17$\\$0.17$\\$0.17$}
    & \makecell[r]{$-0.75$\\$-0.74$\\$-0.72$}
    & \makecell[r]{$-0.61$\\$-0.59$\\$-0.56$}
    & \makecell[r]{$-7.88$\\$-7.73$\\$-7.45$}
    & \makecell[r]{$0.00$\\$0.00$\\$0.00$}
    & \makecell[r]{$0.50$\\$0.50$\\$0.50$} \\
    \Xhline{3\arrayrulewidth}
    \texttt{I-LB} 
    & $1.0$ & $0.0$ & $4$ & $4.75$ & $50\times50$ & $7$ & $245.76$
    & larger box 
    & $0.17$ & $-0.83$ & $-0.71$ & $-8.67$ & $0.00$ & $0.50$ \\[-0.2em]
    \texttt{I-HR} 
    & $1.0$ & $0.0$ & $4$ & $4.75$ & $25\times25$ & $7$ & $491.52$
    & higher res.
    & $0.16$ & $-0.76$ & $-0.66$ & $-8.25$ & $0.00$ & $0.50$ \\[-0.2em]
    \texttt{I-SD} 
    & $1.0$ & $0.0$ & $4$ & $4.75$ & $25\times25$ & $6$ & $245.76$
    & $P_{\rm d}=0.1\Omega_{\rm b}^{-1}$
    & $0.16$ & $-0.79$ & $-0.71$ & $-8.69$ & $0.00$ & $0.50$ \\
    \Xhline{3\arrayrulewidth}
    \texttt{I-r$_{\mathtt{s}}$} 
    & $1.0$ & $0.0$ & \makecell{$2$\\$4$\\$8$} & \makecell{$2.75$\\$4.75$\\$8.75$} 
    & $25\times25$ & $6$ & $245.76$
    & \makecell{-\\fiducial\\-}
    & \makecell[r]{$0.10$\\$0.17$\\$0.41$}
    & \makecell[r]{$-0.88$\\$-0.75$\\$-0.53$}
    & \makecell[r]{$-1.63$\\$-0.61$\\$0.18$}
    & \makecell[r]{$-16.01$\\$-7.88$\\$-1.60$}
    & \makecell[r]{$0.00$\\$0.00$\\$0.00$}
    & \makecell[r]{$0.50$\\$0.50$\\$0.50$} \\
    \Xhline{3\arrayrulewidth}
    \texttt{I-q$_{\mathtt{b}}$} 
    & \makecell{$0.1$\\$0.2$\\$0.3$\\$0.4$\\$0.5$\\$0.6$\\$0.7$\\$0.8$\\$0.9$\\$1.0$} 
    & $0.0$ & $4$ & $4.75$ & $25\times25$ & $6$ & $245.76$
    & \makecell{-\\-\\-\\-\\-\\-\\-\\-\\-\\fiducial}
    & \makecell[r]{$0.31$\\$0.23$\\$0.20$\\$0.17$\\$0.17$\\$0.17$\\$0.18$\\$0.17$\\$0.17$\\$0.17$}
    & \makecell[r]{$-0.43$\\$-0.45$\\$-0.44$\\$-0.55$\\$-0.58$\\$-0.62$\\$-0.65$\\$-0.69$\\$-0.74$\\$-0.75$}
    & \makecell[r]{$0.53$\\$0.27$\\$0.11$\\$-0.18$\\$-0.26$\\$-0.34$\\$-0.40$\\$-0.49$\\$-0.58$\\$-0.61$}
    & \makecell[r]{$-1.78$\\$-2.93$\\$-3.42$\\$-5.44$\\$-5.77$\\$-6.13$\\$-6.41$\\$-7.04$\\$-7.68$\\$-7.88$}
    & \makecell[r]{$0.00$\\$0.00$\\$0.00$\\$0.00$\\$0.00$\\$0.00$\\$0.00$\\$0.00$\\$0.00$\\$0.00$}  
    & \makecell[r]{$0.68$\\$0.56$\\$0.51$\\$0.45$\\$0.46$\\$0.48$\\$0.48$\\$0.49$\\$0.50$\\$0.50$} \\
    \Xhline{3\arrayrulewidth}
    \texttt{I-e$_{\mathtt{b}}$} 
    & $1.0$ & \makecell{$0.0$\\$0.1$\\$0.2$\\$0.3$\\$0.4$\\$0.5$} 
    & $4$ & $4.75$ & $25\times25$ & $6$ & $245.76$
    & \makecell{fiducial\\-\\-\\-\\-\\-}
    & \makecell[r]{$0.17$\\$0.17$\\$0.17$\\$0.17$\\$0.18$\\$0.20$}
    & \makecell[r]{$-0.75$\\$-0.69$\\$-0.66$\\$-0.65$\\$-0.64$\\$-0.60$}
    & \makecell[r]{$-0.61$\\$-0.51$\\$-0.45$\\$-0.38$\\$-0.23$\\$0.04$}
    & \makecell[r]{$-7.88$\\$-7.09$\\$-6.68$\\$-6.47$\\$-6.16$\\$-4.89$}
    & \makecell[r]{$0.00$\\$0.02$\\$-0.03$\\$-0.24$\\$-1.00$\\$-1.71$}
    & \makecell[r]{$0.50$\\$0.50$\\$0.50$\\$0.50$\\$0.50$\\$0.50$} \\
    \Xhline{3\arrayrulewidth}
    \texttt{I-ret} 
    & $1.0$ & $0.0$ & $4$ & $4.75$ & $25\times25$ & $6$ & $245.76$
    & retrograde
    & $1.10$ & $-5.38$ & $0.72$ & $-8.75$ & $0.00$ & $0.50$ \\
    \hline
  \end{tabular} \\
  \begin{flushleft}
    {\large N}OTE --- All simulations in this table (i.e. \texttt{Run I} series) adopt $q=1$e-$6$, $h=0.01$, and $\lambda=2.5$ (or equivalently, $c_{\rm s,\infty}/v_{\rm b}=0.633$, $V_{\rm s}/v_{\rm b}=0.380$, $\Delta V_{\rm K}/v_{\rm b}=6.33$e-$3$; see Section \ref{subsec:schemes}).  All runs end at $500\Omega_{\rm b}^{-1}$ and the results in the rightmost six columns are time-averaged over the last $300\Omega_{\rm b}^{-1}$ ($\simeq 38P_{\rm b}'$ in the fiducial case, where $P_{\rm b}'=7.90/\Omega_{\rm b}$).  The results of the fiducial\texttt{-I} run are repeated in some rows to show the trends. \\[0.5em]
    {\large N}OTE ---Columns: 
    (1) run names; 
    (2) binary mass ratio $m_2/m_1$;
    (3) binary eccentricity;  
    (4) sink radius around each accretor;
    (5) evaluation radius;
    (6) computational domain size;
    (7) the number of refinement levels;
    (8) The resolution at the finest refinement level (measured by the number of cells across $a_{\rm b}$); 
    (9) Remarks to explain run names;
    (10) time-averaged accretion rate;
    (11) time-averaged rate of change in binary specific energy;
    (12) accretion eigenvalue;
    (13) binary semimajor axis change rate or migration rate;
    (14) binary eccentricity change rate;
    (15) ratio between the secondary accretion rate and the total accretion rate.
    \end{flushleft}
\end{table*}
\endgroup

\begingroup 
\setlength{\medmuskip}{0mu} 
\setlength\tabcolsep{4pt} 
\setcellgapes{3pt} 
\begin{table*}
  \nomakegapedcells
  \caption{Simulation Setups and Results for \texttt{Run II} Series 
  ($q=1$e-$6$, $h=0.01$, $\lambda=5$)}\label{tab:run_II}
  \makegapedcells 
  \linespread{1.025} 
  \begin{tabular}{lcccccccc|rrrrrr}
    \hline
    \texttt{Run}
    & $q_{\rm b}$
    & $e_{\rm b}$
    & $\displaystyle \frac{r_{\rm s}}{a_{\rm b}}$
    & $\displaystyle \frac{r_{\rm e}}{a_{\rm b}}$
    & $L_X\times L_Y$
    & \footnotesize $N_{\rm SMR}$
    & $\displaystyle \frac{a_{\rm b}}{\delta_{\rm fl}}$
    & Remarks
    & \makecell[c]{$\langle\dot{m}_{\rm b}\rangle$}
    & \makecell[c]{$\langle\dot{\mathcal{E}}_{\rm b}\rangle$}
    & \makecell[c]{$\ell_0$}
    & \makecell[c]{$\displaystyle \frac{\langle\dot{a}_{\rm b}\rangle}{a_{\rm b}}$}
    & \makecell[c]{$\langle\dot{e_{\rm b}^2}\rangle$}
    & \makecell[c]{$\eta$} \\

    &
    &
    & $[\%]$
    & $[\%]$
    & $[a_{\rm b}^2]$
    &
    &
    &
    & \makecell[c]{\footnotesize $[\Sigma_{\infty} v_{\rm b} a_{\rm b}]$}
    & \makecell[c]{\footnotesize $\displaystyle \left[\frac{\Sigma_\infty v_{\rm b}^3 a_{\rm b}}{m_{\rm b}}\right]$}
    & \makecell[c]{$[v_{\rm b} a_{\rm b}]$}
    & \makecell[c]{\small $\displaystyle \left[\frac{\langle\dot{m}_{\rm b}\rangle}{m_{\rm b}}\right]$}
    & \makecell[c]{\small $\displaystyle \left[\frac{\langle\dot{m}_{\rm b}\rangle}{m_{\rm b}}\right]$}
    & \\
    \hline\hline
    \texttt{II-FID}
    & $1.0$ & $0.0$ & $4$ & \makecell{$4.75$\\$5.5$\\$6.5$} & $50\times50$ & $7$ & $245.76$
    & \makecell{fiducial\\-\\-}
    & \makecell[r]{$0.17$\\$0.17$\\$0.17$}
    & \makecell[r]{$-0.48$\\$-0.45$\\$-0.44$}
    & \makecell[r]{$-0.23$\\$-0.17$\\$-0.14$}
    & \makecell[r]{$-4.81$\\$-4.37$\\$-4.15$}
    & \makecell[r]{$0.00$\\$0.00$\\$0.00$}
    & \makecell[r]{$0.49$\\$0.49$\\$0.49$} \\
    \Xhline{3\arrayrulewidth}
    \texttt{II-q$_{\mathtt{b}}$} 
    & \makecell{$0.1$\\$0.2$\\$0.3$\\$0.4$\\$0.5$\\$0.6$\\$0.7$\\$0.8$\\$0.9$\\$1.0$} 
    & $0.0$ & $4$ & $4.75$ & $50\times50$ & $7$ & $245.76$
    & \makecell{-\\-\\-\\-\\-\\-\\-\\-\\-\\fiducial}
    & \makecell[r]{$0.15$\\$0.15$\\$0.14$\\$0.13$\\$0.14$\\$0.15$\\$0.16$\\$0.17$\\$0.17$\\$0.17$}
    & \makecell[r]{$0.40$\\$0.01$\\$-0.02$\\$-0.18$\\$-0.33$\\$-0.37$\\$-0.43$\\$-0.48$\\$-0.48$\\$-0.48$}
    & \makecell[r]{$0.76$\\$0.51$\\$0.49$\\$0.21$\\$-0.02$\\$-0.09$\\$-0.16$\\$-0.21$\\$-0.22$\\$-0.23$}
    & \makecell[r]{$6.46$\\$1.17$\\$0.76$\\$-1.70$\\$-3.59$\\$-4.00$\\$-4.45$\\$-4.78$\\$-4.77$\\$-4.81$}
    & \makecell[r]{$0.00$\\$0.00$\\$0.00$\\$0.00$\\$0.00$\\$0.00$\\$0.00$\\$0.00$\\$0.00$\\$0.00$}
    & \makecell[r]{$0.55$\\$0.50$\\$0.51$\\$0.46$\\$0.47$\\$0.51$\\$0.49$\\$0.50$\\$0.50$\\$0.49$} \\
    \Xhline{3\arrayrulewidth}
    \texttt{II-e$_{\mathtt{b}}$} 
    & $1.0$ & \makecell{$0.0$\\$0.1$\\$0.2$\\$0.3$\\$0.4$\\$0.5$} 
    & $4$ & $4.75$ & $50\times50$ & $7$ & $245.76$
    & \makecell{fiducial\\-\\-\\-\\-\\-}
    & \makecell[r]{$0.17$\\$0.16$\\$0.16$\\$0.17$\\$0.21$\\$0.24$}
    & \makecell[r]{$-0.48$\\$-0.47$\\$-0.42$\\$-0.36$\\$-0.33$\\$-0.33$}
    & \makecell[r]{$-0.23$\\$-0.20$\\$-0.08$\\$0.13$\\$0.40$\\$0.54$}
    & \makecell[r]{$-4.81$\\$-4.67$\\$-4.12$\\$-3.35$\\$-2.21$\\$-1.80$}
    & \makecell[r]{$0.00$\\$-0.09$\\$-0.45$\\$-1.28$\\$-2.30$\\$-2.87$}
    & \makecell[r]{$0.49$\\$0.50$\\$0.49$\\$0.50$\\$0.50$\\$0.49$} \\
    \Xhline{3\arrayrulewidth}
    \texttt{II-ret} 
    & $1.0$ & $0.0$ & $4$ & $4.75$ & $100\times100$ & $8$ & $245.76$
    & retrograde
    & $0.78$ & $-4.22$ & $0.85$ & $-9.79$ & $0.00$ & $0.50$ \\
    \hline    
  \end{tabular} \\
  \begin{flushleft}
    {\large N}OTE --- All simulations in this table (i.e. \texttt{Run II} series) adopt $q=1$e-$6$, $h=0.01$, and $\lambda=5$ (or equivalently, $c_{\rm s,\infty}/v_{\rm b}=0.447$, $V_{\rm s}/v_{\rm b}=0.134$, $\Delta V_{\rm K}/v_{\rm b}=4.47$e-$3$; see Section \ref{subsec:schemes}).  All prograde runs end at $500\Omega_{\rm b}^{-1}$ and the results in the rightmost six columns are time-averaged over the last $240\Omega_{\rm b}^{-1}$ ($\simeq 35P_{\rm b}'$ in the fiducial case, where $P_{\rm b}'=6.86/\Omega_{\rm b}$).  \texttt{Run II-ret} ends at $1000\Omega_{\rm b}^{-1}$ and its results are time-averaged over the last $720\Omega_{\rm b}^{-1}$ to accommodate the violent accretion flows (see Sections \ref{subsec:setups} and \ref{subsec:result_retro}).  The results of the fiducial run are repeated in some rows to show the trends. \\[0.5em]
    {\large N}OTE --- Column names and units are the same as those in Table \ref{tab:run_I}.
  \end{flushleft}
\end{table*}
\endgroup

\subsection{Calculations of Accretion Rate, Torque, Energy Transfer Rate, and Orbital Evolution}
\label{subsec:acc_T}

To evaluate the accretion onto the binary components and to calculate the torque, energy transfer rate and orbital evolution of the binary, we adopt a similar approach as used in \citet[][see their Sections 2 and 4]{Munoz2019}, but with several modifications.

Each binary component in our numerical model is treated as an absorbing sphere (circular boundary) with a sink radius of $r_{\rm s}$.  After integrating the hydrodynamics in each time step, we identify all cells with $|\bm{r}_i - \bm{r}_k| < r_{\rm s}$ (where $i=1,2$ labels the two accretors, and $k$ labels the cell) and set the velocity to zero and density and pressure to tiny values (e.g., $10^{-20} \Sigma_{\infty}$ for $\Sigma_{\rm g}$) in these sink cells.  In this way, we can safely use a very small gravitational softening length ($\xi_{\rm s} = 10^{-8} a_{\rm b}$) and accurately model the accretion flow under the binary potential.

To properly resolve the accretor and its accretion flows, we refine the mesh towards the COM of the binary through multiple levels (each refinement doubles the resolution of the last level) and the finest level encloses the entire binary orbit.  Each accretor is well resolved with $r_{\rm s}/\delta_{\rm fl} = 9.83$ in our fiducial models, where $\delta_{\rm fl}$ is the cell size of the finest level, meeting the resolution criterion suggested by \citet{Xu2019}, i.e., $r_{\rm s}/\delta_{\rm fl} \gtrsim 10$.

To evaluate the accretion-related quantities, we linearly interpolate the conservative variables of \texttt{ATHENA} onto structured polar grid points around each accretor at an evaluation radius $r_{\rm e}$ in each time step, again after the integration of hydrodynamics.  We requires $r_{\rm e}$ to be slightly larger than $r_{\rm s}$ such that the interpolation does not use any sink cells.  For linear interpolation, $r_{\rm e}$ should be larger than $r_{\rm s} + \sqrt{2} \delta_{\rm fl}$.  Section \ref{subsubsec:deps_numParas} and Appendix \ref{appsec:vali_BHL} demonstrate that our evaluations of various quantities have little dependence on the choice of $r_{\rm e}$ as long as it meets the aforementioned requirements.

Along the constructed polar grid points at $r_{\rm e}$ around each accretor (labelled by $i=1,2$), we perform the following integrations to obtain the accretion rate and the specific force due to accretion and pressure:
\begin{align}
  \dot{m}_i &= \oint d\dot{m}_i = \oint (-\Sigma_{\rm g} \bm{u}) \cdot \textnormal{d}\bm{A}, \label{eq:m_dot_i} \\
  \bm{f}_{\mathrm{acc},i} &= \frac{1}{m_i} \oint \textnormal{d}\dot{m}_i (\bm{u} - \bm{v}_{i,\mathrm{SB}}), \label{eq:f_hydro}\\
  \bm{f}_{\mathrm{pres},i} &= \frac{1}{m_i} \oint (-P) \ \textnormal{d}\bm{A}, \label{eq:f_pres}
\end{align}
where $\textnormal{d}\bm{A}$ is the area (line) element around each accretor, $\bm{v}_{i,\mathrm{SB}} = \bm{v}_i + \bm{\Omega}_{\rm pre}\times \bm{r}_i$ is its velocity in the shearing box reference frame. The inclusion of the relative velocity $(\bm{u} - \bm{v}_{i,\mathrm{SB}})$ ensures that $\bm{f}_{\mathrm{acc},i}$ is readily used for calculating the orbital evolution of the binary in the inertial frame (see below).  We store these quantities at the end of each time step throughout the entire simulation.

To calculate the dynamical friction on each accretor (per unit mass), we sum over the specific gravitational forces from all cells outside the evaluation radius, i.e., 
\begin{equation}
  \bm{f}_{\mathrm{grav}, i} = -\sum\limits_{k} G m_k \frac{\bm{r}_i - \bm{r}_k}{|\bm{r}_i - \bm{r}_k|^3}, \label{eq:f_grav}
\end{equation}
where $m_k = \Sigma_{\rm g} \delta^2$ is the gas mass in the $k$-th cell and $\delta$ is the cell size of the finest available level at $\bm{r}_k$ (in other words, in the regions where multiple refinement levels overlap, only the finest level is used in this summation).  For cells intersecting with $r_{\rm e}$, we carefully calculate their sub-cell contributions based on the area percentage that is outside $r_{\rm e}$ (see Appendix \ref{appsec:vali_BHL} for more details).

In practice, to account for the gravitational force to a second-order accuracy, we evaluate $\bm{f}_{\mathrm{grav}, i}$ twice per time step --- before and after updating the binary's positions --- to mimic the kick-drift-kick scheme often used in orbital integration.  We then store the cumulative velocity changes $\Delta \bm{v}_{\mathrm{grav},i}$ from a series of $\bm{f}_{\mathrm{grav}, i} \Delta t/2$, where the time step $\Delta t$ in \texttt{ATHENA} is a varying quantity that satisfies the Courant-Friedrichs-Lewy (CFL) stability condition based on wavespeeds (i.e., of sound waves).  From momentum conservation, the specific gravitational force for each time step can be then obtained with $\bm{f}_{\mathrm{grav}, i} = \textnormal{d} (\Delta \bm{v}_{\mathrm{grav},i}) / \textnormal{d}t$.

The orbital evolution of the binary is governed by the torque and energy transfer rate associated with accretion, pressure and gravitational force.  The net hydrodynamical force (per unit mass) from the gas on each binary component is
\begin{equation}
  \bm{f}_i \equiv \bm{f}_{\mathrm{acc},i} + \bm{f}_{\mathrm{pres},i} + \bm{f}_{\mathrm{grav}, i}.
\end{equation}
The equation of motion for $m_1$ and $m_2$ are $\textnormal{d} \bm{v}_1 / \textnormal{d}t = \bm{F}_{12}/m_1 + \bm{f}_1$ and $\textnormal{d} \bm{v}_2 / \textnormal{d}t = -\bm{F}_{12}/m_2 + \bm{f}_2$, where $\bm{F}_{12}= -G m_1 m_2 \bm{r}_{\rm b} / r_{\rm b}^3$.  
Thus, the time derivatives of the specific binary angular momentum $\bm{\ell}_{\rm b} = \bm{r}_{\rm b} \times \dot{\bm{r}}_{\rm b}$ and energy $\mathcal{E}_{\rm b} = \dot{\bm{r}}_{\rm b}^2 / 2 - G m_{\rm b} / r_{\rm b}$ are \citep{Munoz2019}
\begin{align}
  \dot{\bm{\ell}}_{\rm b} &= \bm{r}_{\rm b} \times (\bm{f}_1 - \bm{f}_2), \label{eq:dot_ell} \\
  \dot{\mathcal{E}}_{\rm b} &= -\frac{G \dot{m}_{\rm b}}{r_{\rm b}} + \dot{\bm{r}}_{\rm b} \cdot (\bm{f}_1 - \bm{f}_2) \label{eq:dot_E}
\end{align}
where $\dot{m}_{\rm b}=\dot{m}_1+\dot{m}_2$. It is also of interest to compute the rate of change of the total binary angular momentum:
\begin{equation}
  \dot{\bm{L}}_{\rm b} = \mu_{\rm b} \dot{\bm{\ell}}_{\rm b} + \dot{\mu}_{\rm b} \bm{\ell}_{\rm b} \equiv \dot{\bm{L}}_{\rm b,acc} + \dot{\bm{L}}_{\rm b,pres} + \dot{\bm{L}}_{\rm b,grav}, \label{eq:dotL_breakdown}
\end{equation}
where we decompose $\dot{\bm{L}}_{\rm b}$ into three parts:
\begin{align}
  \dot{\bm{L}}_{\rm b,acc} &\equiv \mu_{\rm b} \bm{r}_{\rm b} \times (\bm{f}_{\mathrm{acc},1} - \bm{f}_{\mathrm{acc},2}) + \dot{\mu}_{\rm b} (\bm{r}_{\rm b} \times \dot{\bm{r}}_{\rm b}), \label{eq:dotL_acc} \\
  \dot{\bm{L}}_{\rm b,pres} &\equiv \mu_{\rm b} \bm{r}_{\rm b} \times (\bm{f}_{\mathrm{pres},1} - \bm{f}_{\mathrm{pres},2}), \label{eq:dotL_pres} \\
  \dot{\bm{L}}_{\rm b,grav} &\equiv \mu_{\rm b} \bm{r}_{\rm b} \times (\bm{f}_{\mathrm{grav},1} - \bm{f}_{\mathrm{grav},2}). \label{eq:dotL_grav}
\end{align}

Using $\mathcal{E}_{\rm b} = -Gm_{\rm b} / (2 a_{\rm b})$ and $\ell_{\rm b} = \pm \sqrt{G m_{\rm b} a_{\rm b} (1 - e_{\rm b}^2)}$ (the $\pm$ are for prograde/retrograde orbits), we compute the secular rates of change in $a_{\rm b}$ and $e_{\rm b}$ via
\begin{align}
  \frac{\langle\dot{a}_{\rm b}\rangle}{a_{\rm b}} &= -\frac{\langle\dot{\mathcal{E}}_{\rm b}\rangle}{\mathcal{E}_{\rm b}} + \frac{\langle\dot{m}_{\rm b}\rangle}{m_{\rm b}}, \label{eq:aoa_dotE} \\[4pt]
  \frac{2 e_{\rm b} \langle\dot{e}_{\rm b}\rangle}{1 - e_{\rm b}^2} &= -2\frac{\langle\dot{\ell}_{\rm b}\rangle}{\ell_{\rm b}} - \frac{\langle\dot{\mathcal{E}}_{\rm b}\rangle}{\mathcal{E}_{\rm b}} + 2\frac{\langle\dot{m}_{\rm b}\rangle}{m_{\rm b}},
\end{align}
where $\langle\cdots\rangle$ stands for time-averaged quantities.  It is also convenient to define the accretion ``eigenvalue''
\begin{equation}
  \ell_0 \equiv \frac{\langle\dot{L}_{\rm b}\rangle}{\langle\dot{m}_{\rm b}\rangle}
\end{equation}
to represent the accreted angular momentum per unit of accreted mass (see also Equations 7-9 in \citealt{Munoz2020}).  For \textit{circular} binaries, the rate of change in $a_{\rm b}$ can be written as
\begin{equation}
  \frac{\langle\dot{a}_{\rm b}\rangle}{a_{\rm b}} = 2\frac{\langle\dot{m}_{\rm b}\rangle}{m_{\rm b}} \frac{(1 + q_{\rm b})^2}{q_{\rm b}}\frac{\ell_0 - \ell_{0,\mathrm{crit}}(\eta)}{\ell_{\rm b}}, \label{eq:dot_a/a_eta}
\end{equation}
where
\begin{align}
  \eta &\equiv \langle\dot{m}_{\rm 2}\rangle / \langle\dot{m}_{\rm b}\rangle,  \label{eq:eta} \\
  \ell_{0,\mathrm{crit}}(\eta) &\equiv \left[(1-\eta)q_{\rm b} + \eta \right]\frac{\ell_{\rm b}}{1 + q_{\rm b}} - \frac{\ell_{\rm b} q_{\rm b}}{2(1 + q_{\rm b})^2}  \label{eq:ell_0_crit_eta}
\end{align}
is a critical threshold for $\ell_0$.  Therefore, for prograde orbits ($\ell_{\rm b}>0$), $\ell_0 < \ell_{0,\mathrm{crit}}$ is required for binary contraction, and vice versa.

For equal-mass circular binaries, the long-term symmetry results in $\eta = 0.5$ and thus $\ell_{0,\mathrm{crit}} = 3 \ell_{\rm b}/8$.  Eq. \ref{eq:dot_a/a_eta} becomes
\begin{equation}
  \frac{\langle\dot{a}_{\rm b}\rangle}{a_{\rm b}} = 8 \left( \frac{\ell_0}{\ell_{\rm b}} - \frac{3}{8}\right) \frac{\langle\dot{m}_{\rm b}\rangle}{m_{\rm b}}. \label{eq:dot_a/a_l0}
\end{equation}
Furthermore, for equal-mass \textit{eccentric} binaries, we have
\begin{align}
  \frac{\langle\dot{a}_{\rm b}\rangle}{a_{\rm b}} &= 8\frac{\langle\dot{m}_{\rm b}\rangle}{m_{\rm b}}  \frac{\ell_0 - \ell_{0,\mathrm{crit}}(e_{\rm b})}{\ell_{\rm b}}, \label{eq:dot_a/a_e2} \\
  \ell_{0,\mathrm{crit}}(e_{\rm b}) &\equiv \frac{3\ell_{\rm b}}{8} - \frac{\langle\dot{e^2_{\rm b}}\rangle}{\langle\dot{m}_{\rm b}\rangle/m_{\rm b}} \frac{\ell_{\rm b}}{8(1 - e_{\rm b}^2)}.
\end{align}

\begin{figure*}
  \centering
  \includegraphics[width=\linewidth]{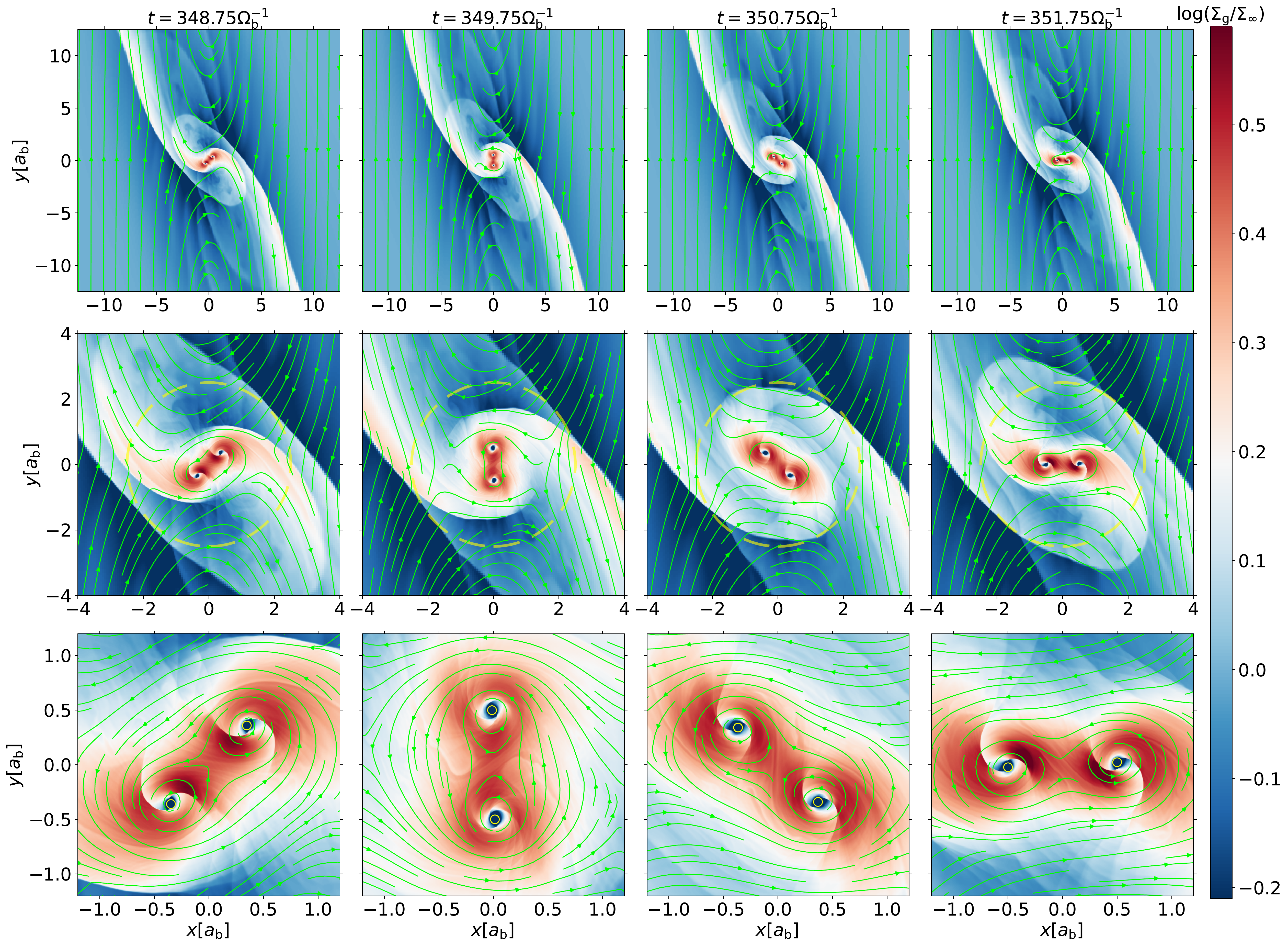}
  \caption{Snapshots for four key phases (from \textit{left} to \textit{right}: $\pi/4$, $\pi/2$, $3\pi/4$, and $\pi$) in the fiducial \texttt{Run I-FID} (see Table \ref{tab:run_I}), where the mesh is refined progressively towards the binary (zooming in from \textit{top} to \textit{bottom}) and the green streamlines show the detailed flow structure.  The \textit{yellow dashed} circles in the middle row with a radius of $R_{\rm H}$ denotes the Hill radius of the binary.  The \textit{yellow solid} circles in the bottom row with a radius of $r_{\rm s} = 0.04 a_{\rm b}$ represent the positions of the binary components.  \label{fig:runI_snapshot}}
\end{figure*}

\begin{figure*}
  \centering
  \includegraphics[width=\linewidth]{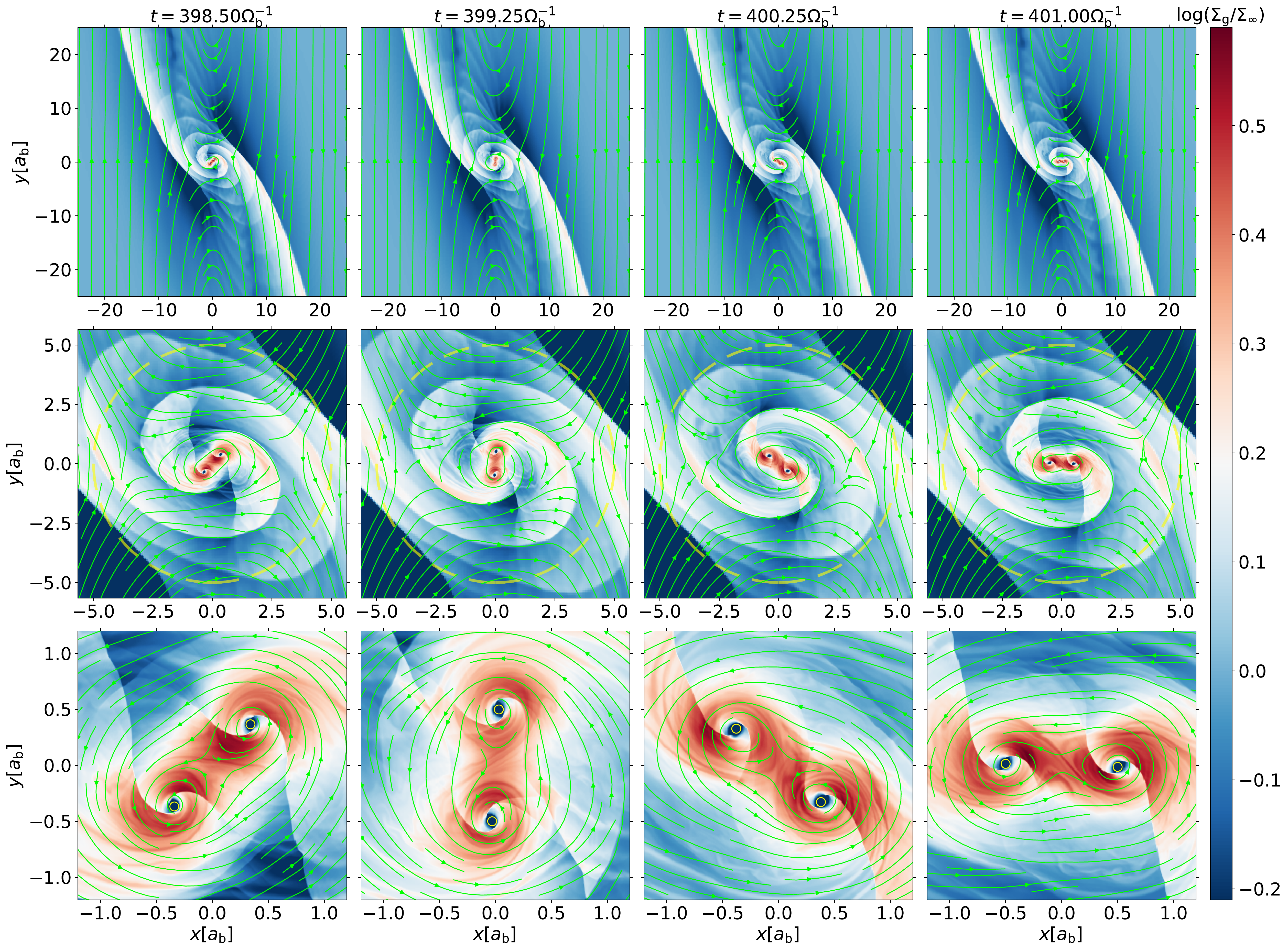}
  \caption{Similar to Fig. \ref{fig:runI_snapshot} but for the fiducial \texttt{Run II-FID} (see Table \ref{tab:run_II}).  The scales of the first two rows are adjusted to accommodate the larger simulation domain, whereas the bottom row has the same scale as that in Fig. \ref{fig:runI_snapshot}.  \label{fig:runII_snapshot}}
\end{figure*}

\subsection{Numerical Parameters}
\label{subsec:setups}

The flow dynamics and results of our simulations depend on the dimensionless parameters ($q$, $h$, and $\lambda$; see Eqs. \ref{eq:q}, \ref{eq:h}, and \ref{eq:lambda}) and the binary parameters ($q_{\rm b}$, $e_{\rm b}$, $\hatomega_{\rm b}$).  In this work, we fix $q=10^{-6}$ and $h=0.01$ and explore $\lambda = 2.5$ in \texttt{Run I} series and $\lambda = 5$ in \texttt{Run II} series.  The corresponding characteristic velocities are $(c_{\rm s,\infty}, V_{\rm s}, \Delta V_{\rm K}) = (0.633, 0.380, 6.33$e-$3)[v_{\rm b}]$ in \texttt{Run I} series and $(0.447, 0.134, \mbox{$4.47$e-$3$})[v_{\rm b}]$ in \texttt{Run II} series.  We hereafter refer to these series directly by \texttt{Run I} and \texttt{Run II}, where the setups below apply to the entire series if not otherwise specified.  \rlr{In \citet[][hereafter \citetalias{Li2022b}]{Li2022b}, we perform more surveys on the EOS, $q/h^3$, and $\lambda$.} 

Tables \ref{tab:run_I} and \ref{tab:run_II} summarize the parameters for \texttt{Run I} and \texttt{Run II}, respectively.  For each series, we first conduct a fiducial run with a pair of prograde equal-mass circular binary (i.e., $q_{\rm b}=1$, $e_{\rm b}=0$, $\hatomega_{\rm b}=\hat{\bm{z}}$; \texttt{Run I-FID} and \texttt{Run II-FID}).  These fiducial simulations adopt a sink radius of $r_{\rm s}=0.04 a_{\rm b}$ with an evaluation radius of $r_{\rm e}=0.0475 a_{\rm b}$ (slightly larger than $r_{\rm s}+\sqrt{2} \delta_{\rm fl}$, see Section \ref{subsec:acc_T}).

The gas in all of our simulations are initialized with $\Sigma_{\rm g} = \Sigma_{\infty}$ and with the velocity given by the background wind profile $\bm{V}_{\rm w}$ (see Eq. \ref{eq:v_wind}).  We set the root computational domain size to $10 R_{\rm H}$ in both $x$ and $y$ directions, i.e., $L_X\times L_Y = (25 a_{\rm b})^2$ in \texttt{Run I} and $(50 a_{\rm b})^2$ in \texttt{Run II}.  Such box sizes ensure that the outer boundaries are sufficiently far away from the binary and the wind profile is only modestly affected by the flow structures near the binary.  Furthermore, these domain sizes are sufficiently small to ensure the good approximation of the local shearing box.

Six SMR levels are employed in \texttt{Run I} to refine the mesh progressively towards the binary, where non-root levels cover $(25a_{\rm b}/2)^2$, $(25 a_{\rm b}/4)^2$, $(25a_{\rm b}/6)^2$, $(25 a_{\rm b}/8)^2$, $(25a_{\rm b}/12)^2$, respectively.  The resolutions are $a_{\rm b}/\delta_{\rm root} = 7.68$ at the root level and $a_{\rm b}/\delta_{\rm fl} = 245.76$ at the finest level.  \texttt{Run II} have the same finest resolution and a similar domain setup with an additional outermost level because of the larger root domain.

We adopt a wave-damping open boundary condition (BC) to handle gas flow at all boundaries and to minimize the effect of artificial boundaries on the flow.  We follow the standard open BC to copy gas quantities from the boundary cells into ghost zones along the normal direction of the boundary.  We impose the shear velocity difference $-q_{\rm sh}\Omega_{\rm K}\Delta x$ outside the radial boundaries, where $\Delta x$ is the directional distance between the ghost zone cell and the boundary cell.  In the wave-damping zone, any hydrodynamical quantity ($\mathcal{U}$) is damped towards its initial value according to
\begin{equation}
  \fracd{}{t} \mathcal{U} = - \frac{\mathcal{U} - \mathcal{U}(t = 0)}{ P_{\rm d}},
\end{equation}
and we choose the wave-damping timescale $P_{\rm d} = 0.02\Omega_{\rm b}^{-1}$ in our canonical runs.

In each simulation, we prescribe the binary orbital motion and evolve the flow dynamics for $500\Omega_{\rm b}^{-1}$, where the accretion rates and torques are measured on-the-fly in each time-step.  The binary orbital evolution is determined by the time-averaged long-term measurements in the post-processing analyses (see Section \ref{subsec:acc_T}).

To examine the numerical robustness of our simulations, we conduct various experiments on our fiducial runs.  We first test the dependence of the evaluation radius with $r_{\rm e}=0.055 a_{\rm b}$ and $0.065 a_{\rm b}$ in both fiducial runs.  Note that $r_{\rm e}$ should not be too far away from $r_{\rm s}$ since the flow dynamics near the accretor can strongly influence the binary orbital evolution.  In addition, we experiment various setups on \texttt{Run I}, including a double-sized root domain in \texttt{Run I-LB}, double resolution in \texttt{Run I-HR}, a five times slower wave-damping time-scale in \texttt{Run I-SD}, and half/double the fiducial sink radius in \texttt{Run I-r$_{\mathtt{s}}$}.
\footnote{The SMR setup in \texttt{Run I-LB} is the same as that in \texttt{Run II-FID} due to the same box size.  \texttt{Run I-HD} has an extra SMR level as the the finest level, which covers $(125a_{\rm b}/96)^2$.}

Based on the fiducial runs, we then survey prograde circular binaries ($e_{\rm b}=0$, $\hatomega_{\rm b}=\hat{\bm{z}}$) with a range of mass ratios ($q_{\rm b}$ from $0.1$ to $1.0$; \texttt{Run I-q$_{\tt b}$} and \texttt{Run II-q$_{\tt b}$}; see Tables \ref{tab:run_I} and \ref{tab:run_II}) and survey prograde equal-mass binaries ($q_{\rm b}=1$, $\hatomega_{\rm b}=\hat{\bm{z}}$) with a series of eccentricities ($e_{\rm b}$ from $0.0$ to $0.5$; \texttt{Run I-e$_{\tt b}$} and \texttt{Run II-e$_{\tt b}$}).  The eccentricity considered here is limited to $0.5$ and below so that the binary orbit is fully covered by the finest mesh block.

Furthermore, we perform two experiments that are similar to our fiducial runs but with a retrograde circular binary ($\hatomega_{\rm b}=-\hat{\bm{z}}$; \texttt{Run I-ret} and \texttt{Run II-ret}).  The apparent orbital velocity of a retrograde binary in the rotating frame is $(\Omega_{\rm b} + \dot{\varpi} + \Omega_{\rm K}) a_{\rm b}$, which is higher than that of its prograde counterpart, $(\Omega_{\rm b} + \dot{\varpi} - \Omega_{\rm K}) a_{\rm b}$.  Thus, the accretion flows in retrograde systems are much more dynamic and turbulent.  Moreover, the binary orbital velocity in \texttt{Run II} series is already supersonic (i.e., $c_{\rm s,\infty}/v_{\rm b} = 0.447$), making the accretion flows in \texttt{Run II-ret} vastly more violent.  To accommodate this situation, we perform \texttt{Run II-ret} in a larger root domain $(100 a_{\rm b})^2$ with a longer evolution time $1000\Omega_{\rm b}^{-1}$ so that reliable time averaging of various quantities can be achieved.

\section{Results}
\label{sec:results}

Tables \ref{tab:run_I} and \ref{tab:run_II} summarize the key parameters and results of our simulation suite.
In Section \ref{subsec:result_fiducial}, we present a detailed analysis of our fiducial runs with prograde equal-mass circular binaries.  Sections \ref{subsec:result_ecc}, \ref{subsec:result_retro}, and \ref{subsec:result_qb} then describe our results for eccentric binaries, retrograde binaries, and unequal-mass binaries, respectively.

\subsection{Prograde Equal-mass Circular Binaries}
\label{subsec:result_fiducial}

Here, we focus on our fiducial runs, \texttt{Run I-FID} and \texttt{Run II-FID}.  These two runs have the same $q$ ($=10^{-6}$) and $h$ ($=0.01$), with the only difference being the value of $\lambda = R_{\rm H}/a_{\rm b}$ ($2.5$ for \texttt{Run I-FID} and $5$ for \texttt{Run II-FID}).  Section \ref{subsubsec:flow_field} describes the accretion flow morphologies.  Section \ref{subsubsec:orbital_evolution} then presents the orbital evolution results, followed by the investigations on how our results depend on various numerical parameters (e.g. resolution) in Sections \ref{subsubsec:deps_numParas} and \ref{subsubsec:r_sink_deps}.

\subsubsection{Flow Structure}
\label{subsubsec:flow_field}

Figs. \ref{fig:runI_snapshot} and \ref{fig:runII_snapshot} show the snapshots of our fiducial runs in the quasi-steady state, where the binaries are at four key orbital phases, namely $\pi/4$, $\pi/2$, $3\pi/4$, and $\pi$.  The snapshots at the next key phase $5\pi/4$ are similar to those at $\pi/4$ due to the symmetry of the equal-mass circular binaries.  For each orbital phase, we progressively zoom into the binary to show the detailed accretion flows and demonstrate that multiple SMR levels are employed to refine the mesh towards the binary (see Section \ref{subsec:setups}).

The quasi-steady state flows in the two fiducial cases are overall similar, where prograde circum-single discs (CSDs) form around each binary component and are embedded in the prograde circumbinary flows.  Grand spirals originated from the circumbinary flows extend all the way to the $\pm y$ boundaries along the shear flow directions.  They are in fact large half bow shocks of the binary accretion with upstream gradients due to the shear.  Although $V_{\rm s}$ is sub-sonic in both fiducial runs, both $v_{\rm b}$ and $w_{\rm b}'$ are supersonic (see Eqs. \ref{eq:c_s_v_b} and \ref{eq:w_bprime}).  Specifically, the Mach number ($\mathcal{M}a$) for the relative velocity of binary components with respect to the shear flow in the rotating frame at phase $\pi$ is $0.93$ and $1.17$ in \texttt{Run I-FID} and \texttt{Run II-FID}, respectively.  The accretion flows near the binary in the latter run are thus more chaotic, leading to slightly smaller CSDs.

Besides the grand spirals, there are horseshoe flows and the inner/outer shear flows (also known as disc flows) around the binary.  Such flow structures are similar to those observed in previous studies of a single accretor, i.e., circumplanetary discs \citep[e.g.,][]{Fung2015, Zhu2016, Kuwahara2019, Bailey2020}.  That said, the flow close to the binary is much more dynamic.  Both CSDs contain two spiral shocks that drive accretion throughout the disc.  Each CSD is then encompassed and attached by a small half bow shock, the tail of which is slingshot away along each grand spiral once a binary orbit.  Fig. \ref{fig:runI_snapshot} shows the propagation of such waves, which gradually fade away in the horseshoe streams.  Similar waves are more prominent in Fig. \ref{fig:runII_snapshot} due to the higher binary orbital frequency in \texttt{Run II-FID} and the larger domain available for wave propagation.

\begin{figure*}
  \centering
  \includegraphics[width=0.951\linewidth]{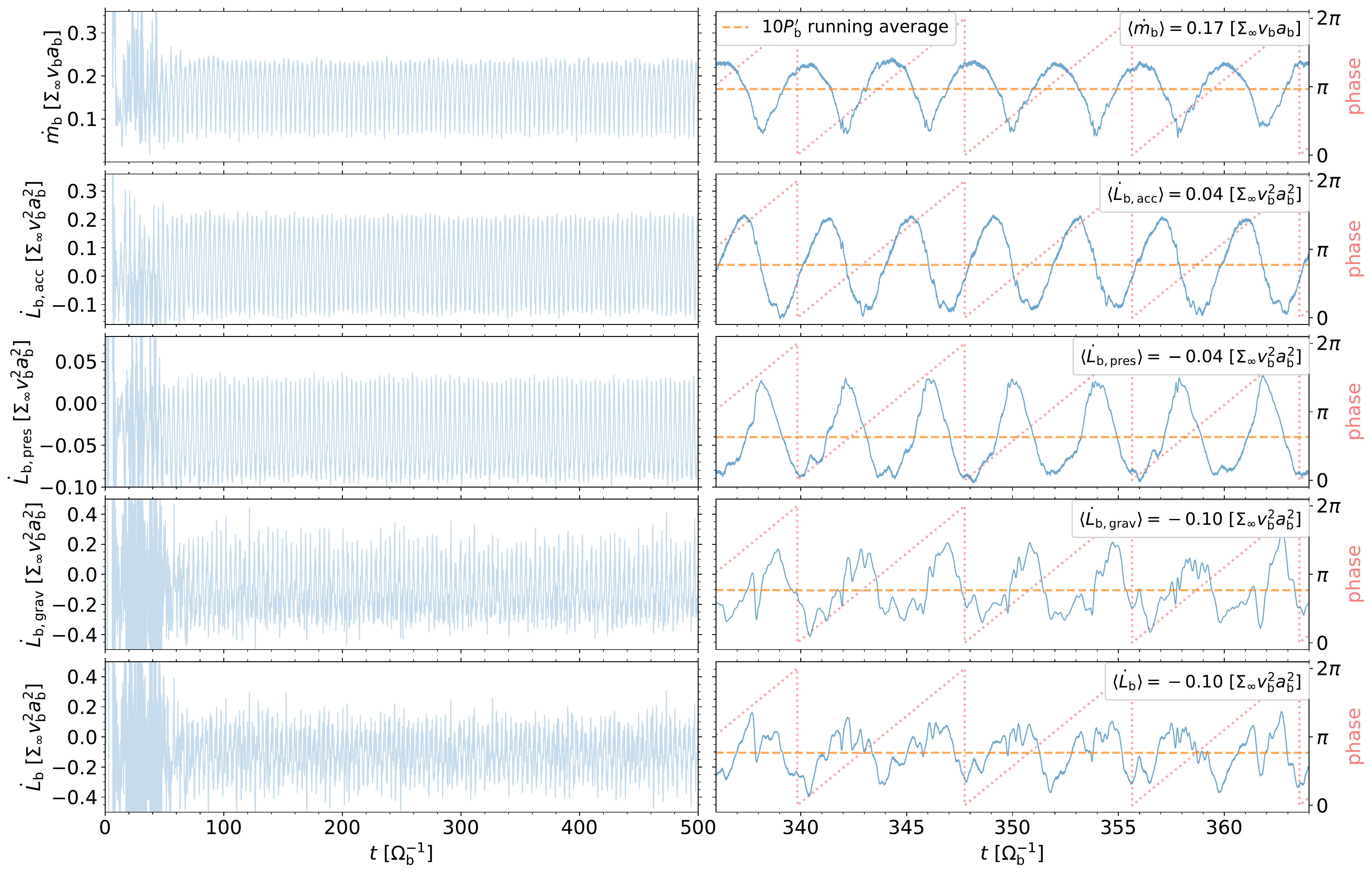}
  \includegraphics[width=0.951\linewidth]{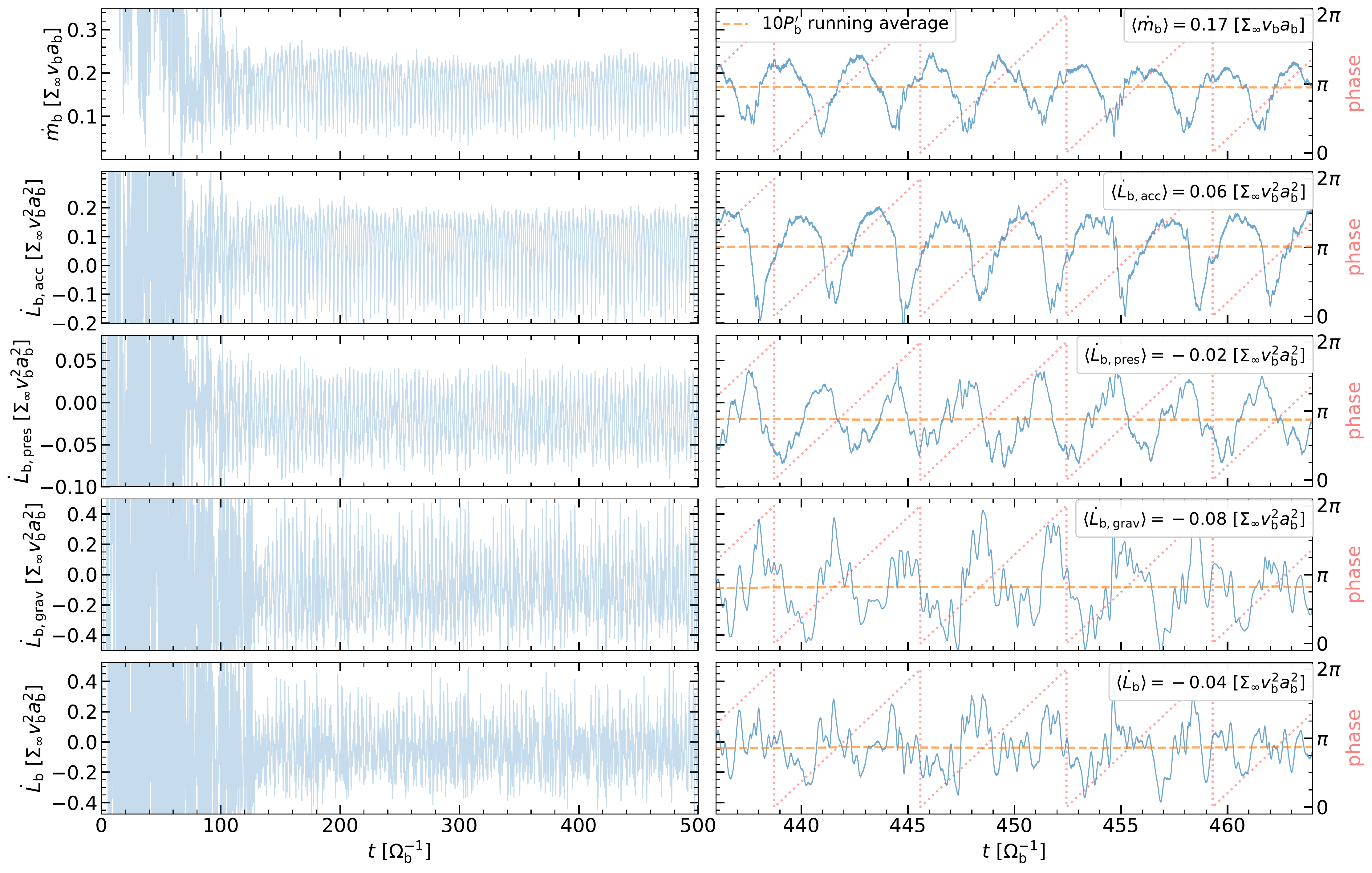}
  \caption{Full (\textit{left}) and a slice of (\textit{right}) time series of accretion and torques (from \textit{top} to \textit{bottom}: $\dot{m}_{\rm b}$, $\dot{L}_{\rm b,acc}$, $\dot{L}_{\rm b,pres}$, $\dot{L}_{\rm b,grav}$, $\dot{L}_{\rm b,\dot{\mu}_{\rm b}}$, and $\dot{L}_{\rm b}$; see Section \ref{subsec:acc_T}) for the prograde equal-mass circular binary in our fiducial \texttt{Run I-FID} (\textit{upper}; see Table \ref{tab:run_I}) and \texttt{Run II-FID} (\textit{lower}; see Table \ref{tab:run_II}).  Each slice of time series (\textit{blue solid}) is accompanied by the binary orbital phase curve with period $P_{\rm b}' = 2\pi / \Omega_{\rm b}'$   
  (\textit{pink dotted}) and the $10 P_{\rm b}'$ running average (\textit{orange dashed}), with the time-averaged value in the legend (averaged over the last $300\Omega_{\rm b}^{-1}$ or $\approx 38P_{\rm b}'$ for \texttt{Run I-FID} and over the last $240\Omega_{\rm b}^{-1}$ or $\approx 35P_{\rm b}'$ for \texttt{Run II-FID}).  The running average curves are in good agreement with the time-averaged results, indicating that all quantities have reached their quasi-steady state values. 
  \label{fig:runI+II_breakdown}}
\end{figure*}

\begin{figure*}
  \centering
  \includegraphics[width=0.8\linewidth]{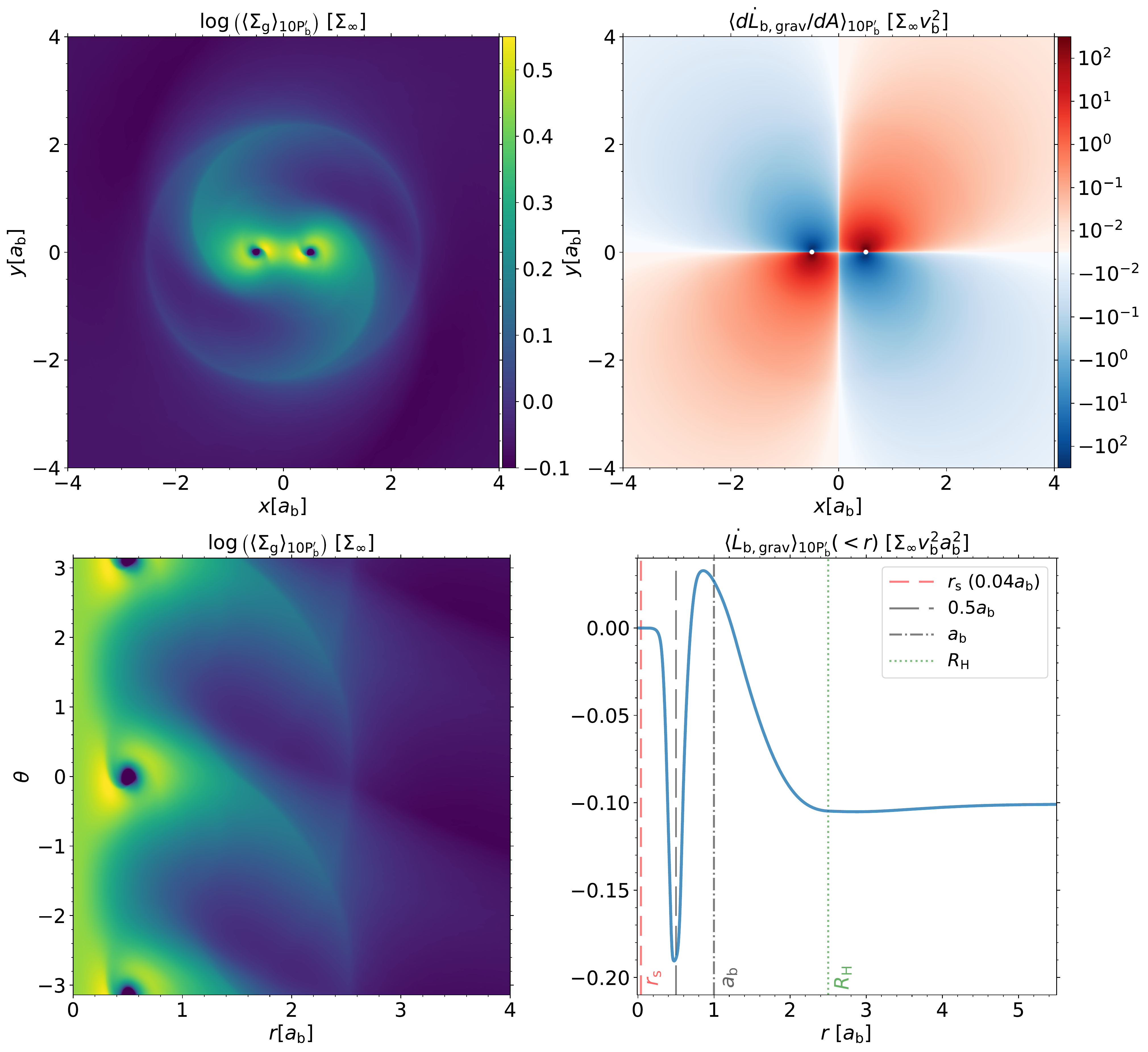}
  \caption{Time-averaged maps of gas density in the Cartesian coordinates (\textit{upper left}) and in the polar coordinates (\textit{lower left}), the gravitational torque surface density (\textit{upper right}), and the cumulative gravitational torque as a function of radius (\textit{lower right}) for our fiducial run \texttt{Run I-FID}.  Characteristic radial scales are marked by the vertical lines in the last panel, indicating the sink radius (\textit{red short dashed}), half binary separation (\textit{black medium dashed}), binary separation (\textit{black long dashed}), and Hill radius (\textit{green dotted}).
  \label{fig:runI_Tgrav_map}}
\end{figure*}

\begin{figure}
  \centering
  \includegraphics[width=\linewidth]{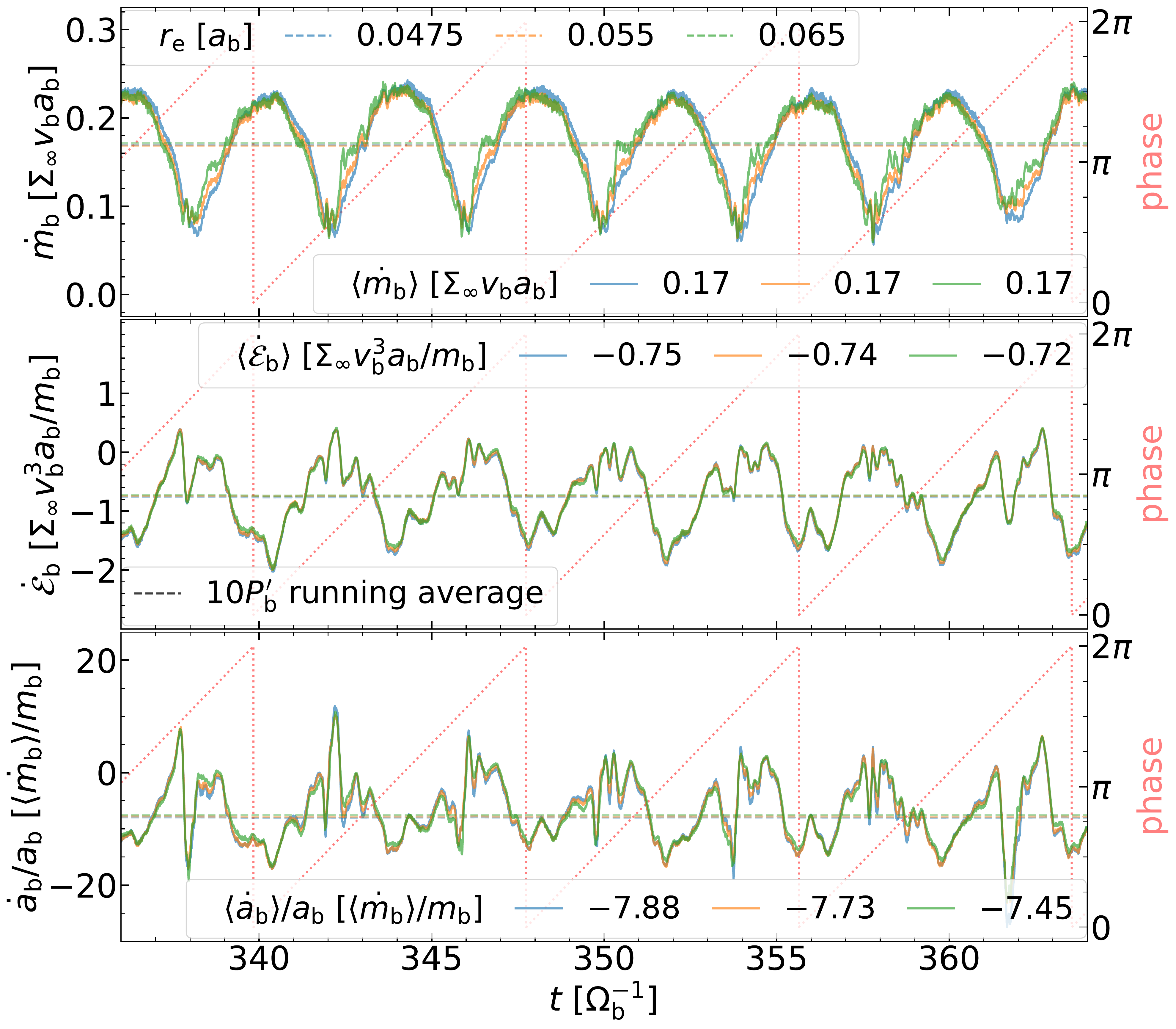}
  \caption{Similar to Fig. \ref{fig:runI+II_breakdown} but focus on the accretion rate ($\dot{m}_{\rm b}$; \textit{upper}), rate of change in specific binary energy ($\dot{\mathcal{E}}_{\rm b}$; \textit{middle}), and orbital decay rate ($\dot{a}_{\rm b}$; \textit{lower}) measured at different evaluation radii ($r_{\rm e}$).  Each time series (\textit{solid}) are color-coded by $r_{\rm e}$, with the corresponding running averages (\textit{dashed}) overplotted and time-averaged values indicated in the legend (averaged over the last $300\Omega_{\rm b}^{-1}$).  The relative changes in the time-averaged values of $\dot{m}_{\rm b}$, $\dot{\mathcal{E}}_{\rm b}$, and $\dot{a}_{\rm b}/a_{\rm b}$ for different $r_{\rm e}$ are $\lesssim 2$ per cent of their average peak-to-trough depths (or variation amplitudes), indicating that the results are robust against different $r_{\rm e}$.
  \label{fig:runI_r_eval}}
\end{figure}

\begin{figure*}
  \centering
  \includegraphics[width=\linewidth]{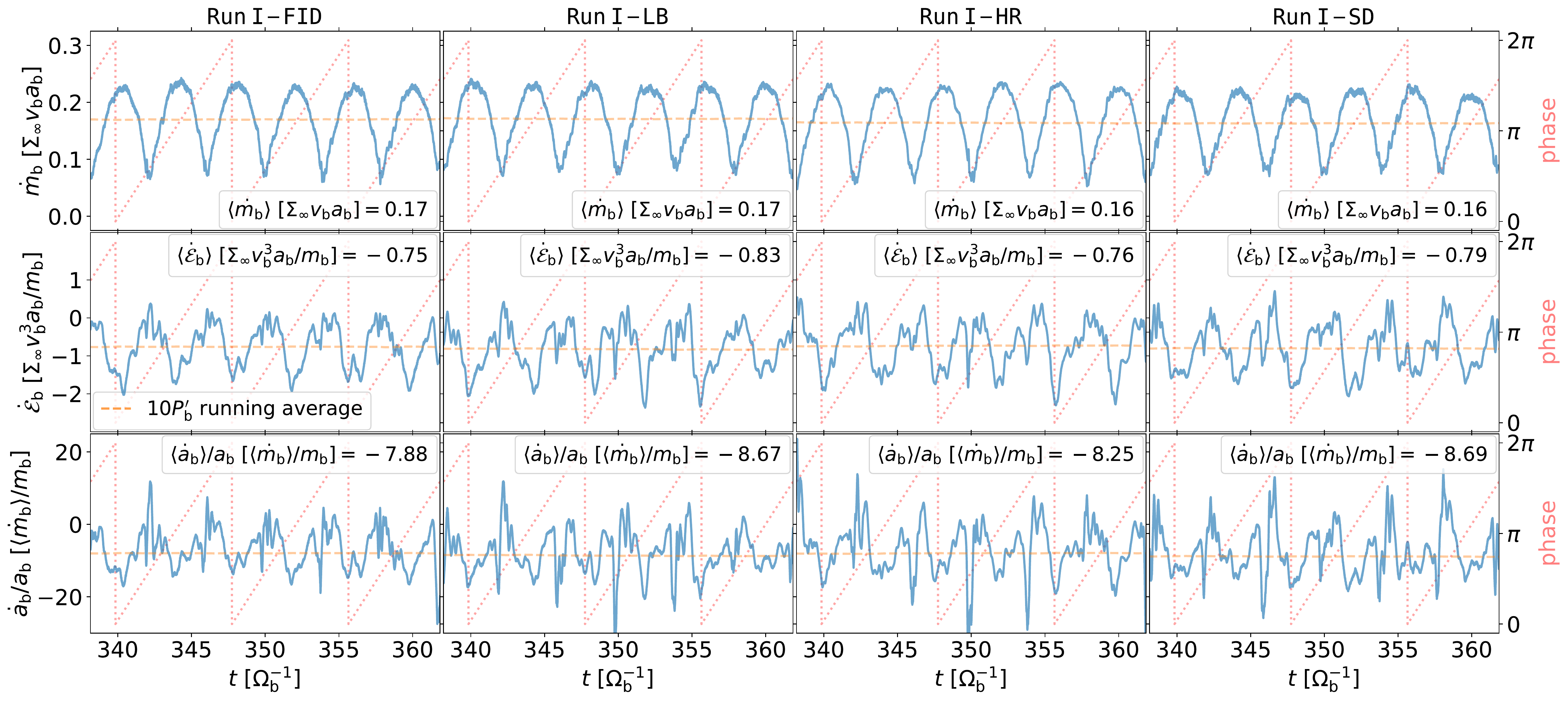}
  \caption{Similar to Fig. \ref{fig:runI_r_eval} but compares the fiducial \texttt{Run I-FID} (\textit{left}) to runs with different numerical setups: a double-sized root domain (\textit{middle left}, \texttt{Run I-LB}), double resolution in the finest SMR level (\textit{middle right}, \texttt{Run I-HR}), and a five times slower wave-damping time-scale (\textit{right}, \texttt{Run I-SD}), all using the fiducial evaluation radius ($r_{\rm e}=0.0475 a_{\rm b}$).  Each time-series (\textit{blue solid}) is overplotted with the corresponding running averages (\textit{orange dashed}), and again with the time-averaged value shown in the legend (averaged over the last $300\Omega_{\rm b}^{-1}$).  The relative changes in $\langle\dot{m}_{\rm b}\rangle$, $\langle\dot{\mathcal{E}}_{\rm b}\rangle$, and $\langle\dot{a}_{\rm b}\rangle/a_{\rm b}$ between these runs are $\lesssim 3$ per cent of their average variation amplitudes, indicating that our results have little dependency on these numerical aspects.
  \label{fig:runI_num_tests}}
\end{figure*}

\subsubsection{Secular Evolution of Binary}
\label{subsubsec:orbital_evolution}

As discussed in Section \ref{subsec:acc_T}, we compute the binary orbital evolution by recording the time series of accretion rate and hydrodynamical and gravitational forces for each accretor.  The derived total accretion rate ($\dot{m}_{\rm b}$), rate of change in mass ratio ($\dot{q}_{\rm b}$), and total and decomposed torques ($\dot{L}_{\rm b}$, $\dot{L}_{\rm b,acc/pres/grav}$; see Eqs. \ref{eq:dotL_acc}, \ref{eq:dotL_pres}, and \ref{eq:dotL_grav}) of \texttt{Run I-FID} are show in Fig. \ref{fig:runI+II_breakdown} for both the full simulation and a selected short period.

We find that all these quantities exhibit stable periodic variations after $\sim 60\Omega_{\rm b}^{-1}$, implying that the flow in \texttt{Run I-FID} has become quasi-steady after that time.  Fig. \ref{fig:runI+II_breakdown} further shows that these variations rigorously follow the orbital phase determined by $\Omega_{\rm b}'$ ($P'=7.90 \Omega_{\rm b}^{-1}$; see Eq. \ref{eq:Omega_bprime}; see also periodogram analyses in Section \ref{subsec:result_ecc}) with constant running averages, which are consistent with the quasi-steady state.

The accretion rate $\dot{m}_{\rm b}$ varies relatively smoothly and has two evenly-spaced peaks/troughs within each binary orbit due to the symmetry of the equal-mass circular binaries.  The binary accrete fastest at phase $\pi$ and $2\pi$ with a slight lag.  Such orbital phases correspond to the times when the binary components reach farthest into the shear flow such that more incoming materials are available for accretion (see the last column in Figs. \ref{fig:runI_snapshot} and \ref{fig:runII_snapshot}).  However, the accreted gas first falls into the CSDs before sinking into each accretor, causing the slight phase lag in the peak positions of $\dot{m}_{\rm b}$ relative to the binary orbital phase.  Similarly, at phase $\pi/2$ and $3\pi/2$ with a slight lag, the binary accretes slowest due to the nearly-negligible shear flows along $x = 0$ (see the second column in Figs. \ref{fig:runI_snapshot} and \ref{fig:runII_snapshot}).

Fig. \ref{fig:runI+II_breakdown} also shows that the accretion-related torques ($\dot{L}_{\rm b,acc}$ and $\dot{L}_{\rm b,pres}$) largely share the periodic variations and the smoothness of $\dot{m}_{\rm b}$.  Consequently, the aperiodic fluctuations in $\dot{L}_{\rm b}$ is mainly due to $\dot{L}_{\rm b,grav}$, which is affected by the flow dynamics in the CSDs, the circumbinary flows, and the small half bow shocks.

In addition, there appear to be phase shifts in $\dot{L}_{\rm b,acc}$, $\dot{L}_{\rm b,pres}$, and $\dot{L}_{\rm b,grav}$ with respect to $\dot{m}_{\rm b}$.  Specifically, $\dot{L}_{\rm b,acc}$ lags roughly $\pi/4$ behind $\dot{m}_{\rm b}$ since $f_{\mathrm{acc},i}$ depends on both the accretion rate and the relative velocity $\bm{u} - \bm{v}_{i,\mathrm{SB}}$ (see Eq. \ref{eq:f_hydro}).  We find that faster accretion results in a stronger negative torque originated from the pressure force.  In other words, $\dot{L}_{\rm b,pres}$ lags $\pi/2$ and is visually inverted relative to $\dot{m}_{\rm b}$.  Interestingly, $\dot{L}_{\rm b,grav}$ seems to lag even more.  It becomes most negative between phase $0$ and $\pi/4$, where the two trailing small half bow shocks are the most prominent and extended since the binary is moving against the shear flow.  The gravitational torque becomes most positive after a quarter of the binary orbital period, where the two small half bow shocks are largely suppressed and the CSDs dominate $\dot{L}_{\rm b,grav}$.  Furthermore, we note that all these decomposed torques are comparable in order of magnitude and are therefore non-negligible contributors to the total torque.

To evaluate the secular evolution of the binary, we perform a conservative time-average over the last $300 \Omega_{\rm b}^{-1}$ (hereafter the default time-average period for the \texttt{Run I} Series) for each time series (see Table \ref{tab:run_I}).  We find that $\langle\dot{q}_{\rm b}\rangle = 0.00$, as expected for such an equal-mass binary.  The total accretion rate and eigenvalue of the accretion flow are
\begin{align}
  \langle\dot{m}_{\rm b}\rangle &\simeq 0.17\ \Sigma_{\infty} v_{\rm b} a_{\rm b}, \\
  \ell_0 &= \frac{\langle\dot{L}_{\rm b}\rangle}{\langle\dot{m}_{\rm b}\rangle}
  \simeq -0.61\ v_{\rm b} a_{\rm b},
\end{align}
which indicates an inspiral binary orbit (see Eq. \ref{eq:dot_a/a_l0}) with a decay rate of
\begin{equation}
  \frac{\langle\dot{a}_{\rm b}\rangle}{a_{\rm b}} \simeq -7.88 \frac{\langle\dot{m}_{\rm b}\rangle}{m_{\rm b}} \simeq -1.34 \frac{\Sigma_\infty v_{\rm b} a_{\rm b}}{m_{\rm b}}.  \label{eq:adot_a_RunI-FID}
\end{equation}
For such a circular binary, the same decay rate can be also derived from $\dot{\mathcal{E}}_{\rm b}$ using Eq. \ref{eq:aoa_dotE}, with $\dot{\mathcal{E}}_{\rm b}$ computed from Eq. \ref{eq:dot_E}.  Moreover, the time series of $\dot{\mathcal{E}}_{\rm b}$ (not shown) shares the exact same shape as that of $\dot{L}_{\rm b}$.

We apply similar analysis to \texttt{Run II-FID}.  The flow around the binary becomes quasi-steady after $\sim 120 \Omega_{\rm b}^{-1}$, a bit later than \texttt{Run I-FID}, due to the faster binary orbital frequency ($P'=6.86 \Omega_{\rm b}^{-1}$) and the more chaotic flow dynamics.  Consequently, the accretion rate oscillates faster with conspicuous small fluctuations as well as varying trough depths.  These fluctuations are however stochastic and possess little power in the frequency space (see Section \ref{subsec:result_ecc}).  Still, we find that the accretion-related torques share both the periodic variations and the fluctuations of $\dot{m}_{\rm b}$.  The average variation amplitude of $\dot{L}_{\rm b,grav}$ almost doubles that in \texttt{Run I-FID}, serving as another indication of the turbulent accretion flows.  The phase shift between $\dot{L}_{\rm b,grav}$ and $\dot{m}_{\rm b}$ also differs because of the more complex circumbinary flow and the swirling small half bow shocks.   Nevertheless, the nearly constant running averages in Fig. \ref{fig:runI+II_breakdown} validate that the flow is quasi-steady in the long run despite the short-term aperiodic fluctuations. 

Given the longer initial time needed to reach the quasi-steady time in \texttt{Run II-FID}, we perform the time-average over the last $240 \Omega_{\rm b}^{-1}$ (hereafter the default time-average period for the \texttt{Run II} Series, unless otherwise specified).  We find
\begin{align}
  \langle\dot{m}_{\rm b}\rangle &\simeq 0.17\ \Sigma_{\infty} v_{\rm b} a_{\rm b}, \\
  \ell_0 &\simeq -0.23\ v_{\rm b} a_{\rm b}, \\
  \frac{\langle\dot{a}_{\rm b}\rangle}{a_{\rm b}} &\simeq -4.81 \frac{\langle\dot{m}_{\rm b}\rangle}{m_{\rm b}} \simeq -0.80 \frac{\Sigma_\infty v_{\rm b} a_{\rm b}}{m_{\rm b}}.  \label{eq:adot_a_RunII-FID}
\end{align}
The accretion rate is similar to that in \texttt{Run I-FID}, but the total torque is weaker, leading to a slower orbital decay rate.

\subsubsection{Spatial Distribution of Gravitational Torque}
\label{subsubsec:runI_Tgrav_map}

We are interested in the gravitational torque since $\langle\dot{L}_{\rm b,grav}\rangle$ is the largest contributor to the total torque $\langle\dot{L}_{\rm b}\rangle$ in both fiducial cases.  To better understand how the flow exerts dynamical friction on the binary, we follow \citet{Munoz2019} and construct the time-averaged maps for the gas surface density and gravitational torque surface density for \texttt{Run I-FID} in Fig. \ref{fig:runI_Tgrav_map}.  These maps are only averaged over the last $10 P_{\rm b}'$ ($\approx 79 \Omega_{\rm b}^{-1}$; $4$ snapshots per $\Omega_{\rm b}^{-1}$) because such calculations require angular differential corrections and are thus computational expensive.  However, we argue that these maps are accurate enough to fulfil our purpose given that the $10 P_{\rm b}'$ running average is in good agreement with $\langle\dot{L}_{\rm b,grav}\rangle$ (averaged from the last $300 \Omega_{\rm b}^{-1}$) as shown in Fig. \ref{fig:runI+II_breakdown}.

The time-averaged map for $\Sigma_{\rm g}$ exhibits evident CSDs, circumbinary flow, and small half bow shocks that extend out with increasing pitch angle and eventually form a circle at roughly $R_{\rm H}$ from the COM of the binary, corresponding to the launching points of the grand spirals.  These structures are persistent non-axisymmetric features that have long-term influence on the binary.  On the contrary, the grand spirals are averaged out since they are continuously rotating from the binary's perspective and are not persistent.  

The time-averaged map for $\textnormal{d}\dot{L}_{\rm b,grav}/\textnormal{d}A$, where $\textnormal{d}A$ denotes unit surface area, shows that the torque density is the strongest near each accretor and drops rapidly far away, as one would expect for gravitational torques.  The positive and negative torques are nearly symmetric, with slight negative excesses matching the geometry of the small half bow shocks as seen in the map of $\langle\Sigma_{\rm g}\rangle_{\rm 10P_{\rm b}'}$.

To determine the relative contributions of these persistent non-axisymmetric features,  we further transform these maps into polar coordinates centered at the binary COM and calculate the radially cumulative gravitational torque
\begin{equation}
  \langle\dot{L}_{\rm b,grav}\rangle_{\rm 10P_{\rm b}'}(<r) = \int_0^{r} \left\langle\frac{\textnormal{d}\dot{L}_{\rm b,grav}}{\textnormal{d}A}\right\rangle_{\rm 10P_{\rm b}'} \textnormal{d}A,
\end{equation}
as shown in Fig. \ref{fig:runI_Tgrav_map}.  We find that the persistent structures that immediately lag behind the binary in the azimuthal direction lead to negative torques, including the part of the CSDs inside the binary orbit ($0.2 a_{\rm b} \lesssim r < 0.5 a_{\rm b}$) and the circumbinary flows with the trailing small half bow shocks ($0.8 a_{\rm b} \lesssim r \lesssim R_{\rm H}$).  The part of the CSDs outside the binary orbit ($0.5 a_{\rm b} < r \lesssim 0.8 a_{\rm b}$), however, contributes positive torques due to the azimuthally leading positions.  The net torque of the whole CSDs is slightly positive ($0.2 a_{\rm b} \lesssim r \lesssim 0.8 a_{\rm b}$) and is countered by the torques further out.  Towards larger radii, the cumulative gravitational torque becomes almost constant after a small increase from $\sim R_{\rm H}$ to $\sim 5 a_{\rm b}$, where the final value agrees with $\langle\dot{L}_{\rm b,grav}\rangle = -0.10\ \Sigma_{\infty} v_{\rm b}^2 a_{\rm b}^2$.  Our findings therefore indicate that, it is the circumbinary flows and the small half bow shocks within $R_{\rm H}$ that largely determine the negative time-averaged total gravitational torque.

\begin{figure}
  \centering
  \includegraphics[width=0.99\linewidth]{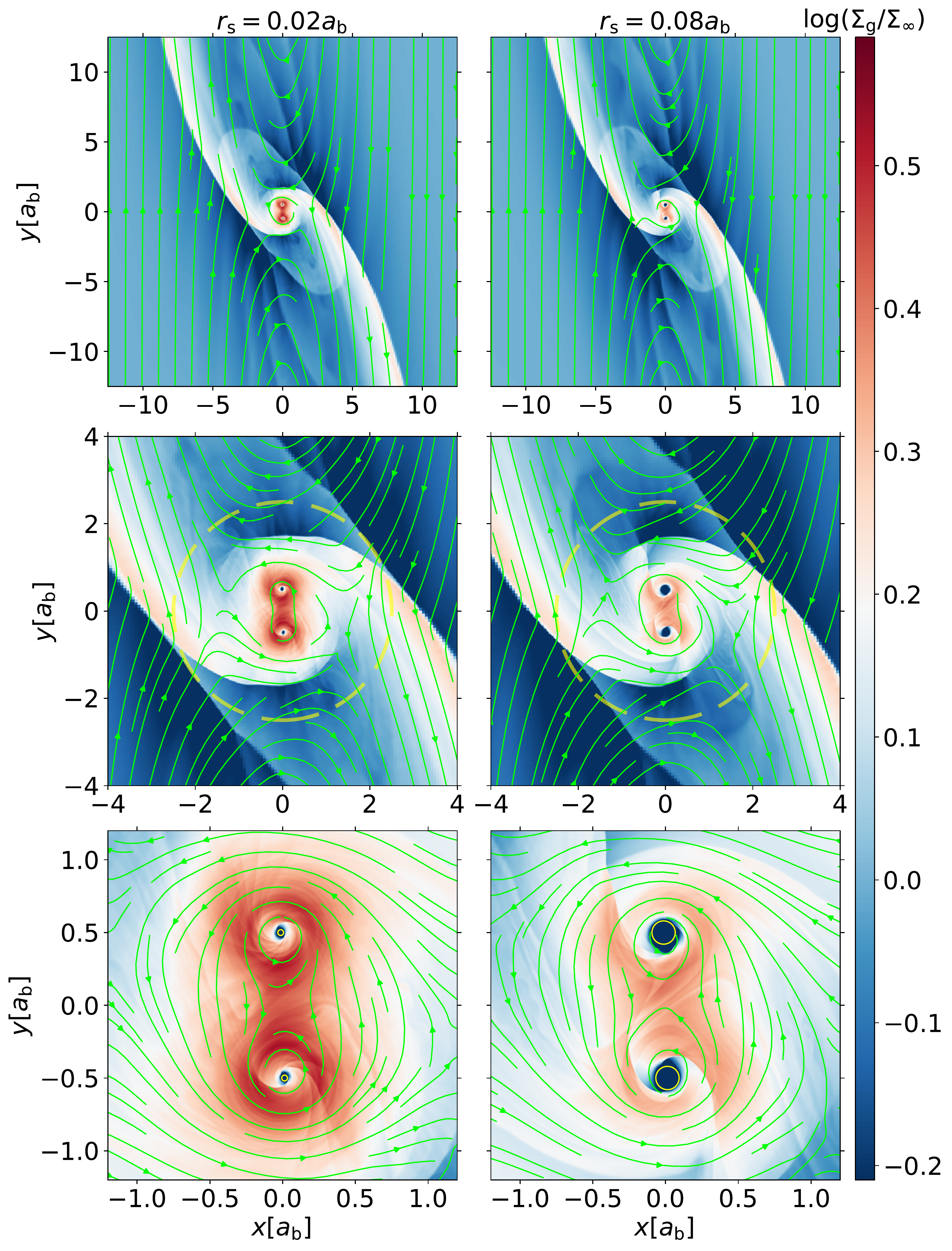}
  \caption{Similar to Fig. \ref{fig:runI_snapshot} but showing the flow structures for \texttt{Run I-r$_\mathtt{s}$} with $r_{\rm s}=0.02 a_{\rm b}$ and $r_{\rm s}=0.08 a_{\rm b}$ at phase $\pi/2$, respectively (see Table \ref{tab:run_I}).
  \label{fig:runI_snapshot_rs}}
\end{figure}

\begin{figure}
  \includegraphics[width=0.945\linewidth]{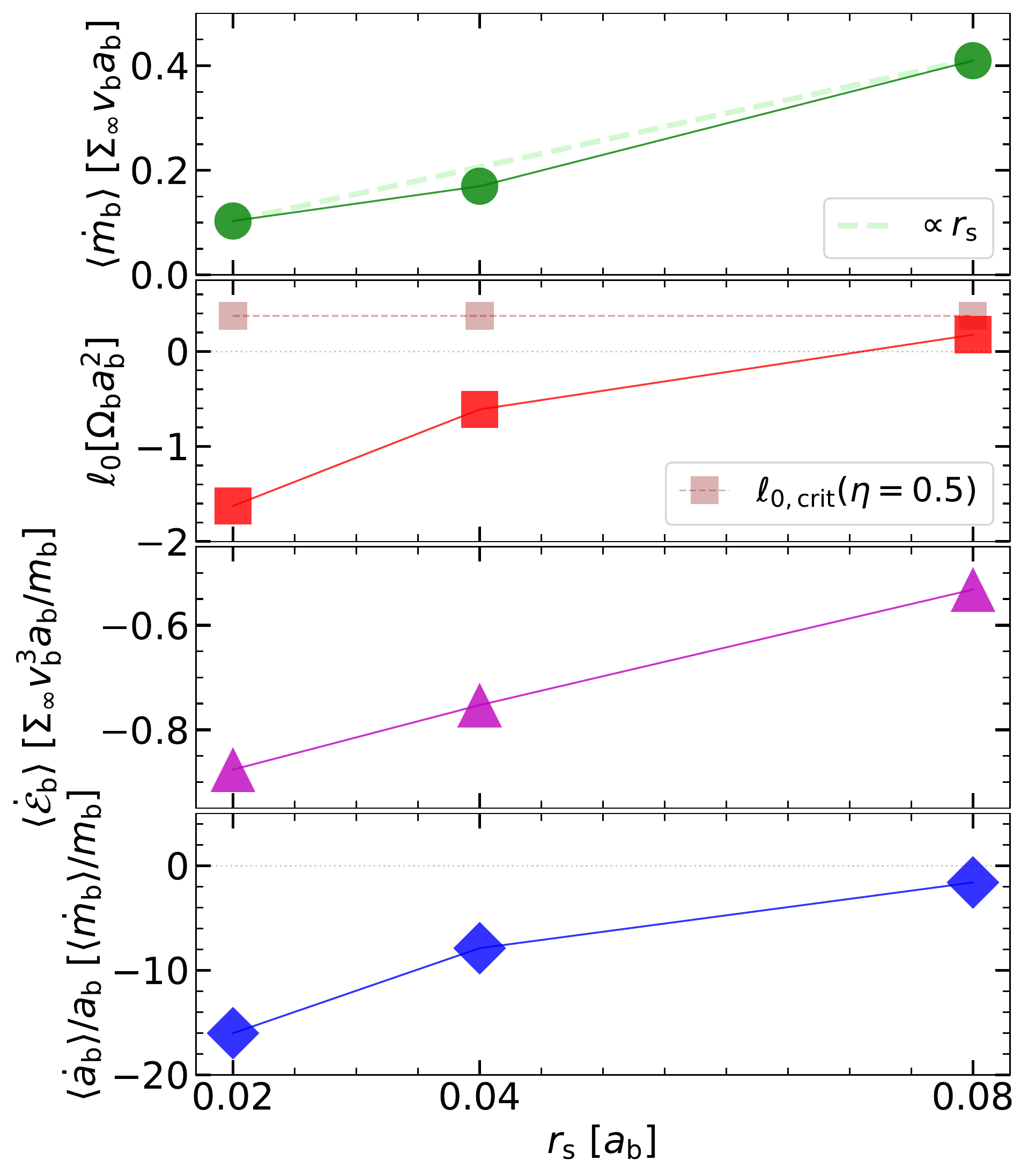}
  \caption{Time averaged (from \textit{top} to \textit{bottom}) accretion rate $\langle\dot{m}_{\rm b}\rangle$, accretion eigenvalue $\ell_0$, rate of change in binary specific energy $\langle\dot{\mathcal{E}}_{\rm b}\rangle$, and binary migration rate $\langle\dot{a}_{\rm b}\rangle/a_{\rm b}$ as a function of sink radius from simulations \texttt{Run I-r$_{\mathtt{s}}$} (see Table \ref{tab:run_I}).
  \label{fig:RunI_trends_r_s}}
\end{figure}

\begin{figure*}
  \centering
  \includegraphics[width=\linewidth]{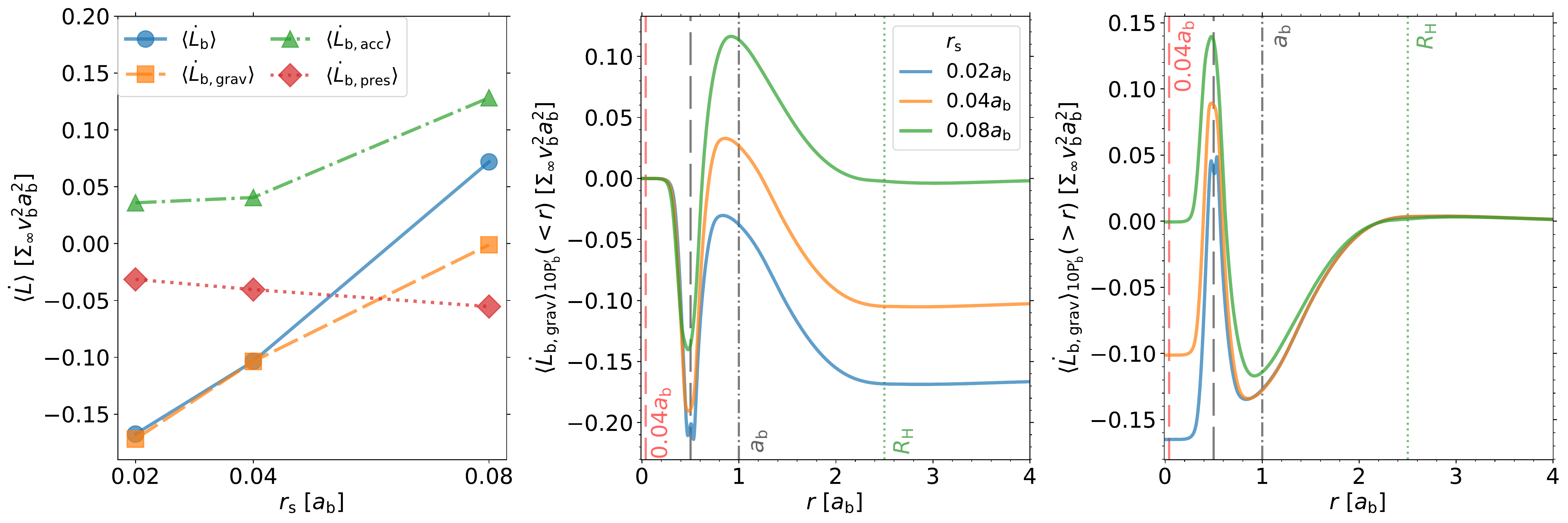}
  \caption{Comparison of time-averaged torques for different sink radii $r_{\rm s}$ (see \texttt{Run I-r$_\mathtt{s}$} in Table \ref{tab:run_I}), showing the total and decomposed torques (\textit{left}), the (complementary) cumulative gravitational torque as a function of radius (\textit{center/right}).
  \label{fig:torque_vs_rs}}
\end{figure*}

\begin{figure*}
  \centering
  \includegraphics[width=\linewidth]{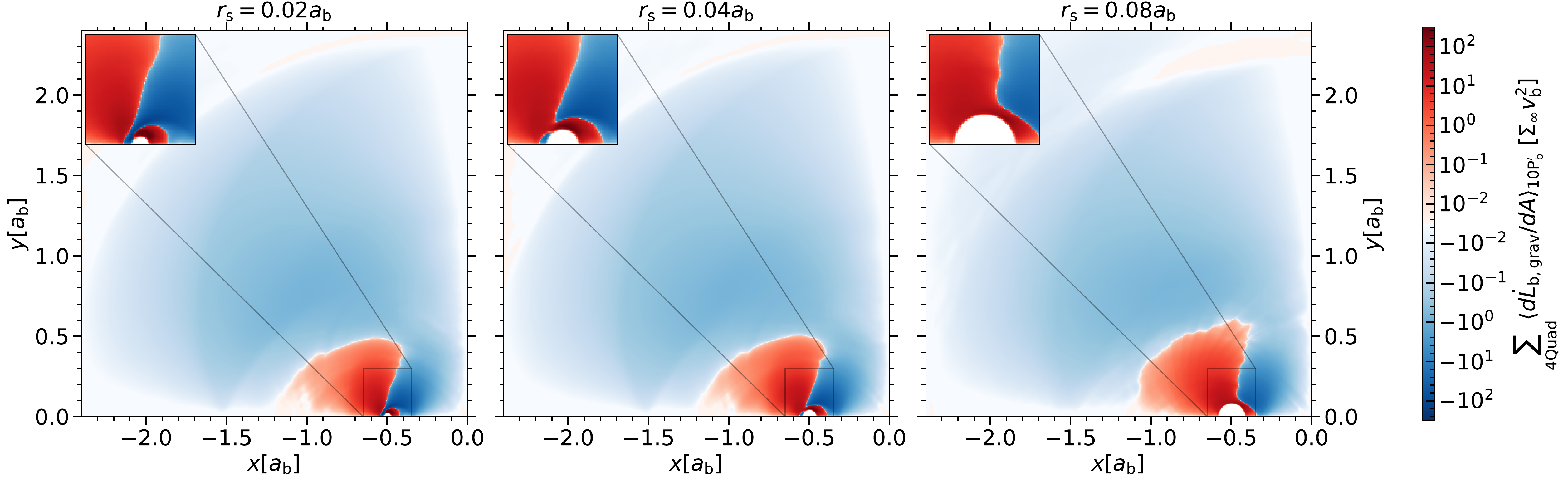}
  \caption{\rlr{Comparison of time-averaged, quadrant-summed maps of the gravitational torque surface density for different sink radii $r_{\rm s}$ (see \texttt{Run I-r$_\mathtt{s}$} in Table \ref{tab:run_I}).  Regions around the sink sphere are zoomed in to show the detailed differences.  In each panel, the narrow white band that divides the red region (i.e., positive torque) and the blue region (i.e., negative torque) represents the \textit{zero-torque curve}. }
  \label{fig:Tgrav_map_vs_rs}}
\end{figure*}

\subsubsection{Dependences on Numerical Parameters}
\label{subsubsec:deps_numParas}

Our fiducial \texttt{Run I-FID} uses the canonical evaluation radius $r_{\rm e}=0.0475 a_{\rm b}$, the root domain size $L_x\times L_Y=(25 a_{\rm b})^2$, the finest level resolution $a_{\rm b}/\delta_{\rm fl}=245.76$, and the wave-damping time-scale of $P_{\rm d}=0.02\Omega_{\rm b}^{-1}$ at the outer boundary.  We have performed extra simulations to test how the orbital evolution results depend on these numerical parameters.

\begin{itemize}
  \item[$\bullet$] \textbf{Evaluation radius}: Fig. \ref{fig:runI_r_eval} compares the time series, running averages, and time-averaged values of $\dot{m}_{\rm b}$, $\dot{\mathcal{E}}_{\rm b}$, and $\dot{a}_{\rm b}$ obtained at the fiducial $r_{\rm e}$ to those obtained at $r_{\rm e} = 0.055a_{\rm b}$ and $r_{\rm e} = 0.065a_{\rm b}$.  We find that $\dot{m}_{\rm b}$ is remarkably insensitive to $r_{\rm e}$ in all cases and in all formats (instantaneous or time-averaged), suggesting that gas falling into the region within a distance of $r_{\rm e}$ from an accretor is almost certainly to be accreted.  Such an independence of the accretion rate from $r_{\rm e}$ is also seen in Appendix \ref{appsec:vali_BHL} but for steady flows.  Similarly, $\dot{\mathcal{E}}_{\rm b}$ and $\dot{a}_{\rm b}/a_{\rm b}$ measured at different $r_{\rm e}$ show great agreement.  The relatively larger differences between the time-averaged values are reconciled by their much larger variation amplitudes.  Compared to the average peak to trough depths, $\langle\dot{m}_{\rm b}\rangle$, $\langle\dot{\mathcal{E}}_{\rm b}\rangle$, and $\langle\dot{a}_{\rm b}\rangle/a_{\rm b}$ all change modestly ($\lesssim 2$ per cent) when $r_{\rm e}$ increases $\approx 37$ per cent.  
  \item[$\bullet$] \textbf{Root domain size}: Fig. \ref{fig:runI_num_tests} compares the fiducial run to \texttt{Run I-LB} with a $\times 2$ larger root domain $L_x\times L_Y=(50 a_{\rm b})^2$ in a similar way to Fig. \ref{fig:runI_r_eval} at the default $r_{\rm e}$.  We find that $\langle\dot{m}_{\rm b}\rangle$, $\langle\dot{\mathcal{E}}_{\rm b}\rangle$, and $\langle\dot{a}_{\rm b}\rangle/a_{\rm b}$ again change modestly ($\lesssim 3$ per cent) when compared to the variation amplitudes, indicating that the canonical root domain size $(10 R_{\rm H})^2$ is large enough for studying the hydrodynamical evolution of the binary.
  \item[$\bullet$] \textbf{Finest level resolution}: Fig. \ref{fig:runI_num_tests} also compares the fiducial run to \texttt{Run I-HR} with $a_{\rm b}/\delta_{\rm fl}=491.52$ (i.e., $\times 2$ the canonical value).  We find that the orbital evolution results vary little, demonstrating that the accretion processes (i.e., the CSDs, the circumbinary flows, and the small half bow shocks) are well resolved in our fiducial simulations.
  \item[$\bullet$] \textbf{Wave-damping time-scale}: Fig. \ref{fig:runI_num_tests} further compares the fiducial run to \texttt{Run I-SD} with $P_{\rm d}=0.1\Omega_{\rm b}^{-1}$.  Following the same standard, we find that $P_{\rm d}$ merely affects the orbital evolution results, once more indicating that the fiducial root domain size is large enough for our study.  Since the time-averaged gravitational torque is mainly determined by the flow structure within $R_{\rm H}$, it is not surprising that our numerical results show little to no dependence on the root domain size and the outer boundary conditions. 
\end{itemize}

Overall, these findings establish the robustness of our accretion prescriptions and our post-processing methods since the orbital evolution results are consistent under various numerical setups, validating our survey results on $r_{\rm s}$, $e_{\rm b}$, and $q_{\rm b}$ in the following sections.  

\subsubsection{Dependence on Sink Radius $r_{\rm s}$}
\label{subsubsec:r_sink_deps}

Fig. \ref{fig:runI_snapshot_rs} shows the quasi-steady state snapshots of \texttt{Run I-r$_{\mathtt{s}}$} with $r_{\rm s}=0.02a_{\rm b}$ and $0.08a_{\rm b}$ at phase $\pi/2$.  Compared to the case with the fiducial sink radius ($0.04 a_{\rm b}$) in Fig. \ref{fig:runI_snapshot}, these runs with different $r_{\rm s}$ present almost identical flow structures except the CSDs.  Particularly, an accretor with a larger $r_{\rm s}$ truncates a larger inner cavity in the surrounding disc.  Such a truncation, if large enough, may disrupt the CSD.  Therefore, we find that the CSDs in the $r_{\rm s}=0.02 a_{\rm b}$ case are similar to those in the fiducial case except the cavity size, while the CSDs in the $r_{\rm s}=0.08 a_{\rm b}$ case appear to be moderately suppressed, both in radial extent and in disc mass.

Fig. \ref{fig:RunI_trends_r_s} shows the secular results for the binary evolution as a function of sink radius.  First, the time-averaged accretion rate $\langle\dot{m}_{\rm b}\rangle$ scales linearly with $r_{\rm s}$, indicating that it is easier for streamlines to intersect \rlr{a} larger accretor.  Previous work on Bondi-Hoyle-Lyttleton accretion with an upstream gradient found that the accretion rate approximately scales with $r_{\rm s}^{1/2}$ when the accretion flow is turbulent \rlr{and contains initial angular momentum} \citep[][\rlr{see also Appendix A in \citealt{Kaaz2021}}]{Xu2019}.  The trend identified in our scenario is similar but steeper, probably due to the much more dynamic accretion flow.  Moreover, our finding that $\langle\dot{m}_{\rm b}\rangle \approxprop r_{\rm s}$ differs from the binary accretion in circumbinary discs, where $\langle\dot{m}_{\rm b}\rangle$ is regulated by viscous disc accretion and does not change with $r_{\rm s}$ \citep{Munoz2019}.

We find that binaries in all three cases are shrinking and the orbital decay rate is faster at a smaller sink radius (for the same $\Sigma_{\infty}$, etc.), as $\ell_0$ (and thus $\langle\dot{{L}_{\rm b}}\rangle$) and $\langle\dot{\mathcal{E}}_{\rm b}\rangle$ become more negative for smaller $r_{\rm s}$.  To better understand this trend, Fig. \ref{fig:torque_vs_rs} breaks down the torque contributions (see Eq. \ref{eq:dotL_breakdown}), where $\langle\dot{L}_{\rm b,acc}\rangle$ increases moderately with $\langle\dot{m}_{\rm b}\rangle$ as anticipated, $\langle\dot{L}_{\rm b,pres}\rangle$ only change slightly, and $\langle\dot{L}_{\rm b,grav}\rangle$ increases prominently with $r_{\rm s}$ and contributes most to the total torque change.

To comprehend the influence of sink radius on gravitational torque, Fig. \ref{fig:torque_vs_rs} also compares the radial profiles of $\langle\dot{L}_{\rm b,grav}\rangle_{\rm 10P_{\rm b}'}(<r)$ for all cases.  These profiles follow a similar trend (see Section \ref{subsubsec:runI_Tgrav_map} for descriptions on persistent non-axisymmetric flow structures), where the negative torques come from the part of the CSDs inside the binary orbit and the flows within $0.8 a_{\rm b} \lesssim r \lesssim R_{\rm H}$, while the positive torques originate from the part of the CSDs outside the binary orbit.  In addition, we find that all the $\langle\dot{L}_{\rm b,grav}\rangle_{\rm 10P_{\rm b}'}(<r)$ profiles are almost identical from the binary COM to $r \simeq 0.4 a_{\rm b}$ and diverge afterwards towards $0.5 a_{\rm b}$.  Specifically, the smaller $r_{\rm s}$ results in a more negative gravitational torque from the part of the CSDs inside the binary orbit.

We further define the complementary cumulative gravitational torque from the flow structures outside $0.5 a_{\rm b}$ (again see Fig. \ref{fig:torque_vs_rs})
\begin{equation}
  \langle\dot{L}_{\rm b,grav}\rangle_{\rm 10P_{\rm b}'}(>r) \equiv \int_r^{\infty} \left\langle\frac{\textnormal{d}\dot{L}_{\rm b,grav}}{\textnormal{d}A}\right\rangle_{\rm 10P_{\rm b}'} \textnormal{d}A,
\end{equation}
where $\langle\dot{L}_{\rm b,grav}\rangle_{\rm 10P_{\rm b}'}(>r)$ = $\langle\dot{L}_{\rm b,grav}\rangle$ - $\langle\dot{L}_{\rm b,grav}\rangle_{\rm 10P_{\rm b}'}(<r)$.  The radial profiles of $\langle\dot{L}_{\rm b,grav}\rangle_{\rm 10P_{\rm b}'}(>r)$ for runs with different $r_{\rm s}$ are almost identical beyond the Hill radius $R_{\rm H}$, implying that flows outside the gravitational sphere of influence of $m_{\rm b}$ do not feel the size of $r_{\rm s}$.  Moreover, the profile for $r_{\rm s}=0.02a_{\rm b}$ closely follows that of the fiducial case from $r\sim R_{\rm H}$ all the way to $r\sim 0.055 a_{\rm b}$, indicating their torque differences reside in the small region close to accretors, consistent with the overall similarity between their CSDs.  On the contrary, the profile for $r_{\rm s}=0.08 a_{\rm b}$ deviates moderately from that of the fiducial case for $r \lesssim R_{\rm H}$, suggesting that the suppressed CSDs also slightly alter the circumbinary flows.  Finally, the part of the CSDs outside the binary orbit ($0.5 a_{\rm b} \lesssim r \lesssim 0.8 a_{\rm b}$) contributes less positive gravitational torque as $r_{\rm s}$ decreases, strengthening the trend identified for the part of the CSDs inside the binary orbit and leading to the positive correlation between $\langle\dot{L}_{\rm b,grav}\rangle$ and $r_{\rm s}$.

\rlr{To visualize the gravitational torque differences caused by the region close to accretors, Fig. \ref{fig:Tgrav_map_vs_rs} plots the maps of the quadrant sum of gravitational torque surface density $\sum_{\rm 4Quad} \langle\textnormal{d}\dot{L}_{\rm b,grav}/\textnormal{d}A\rangle_{\rm 10P_{\rm b}'}$ for all three cases, where
\begin{equation}
  \sum_{\rm 4Quad} \mathcal{U} = \mathcal{U} (x, y) + \mathcal{U} (-x, y) + \mathcal{U} (x, -y) + \mathcal{U} (-x, -y).
\end{equation}
Such a quadrant sum reveals the detailed deviations from the axial-symmetry that are hard to spot in the map of $\langle\textnormal{d}\dot{L}_{\rm b,grav}/\textnormal{d}A\rangle_{\rm 10P_{\rm b}'}$ (see Fig. \ref{fig:runI_Tgrav_map}) and thus unveils the spatial distribution of the ``net'' torque density.  Specifically, the zero-torque curve that divides each CSD into two parts at $r \sim 0.5 a_{\rm b}$ and separates the CSDs from the circumbinary flows at $r \sim a_{\rm b}$ roughly categorizes these maps into three regions, corresponding to the two turnovers in the radial profiles of (complementary) cumulative gravitational torque (see Fig. \ref{fig:torque_vs_rs}).  We notice that the runs with smaller $r_{\rm s}$ ($\lesssim 0.04 a_{\rm b}$) allow the negative torque region extend to the left boundary of the sink sphere, while in the case of $r_{\rm s}=0.08 a_{\rm b}$ the sink sphere is fully surrounded by the positive torque region.  Consequently, the zero-torque curve in the $r_{\rm s}=0.08 a_{\rm b}$ case with suppressed CSDs somewhat differs from those in the other two cases, consistent with the observed modest deviation of the $\langle\dot{L}_{\rm b,grav}\rangle_{\rm 10P_{\rm b}'}(>r)$ around $r \sim a_{\rm b}$.  By comparing the three maps, Fig. \ref{fig:Tgrav_map_vs_rs} demonstrates that the differences of $\langle\dot{L}_{\rm b,grav}\rangle$ can be largely attributed to the regions closely surrounding the accretor.  In short, the simulations with smaller accretors that allow the CSDs extend to smaller radii tend to yield more negative torques.}

\begin{figure}
  \centering
  \includegraphics[width=\linewidth]{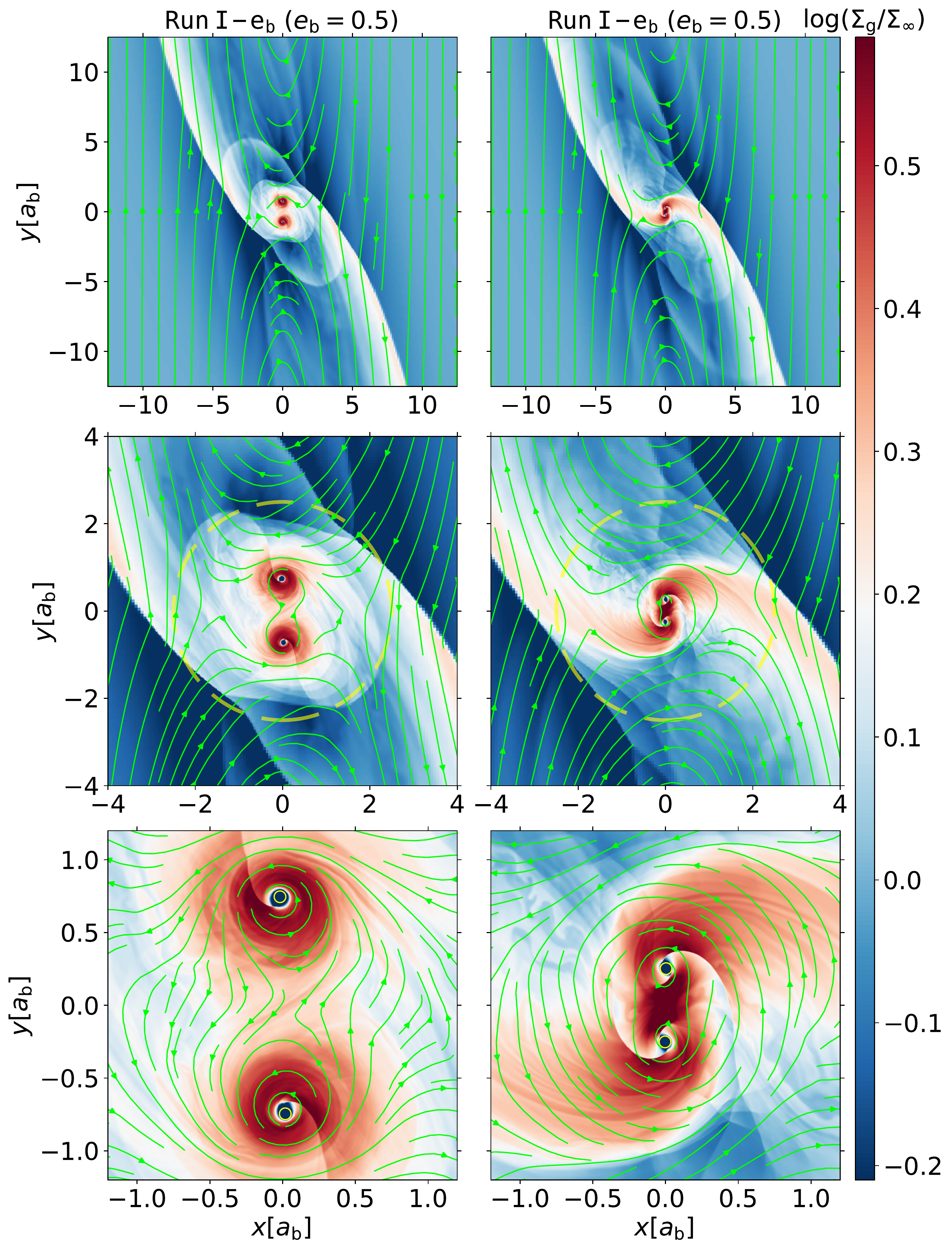}
  \caption{Similar to Fig. \ref{fig:runI_snapshot} but showing the flow structures for \texttt{Run I-e$_\mathtt{b}$} with $e_{\rm b}=0.5$ at near apocentre (\textit{left}) and pericentre (\textit{right}).  The snapshots are chosen such that $\bm{r}_{\rm b}$ is nearly parallel to $\hat{\bm{y}}$(see Table \ref{tab:run_I}).
  \label{fig:runI_snapshot_ecc0.5}}
\end{figure}

\begin{figure}
  \centering
  \includegraphics[width=\linewidth]{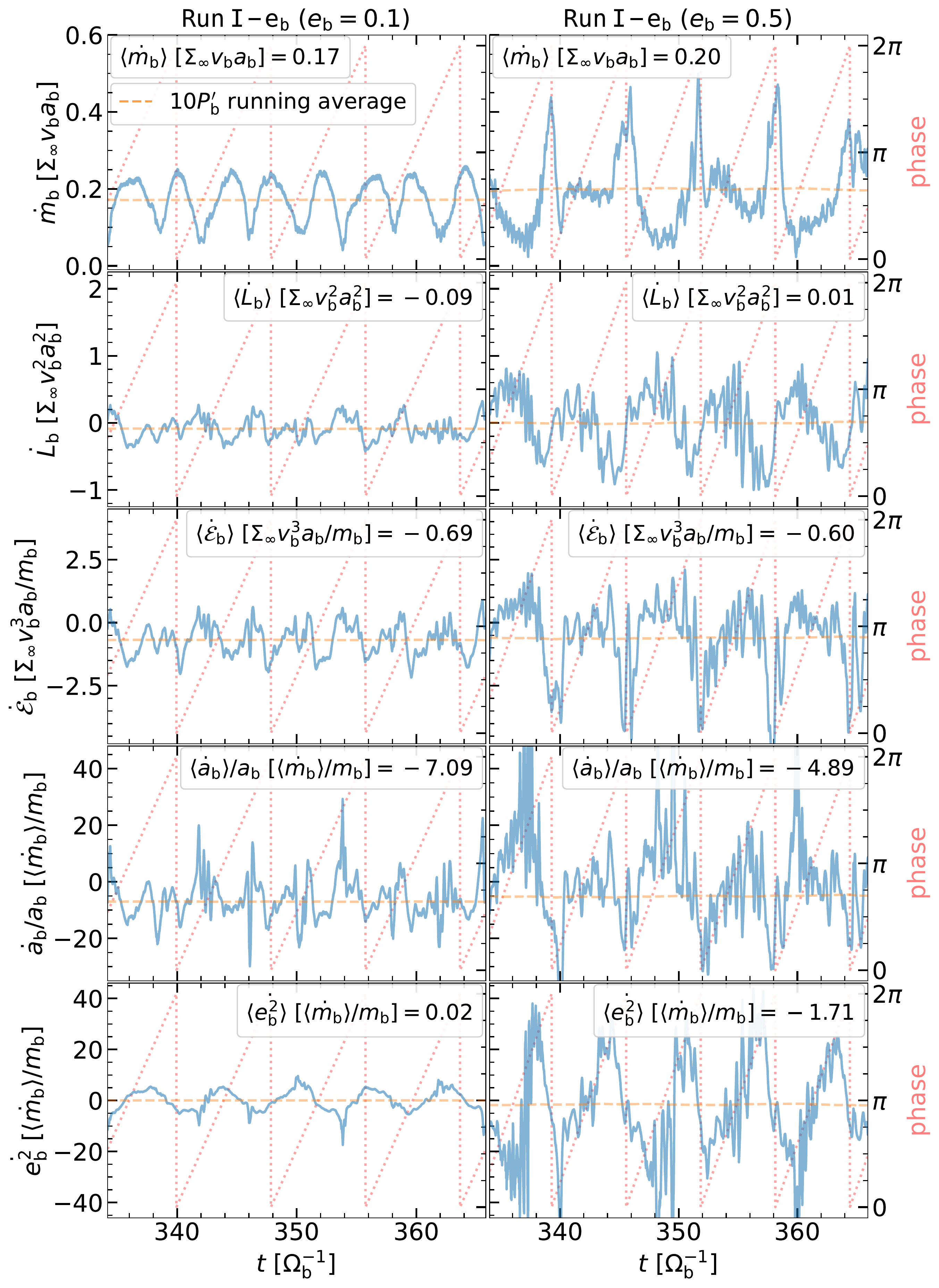}
  \caption{Similar to Fig. \ref{fig:runI_num_tests} but showing the time series of (from \textit{top} to \textit{bottom}) accretion rate ($\dot{m}_{\rm b}$), torque ($\dot{L}_{\rm b}$), rate of change in specific energy ($\dot{\mathcal{E}}_{\rm b}$), in semi-major axis ($\dot{a}_{\rm b}$), and in eccentricity squared ($\dot{e^2_{\rm b}}$) for \texttt{Run I-e$_{\tt b}$} with $e_{\rm b}=0.1$ (\textit{left}) and $e_{\rm b}=0.5$ (\textit{right}).  The time-averaged values shown in legends are averaged over the last $300\Omega_{\rm b}^{-1}$.  The phase curves (\textit{pink dotted}) show the phase of $\Omega_{\rm b}'$ for $e_{\rm b}=0.1$ cases and the phase of $\Omega_{\rm b}$ for $e_{\rm b}=0.5$ (see Section \ref{subsec:result_ecc} for details). 
  \label{fig:runI_ecc_0.1_0.5}}
\end{figure}

\begin{figure}
  \centering
  \includegraphics[width=\linewidth]{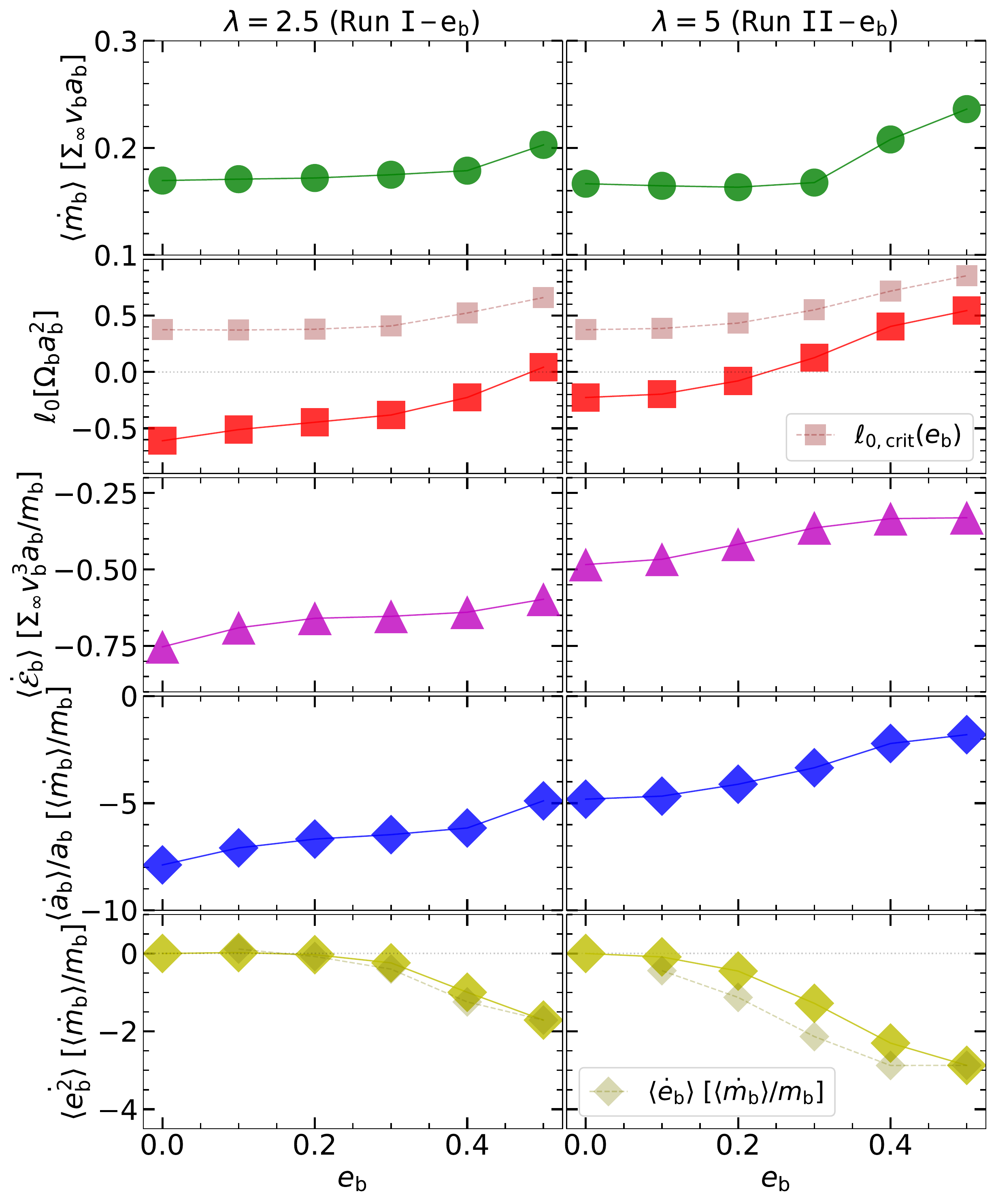}
  \caption{Time averaged measurements of (from \textit{top} to \textit{bottom}) accretion rate $\langle\dot{m}_{\rm b}\rangle$, accretion eigenvalue $\ell_0$, rate of change in binary specific energy $\langle\dot{\mathcal{E}}_{\rm b}\rangle$, binary migration rate $\langle\dot{a}_{\rm b}\rangle/a_{\rm b}$, and binary eccentricity change rates $\langle\dot{e}_{\rm b}^2\rangle$ and $\langle\dot{e}_{\rm b}\rangle$ for a range of eccentricities $e_{\rm b}$ from simulations with $\lambda = 2.5$ (\textit{left}; \texttt{Run I-e$_{\mathtt{b}}$}) and with $\lambda = 5$ (\textit{right}; \texttt{Run II-e$_{\mathtt{b}}$}).  These quantities are averaged over the last $300 \Omega_{\rm b}^{-1}$ for \texttt{Run I-e$_{\mathtt{b}}$} and over the last $240 \Omega_{\rm b}^{-1}$ for \texttt{Run II-e$_{\mathtt{b}}$}.  \label{fig:trends_eccs}}
\end{figure}

\begin{figure*}
  \centering
  \includegraphics[width=0.8\linewidth]{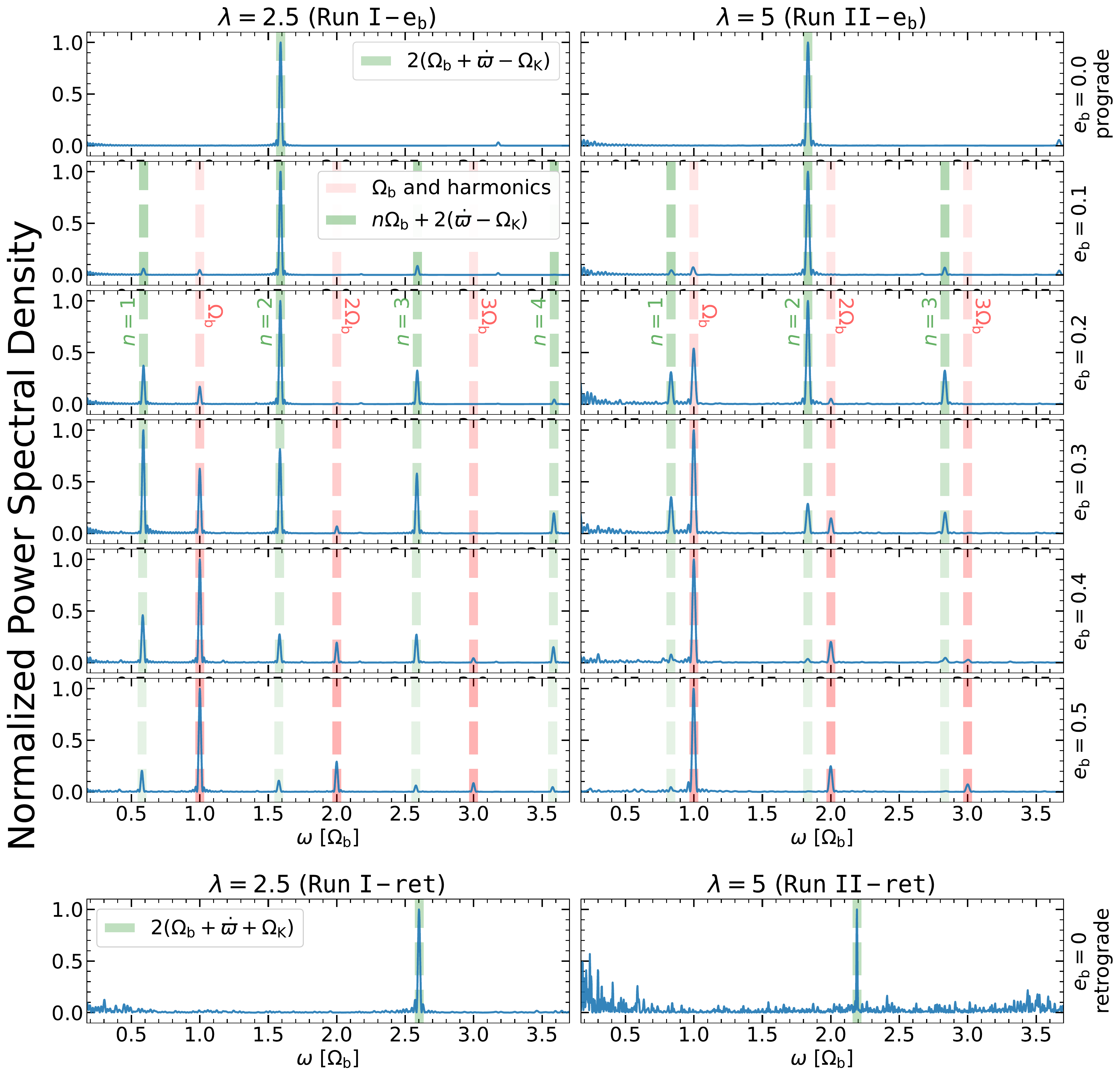}
  \caption{Normalized power spectral density (PSD) of the accretion rate time series for eccentric binary simulations with $\lambda = 2.5$ (\textit{left}; \texttt{Run I-e$_{\mathtt{b}}$}) and with $\lambda = 5$ (\textit{right}; \texttt{Run II-e$_{\mathtt{b}}$}).  These Lomb-Scargle periodogram are computed from the time series over the last $300 \Omega_{\rm b}^{-1}$ for \texttt{Run I-e$_{\mathtt{b}}$} and \texttt{Run I-ret}, over the last $240 \Omega_{\rm b}^{-1}$ for \texttt{Run II-e$_{\mathtt{b}}$}, and over the last $720 \Omega_{\rm b}^{-1}$ for \texttt{Run II-ret}.  For circular binaries, the dominate frequency is $2\Omega_{\rm b}'$, with $\Omega_{\rm b}' = \Omega_{\rm b} + \dot{\varpi} - \Omega_{\rm K}$ for prograde orbits and $\Omega_{\rm b}' = \Omega_{\rm b} + \dot{\varpi} + \Omega_{\rm K}$ for retrograde orbits.
  \label{fig:trends_eccs_PSD}}
\end{figure*}

\begin{figure*}
  \centering
  \includegraphics[width=0.755\linewidth]{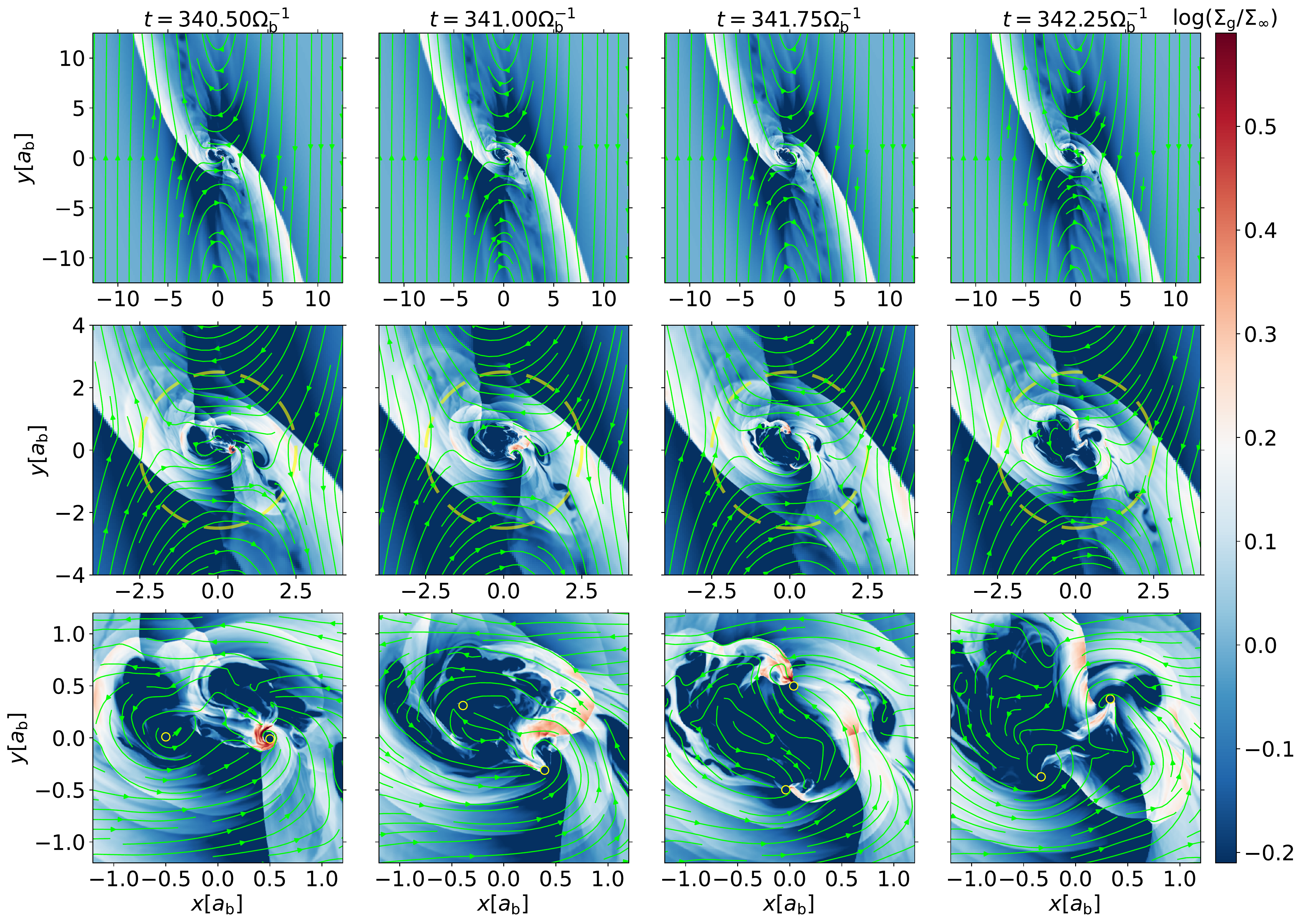}
  \includegraphics[width=0.24\linewidth]{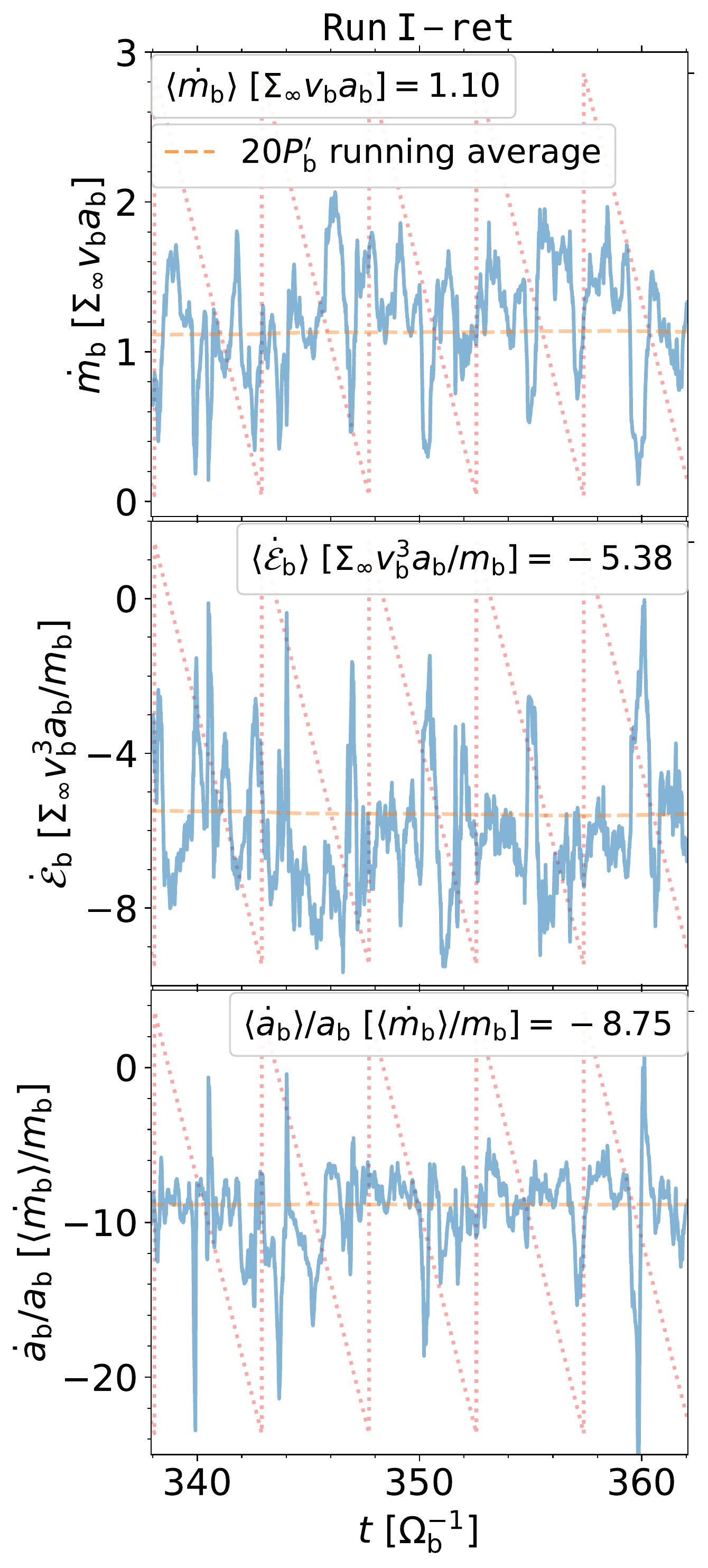}
  \caption{Key phase snapshots (\textit{left}; similar to Fig. \ref{fig:runI_snapshot}) and time series of $\dot{m}_{\rm b}$, $\dot{\mathcal{E}}_{\rm b}$, and $\dot{a}_{\rm b}$ (\textit{right}; similar to Fig. \ref{fig:runI_num_tests}) for a retrograde equal-mass circular binary in \texttt{Run I-ret}.  Compared to Fig. \ref{fig:runI_snapshot}, the binary orbits in the opposite direction and moves from phase $\pi$ to phase $7\pi/4$.  Compared to Fig. \ref{fig:runI_num_tests}, the running averages are carried out over $20 P_{\rm b}'$ because $P_{\rm b}'$ ($=4.83/\Omega_{\rm b}$) in \texttt{Run I-ret} is much shorter than that in \texttt{Run I-FID} ($7.90/\Omega_{\rm b}$).
  \label{fig:RunI-ret}}
\end{figure*}

\subsection{Prograde Equal-mass Eccentric Binaries}
\label{subsec:result_ecc}

The orbits of eccentric binaries are open trajectories in the rotating frame due to precessions (see Eq. \ref{eq:Omega_pre}), with the apsidal period ($2\pi/\Omega_{\rm pre}$) not commensurable with $2\pi/\Omega_{\rm b}$.
Such open orbits fundamentally change the stable periodic behaviours of circular binaries seen in the fiducial runs.  We thus conduct simulations to survey prograde equal-mass binaries with $e_{\rm b}= 0.1, 0.2, \cdots, 0.5$ (see Table \ref{tab:run_I} for \texttt{Run I-e$_{\tt b}$} and Table \ref{tab:run_II} for \texttt{Run II-e$_{\tt b}$}).  

Fig. \ref{fig:runI_snapshot_ecc0.5} presents two sample snapshots from \texttt{Run I-e$_{\tt b}$} with $e_{\rm b}=0.5$ at the times similar to phase $\pi/2$ in the fiducial case, when the binary are at apocentre and pericentre, respectively.  We find that the flow structures outside $R_{\rm H}$ are almost identical to the fiducial case, while the CSDs interfere with and suppress each other at pericentre and become detached and spread out at apocentre.  Consequently, such a constantly changing binary separation induces much more transient fluctuations to the accretion flow and torques.

Fig. \ref{fig:runI_ecc_0.1_0.5} shows the time series of $\dot{m}_{\rm b}$,  $\dot{L}_{\rm b}$,  $\dot{\mathcal{E}}_{\rm b}$,  $\dot{a}_{\rm b}$, and $\dot{e_{\rm b}^2}$($=\textnormal{d}e_{\rm b}^2/\textnormal{d}t$) for \texttt{Run I-e$_{\mathtt{b}}$} with $e_{\rm b}=0.1$ and $0.5$.  We find that the run with $e_{\rm b} = 0.1$ already shows prominent deviations from periodic variations previously seen in Fig. \ref{fig:runI_r_eval}.  For instance, the symmetry of circular orbits (i.e., two evenly-spaced peaks/troughs within one orbital period) disappears.  Moreover, the more eccentric the binary orbit becomes, the more asymmetry and the larger variations/fluctuations the time series present.  The peak accretion rate in the $e_{\rm b} = 0.5$ case is much higher than that for $e_{\rm b} = 0.1$, because binaries with more eccentric orbits can dive farther into the shear flow (i.e., at apocentres when the major axis is roughly in parallel with $\hat{\bm{x}}$), where more materials are available for accretion.  In addition, the time series of $\dot{L}_{\rm b}$ and $\dot{\mathcal{E}}_{\rm b}$ do not share the same curve shape like the circular binary case and their differences lead to finite $\dot{e_{\rm b}^2}$ (see Eq. \ref{eq:aoa_dotE}).

Despite the stochastic short-term fluctuations, the nearly constant running averages in Fig. \ref{fig:runI_ecc_0.1_0.5} are consistent with their time-averaged results, indicating that both runs have achieved quasi-steady.  Fig. \ref{fig:trends_eccs} shows the secular orbital evolution results as a function of eccentricity for our surveys on both run series (including the fiducial $e_{\rm b}=0$ cases).  We find that all runs produce contracting binaries.  Additionally, we identify same trends in both run series, where $\langle\dot{m}_{\rm b}\rangle$, $\ell_0$, $\langle\dot{\mathcal{E}}_{\rm b}\rangle$, and $\langle\dot{a}_{\rm b}\rangle/a_{\rm b}$ generally increase (or become less negative) with $e_{\rm b}$, while $\langle\dot{e^2_{\rm b}}\rangle$ generally decreases with $e_{\rm b}$.  In other words, a higher eccentricity leads to faster and more dynamic accretion, less negative accreted angular momentum per unit of accreted mass, a slower negative energy transfer rate, and a slower orbital decay rate with a faster circularization timescale.  One minor exception is \texttt{Run I-e$_{\mathtt{b}}$} with $e_{\rm b}=0.1$, where $\langle\dot{e^2_{\rm b}}\rangle \simeq 0.02 \langle\dot{m}_{\rm b}\rangle / m_{\rm b}$ is nominally positive and is consistent with zero.

To quantify the variability of the accretion flow around eccentric binaries, we use Lomb-Scargle periodogram to compute the power spectral density (PSD) for the $\dot{m}_{\rm b}$ time series of all cases in \texttt{Run I-e$_{\mathtt{b}}$} and \texttt{Run II-e$_{\mathtt{b}}$} (including the fiducial $e_{\rm b}=0$ cases).  Fig. \ref{fig:trends_eccs_PSD} shows the results, obtained from the same time periods that are used for computing the time-averaged results.  For $e_{\rm b}=0$, the absolute dominant variability frequency is $2\Omega_{\rm b}' = 2(\Omega_{\rm b} + \Omega_{\rm pre})$ (see Eqs. \ref{eq:Omega_pre} and \ref{eq:Omega_bprime}), where $\Omega_{\rm pre} = \dot{\varpi} - \Omega_{\rm K}$ for prograde orbits and the factor $2$ is due to the symmetry of the equal-mass circular binaries (see also Section \ref{subsubsec:orbital_evolution} and Fig. \ref{fig:runI+II_breakdown}).  For $e_{\rm b}=0.1$, $2\Omega_{\rm b}'$ still dominates but local peaks at other frequencies begin to emerge.  As $e_{\rm b}$ increases, the radial epicyclic frequency $\Omega_{\rm b}$ and its harmonics gain more power and $\Omega_{\rm b}$ becomes the dominant frequency when $e_{\rm b} \gtrsim 0.4$.  Meanwhile, the apsidal frequency $\Omega_{\rm pre}$ interferes with $\Omega_{\rm b}$ and its harmonics and the resulting frequencies $n \Omega_{\rm b} + 2\Omega_{\rm pre}$ (with $n=1,2,3,\cdots$) share a moderate fraction of power, where the factor $2$ comes from the symmetry of equal-mass.  However, such interferences are much weaker in \texttt{Run II-e$_{\mathtt{b}}$} series since the apsidal frequency $\Omega_{\rm pre}$ is much smaller than that in \texttt{Run I-e$_{\mathtt{b}}$}.

Our finding that $2\Omega_{\rm b}'$ is the dominant frequency for equal-mass circular binaries differs qualitatively from binary accretion in circumbinary discs, where the dominant frequency is $\sim 0.2 \Omega_{\rm b}$ (see \citealt{Munoz2020} and references therein).

\subsection{Retrograde Equal-mass Circular Binaries}
\label{subsec:result_retro}

In this section, we flip the binary orientation to retrograde ($\hatomega_{\rm b}\cdot \hat{z}=-1$) in the fiducial cases and study the resulting impacts based on \texttt{Run I-ret} and \texttt{Run II-ret}.

Fig. \ref{fig:RunI-ret} shows the quasi-steady state snapshots of \texttt{Run I-ret} at four key orbital phases.  Compared to the prograde fiducial case in Fig. \ref{fig:runI_snapshot}, the flow structure outside the Hill radius is unaffected but changes drastically within $R_{\rm H}$.  There are no persistent CSDs or circumbinary disc,and the binary components accrete directly from the shear flow or horseshoe flow without a coherent pattern.  The retrograde orbit, together with the much shorter apparent orbital period ($P_{\rm b}' = 4.83/\Omega_{\rm b}$), makes each binary component constantly run into the spiral/bow shocks excited by the other component, leading to severe ram pressure stripping and therefore the lack of CSDs.

Fig. \ref{fig:RunI-ret} also shows the time series of $\dot{m}_{\rm b}$, $\dot{\mathcal{E}}_{\rm b}$, and $\dot{a}_{\rm b}$ from \texttt{Run I-ret}.  Compared to the prograde case in Fig. \ref{fig:runI_num_tests}, they appear to be highly aperiodic and somewhat unrelated to the orbital phase due to the scrambled accretion flow.  Nevertheless, we still find that the running averages are nearly constant and are consistent with the final time-averaged results, indicating that the system has reached quasi-steady state.  Another consequence of the retrograde orbit is the fast accretion rate
\begin{equation}
  \langle\dot{m}_{\rm b}\rangle \simeq 1.10 \Sigma_{\infty} v_{\rm b} a_{\rm b},
\end{equation}
much higher than that of the fiducial prograde case.  Part of the reason for the fast accretion is that the relative velocity of binary components with respect to the share flow at phase $\pi$ is lower ($\mathcal{M} = 0.73$).  The lack of CSDs also contribute to the fast accretion since materials can be directly accreted without being processed through the disc.  Regarding the long-term orbital evolution, we find
\begin{align}
  \ell_0 &\simeq 0.72\ v_{\rm b} a_{\rm b}   \\
  \frac{\langle\dot{a}_{\rm b}\rangle}{a_{\rm b}} &\simeq -8.75 \frac{\langle\dot{m}_{\rm b}\rangle}{m_{\rm b}} \simeq -9.66 \frac{\Sigma_\infty v_{\rm b} a_{\rm b}}{m_{\rm b}}.  \label{eq:adot_a_RunI-ret}
\end{align}
Note that the positive $\ell_0$ value implies that the binary receives angular momentum in the $\hat{\bm{z}}$ direction, opposite to $\hatomega_{\rm b}$.  Thus, the retrograde binary contracts at a much faster rate than the corresponding prograde binary (comparing Eqs. \ref{eq:adot_a_RunI-FID} and \ref{eq:adot_a_RunI-ret}).

Following Section \ref{subsec:result_ecc}, we again use Lomb-Scargle periodogram to compute the PSD for the $\dot{m}_{\rm b}$ time series of \texttt{Run I-ret} (see Fig. \ref{fig:trends_eccs_PSD}).  Similar to the prograde case, the absolute dominant frequency is still $2\Omega_{\rm b}'$, although $\Omega_{\rm b}' = \Omega_{\rm b} + \dot{\varpi} + \Omega_{\rm K}$ for retrograde orbits (see Eqs. \ref{eq:Omega_pre} and \ref{eq:Omega_bprime}). 

We apply the same set of analyses on \texttt{Run II-ret} and find 
\begin{align}
  \langle\dot{m}_{\rm b}\rangle &\simeq 0.78 \Sigma_{\infty} v_{\rm b} a_{\rm b} \\
  \ell_0 &\simeq 0.85\ v_{\rm b} a_{\rm b}, \\
  \frac{\langle\dot{a}_{\rm b}\rangle}{a_{\rm b}} &\simeq -9.79 \frac{\langle\dot{m}_{\rm b}\rangle}{m_{\rm b}} \simeq -7.65 \frac{\Sigma_\infty v_{\rm b} a_{\rm b}}{m_{\rm b}}.  \label{eq:adot_a_RunII-ret}
\end{align}
Comparing to the corresponding prograde case (\texttt{Run II-FID}; see Eq. \ref{eq:adot_a_RunII-FID}), the accretion rate is much higher, leading to much faster orbital decay for retrograde binary.  In addition, the PSD for the $\dot{m}_{\rm b}$ time series of \texttt{Run II-ret} in Fig. \ref{fig:trends_eccs_PSD} shows a relatively dominant frequency of $2\Omega_{\rm b}'$, with a forest of small peaks all over other frequencies.  This noisy PSD, despite the longer period adopted for the analysis ($720/\Omega_{\rm b}$), is a manifestation of the vastly more violent accretion flows in \texttt{Run II-ret}, justifying the need for a longer run time in this case (see Table \ref{tab:run_II}).

\begin{figure*}
  \centering
  \includegraphics[width=0.415\linewidth]{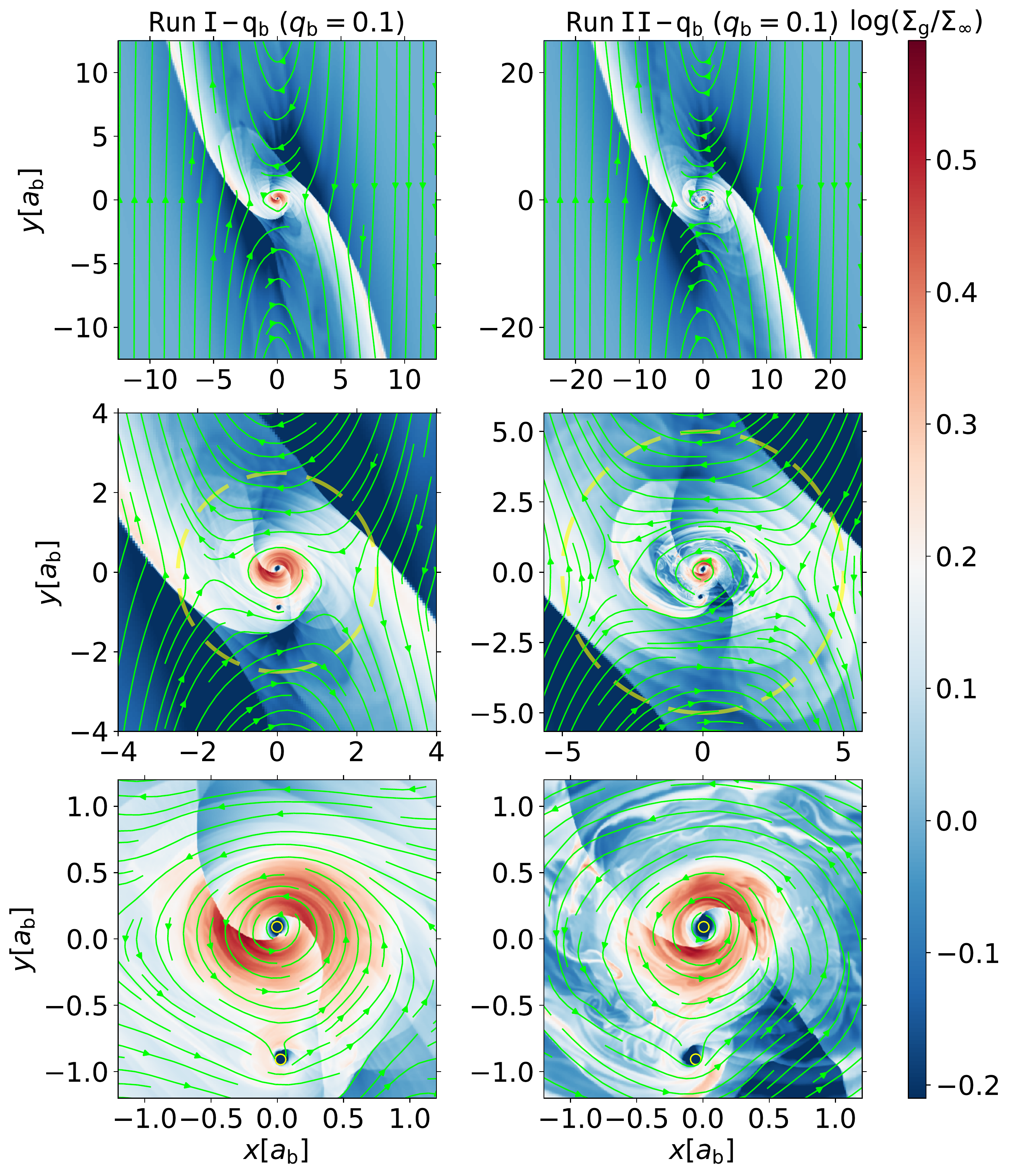}
  \includegraphics[width=0.575\linewidth]{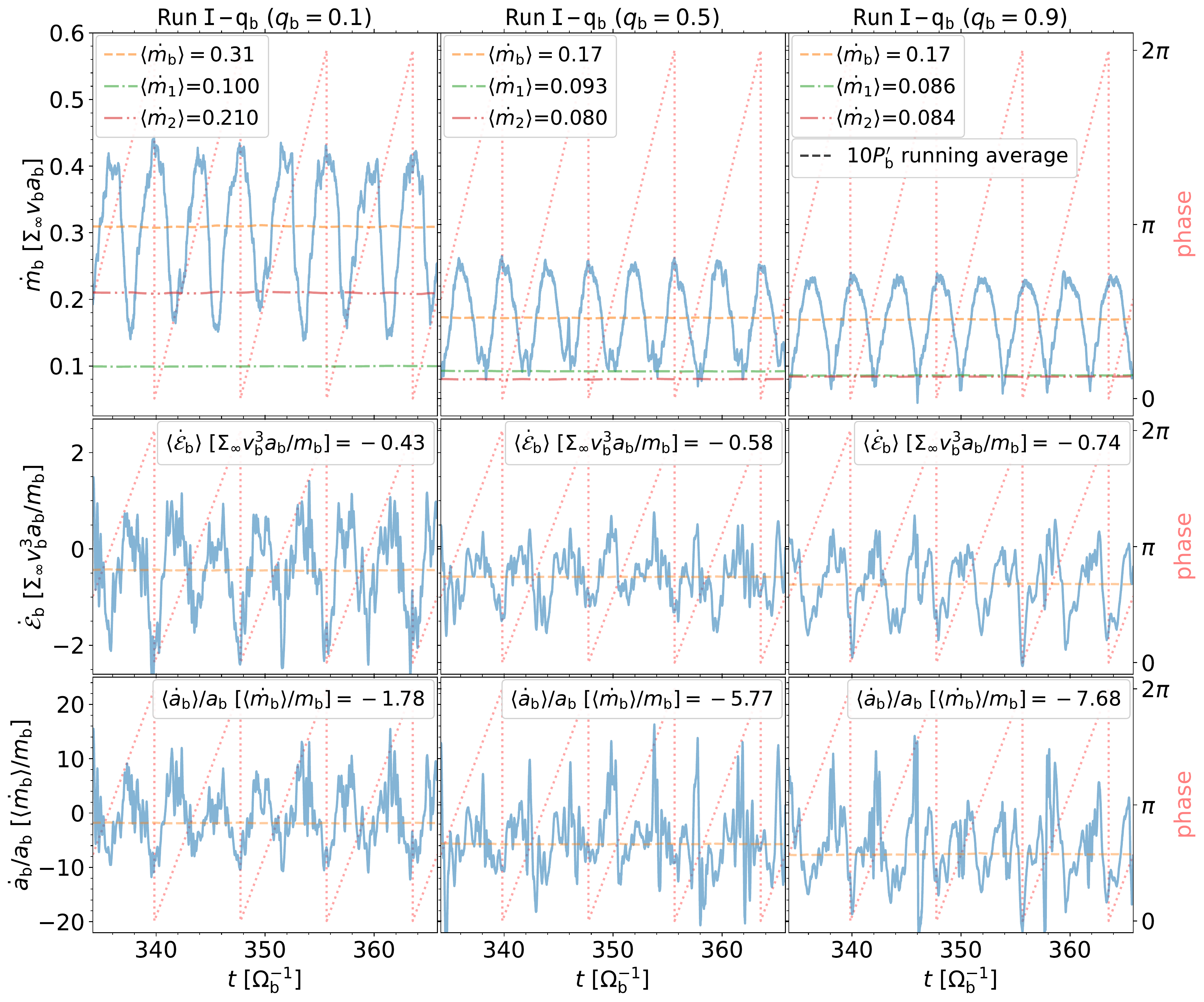}
  \caption{Sample snapshots (\textit{left}; similar to Fig. \ref{fig:runI_snapshot} and Fig. \ref{fig:runII_snapshot}) for \texttt{Run I-q$_\mathtt{b}$} and \texttt{Run II-q$_\mathtt{b}$} with $q_{\rm b}=0.1$ at phase $\pi/2$ and time series of $\dot{m}_{\rm b}$, $\dot{\mathcal{E}}_{\rm b}$, and $\dot{a}_{\rm b}$ (\textit{right}; similar to Fig. \ref{fig:runI_num_tests}) for \texttt{Run I-q$_\mathtt{b}$} with selected $q_{\rm b}$ ($0.1$, $0.5$, and $0.9$).  Compared to Fig. \ref{fig:runI_num_tests}, this figure also shows the running averages and time-averaged values of $\dot{m}_1$ and $\dot{m}_2$.
  \label{fig:runI+II_qbs}}
\end{figure*}

\begin{figure}
  \centering
  \includegraphics[width=\linewidth]{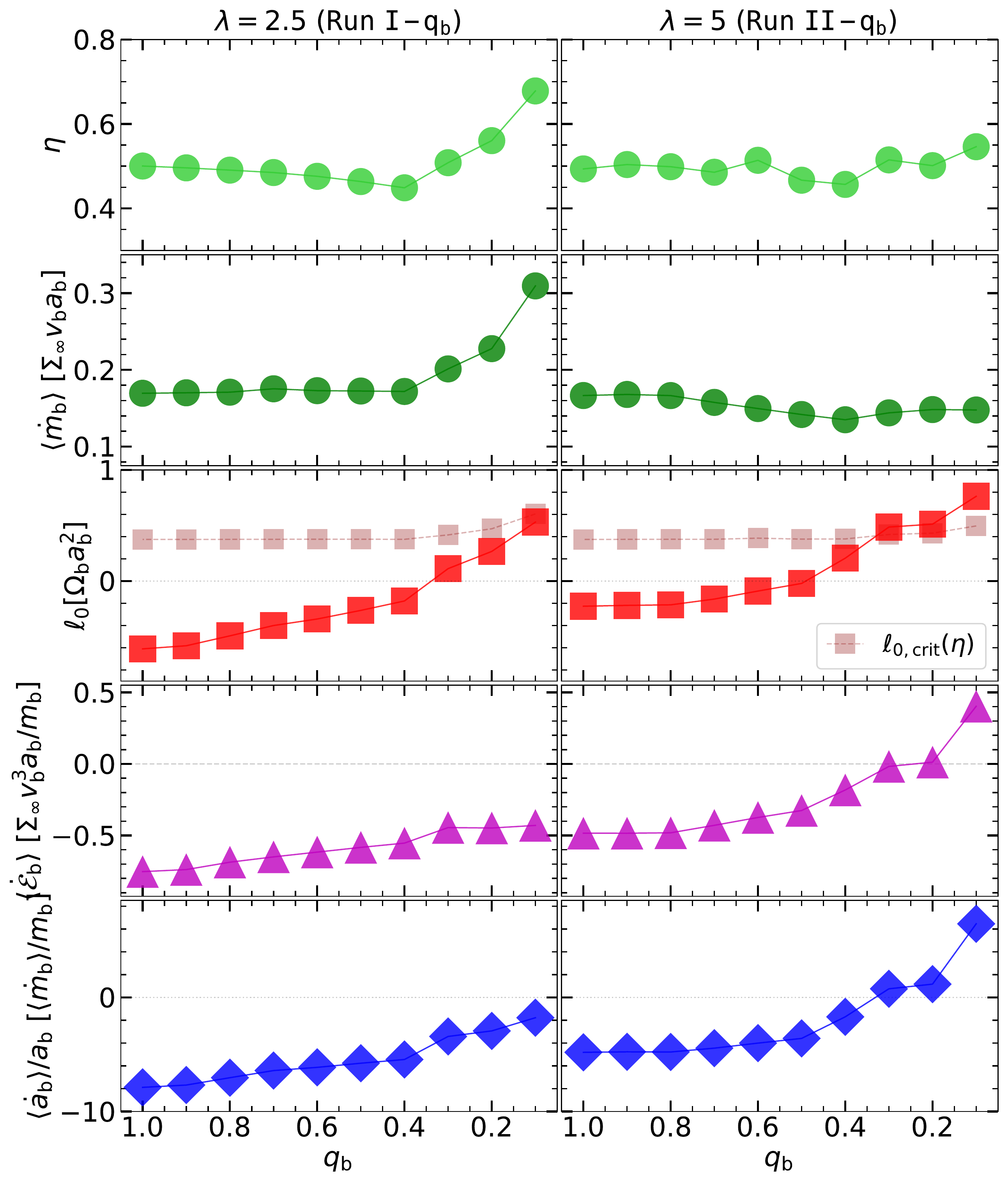}
  \caption{Time averaged measurements of (from \textit{top} to \textit{bottom}) accretion rate ratio $\eta$, accretion rate $\langle\dot{m}_{\rm b}\rangle$, accretion eigenvalue $\ell_0$, rate of change in binary specific energy $\langle\dot{\mathcal{E}}_{\rm b}\rangle$, and binary migration rate $\langle\dot{a}_{\rm b}\rangle/a_{\rm b}$ for a range of binary mass ratios $q_{\rm b}$ from simulations with $\lambda = 2.5$ (\textit{left}; \texttt{Run I-q$_{\mathtt{b}}$}) and with $\lambda = 5$ (\textit{left}; \texttt{Run II-q$_{\mathtt{b}}$}).  These quantities are averaged over the last $300 \Omega_{\rm b}$ for \texttt{Run I-q$_{\mathtt{b}}$} and over the last $240 \Omega_{\rm b}$ for \texttt{Run II-q$_{\mathtt{b}}$}.  Note that $q_{\rm b}$ decreases along the horizontal axis.
  \label{fig:trends_qbs}}
\end{figure}

\subsection{Prograde Unequal-mass Circular Binaries}
\label{subsec:result_qb}

In this section, we investigate how the mass ratio between the binary components, $q_{\rm b}=m_2/m_1$, affects our results.  A non-unity mass ratio is of great interest since it breaks the spatial degeneracy between the individual orbits of binary components in the frame centered at their COM, but preserves the temporal symmetry associated with the closed orbit.  The two binary components would generally have different accretion rate ($\eta = \dot{m}_2/\dot{m}_1 \neq 0.5$), which alters the critical threshold $\ell_{0,\mathrm{crit}}$ (Eq. \ref{eq:ell_0_crit_eta}) for orbital decay/expansion.  We thus conduct simulations to survey prograde unequal-mass circular binaries with $q_{\rm b} = 0.1, 0.2, \cdots, 0.9$ (see Table \ref{tab:run_I} for \texttt{Run I-q$_{\tt b}$} and Table \ref{tab:run_II} for \texttt{Run II-q$_{\tt b}$}).

Fig. \ref{fig:runI+II_qbs} shows the snapshots of the \texttt{Run I-q$_{\tt b}$} and \texttt{Run II-q$_{\tt b}$} with $q_{\rm b}=0.1$ at phase $\pi/2$.  We again find that the flow structures outside $R_{\rm H}$ are almost identical to fiducial cases with $q_{\rm b}=1$.  Within $R_{\rm H}$, the flow is dominated by the gravitational potential of the high-mass component of the binary ($m_1$) that slowly moves along a small circular orbit with a radius of $r_1=a_{\rm b}/11$.  The CSD of $m_1$ and its spiral shocks are therefore larger and steadier than those in the fiducial cases (see Fig. \ref{fig:runI_snapshot} and Fig. \ref{fig:runII_snapshot}).  In contrast, the CSD of $m_2$ is heavily disrupted and suppressed because $m_2$ moves $10$ times faster on a larger circular orbit ($r_2 = 10a_{\rm b}/11$) and experiences regular ram pressure stripping from the spiral shocks of the CSD of $m_1$.

In such flow structure, even streamlines close to $m_2$ are easily deflected away, reducing the accretion onto $m_2$.  On the other hand, if $r_2$ is large enough, $m_2$ would be able to dive farther into the shear flow at phase $\pi$ or $2\pi$ to accrete more materials (see also Sections \ref{subsubsec:orbital_evolution} and \ref{subsec:result_ecc}).  The latter effect, however, is only prominent for small $q_{\rm b}$ and for relatively steady accretion flows.  Fig. \ref{fig:runI+II_qbs} shows the time series of $\dot{m}_{\rm b}$, $\dot{\mathcal{E}}_{\rm b}$, and $\dot{a}_{\rm b}$ for \texttt{Run I-q$_{\mathtt{b}}$} with $q_{\rm b}=0.1$, $0.5$, and $0.9$.  We find that $\langle\dot{m}_2\rangle$ is much larger than $\langle\dot{m}_1\rangle$ in the $q_{\rm b}=0.1$ case, as expected.  For $q_{\rm b}=0.5$, $\langle\dot{m}_2\rangle$ instead becomes slightly smaller than $\langle\dot{m}_1\rangle$ since $r_2$ is not large enough to counterbalance the disrupted and suppressed accretion.  For $q_{\rm b}=0.9$, the gravitational potential is no longer dominated by $m_1$ and the accretion rates of the two binary components are very close to each other, similar to the accretion in the fiducial case.  Furthermore, all the time series of $\dot{m}_{\rm b}$ in Fig. \ref{fig:runI+II_qbs} exhibit periodic variations and the resulting PSDs (not shown) are similar to the PSDs of the fiducial run, with a single dominant frequency at $2\Omega_{\rm b}'$ (see Section \ref{subsec:result_ecc} and Fig. \ref{fig:trends_eccs_PSD}).

Fig. \ref{fig:trends_qbs} presents the secular orbital evolution results as a function of $q_{\rm b}$ for both run series.  Similar to how $\langle\dot{m}_2\rangle$ varies relative to $\langle\dot{m}_1\rangle$ in Fig. \ref{fig:runI+II_qbs}, we find that $\eta$ for \texttt{Run I-q$_{\tt b}$} first declines slightly with decreasing $q_{\rm b}$ and then rises prominently when $q_{\rm b} \lesssim 0.3$.  The total accretion rate $\langle\dot{m}_{\rm b}\rangle$ first remains approximately constant and then increases noticeably at small $q_{\rm b}$.  For \texttt{Run II-q$_{\tt b}$}, the flow is more chaotic.  We find that both $\langle\dot{m}_{\rm b}\rangle$ and $\eta$ exhibit only small/moderate changes as $q_{\rm b}$ decreases from $1$ to $0.1$.  Overall, the dependence of $\eta$ on $q_{\rm b}$ in our simulations is qualitatively different from the monotonic increase of $\eta$ with decreasing $q_{\rm b}$ found in viscous circumbinary accretion \citep[e.g., see Fig. 7 of][]{Munoz2020}.

Fig. \ref{fig:trends_qbs} further shows that $\ell_0$, $\langle\dot{\mathcal{E}}_{\rm b}\rangle$, and $\langle\dot{a}_{\rm b}\rangle/a_{\rm b}$ generally increase (or become less negative) with decreasing $q_{\rm b}$ in both run series.  In other words, a smaller mass ratio leads to less negative accreted angular momentum per unit of accreted mass, a slower negative energy transfer rate, and a slower orbital decay rate.  Moreover, the binary gains angular momentum (i.e., $\ell_0$ transitions from negative to positive) when $q_{\rm b} \leqslant 0.3$ in \texttt{Run I-q$_{\tt b}$} and $q_{\rm b} \leqslant 0.4$ in \texttt{Run II-q$_{\tt b}$}.  This transition occurs roughly when $r_2$ becomes large enough such that $\dot{m}_2$ and $\dot{m}_{\rm b}$ start to increase.  However, since $\ell_{0,\mathrm{crit}}$ increases with $\eta$, we find that all binaries in \texttt{Run I-q$_{\tt b}$} are contracting, whereas only binaries with $q_{\rm b} \geqslant 0.4$ in \texttt{Run II-q$_{\tt b}$} are contracting.

\section{Comparisons with Previous Studies}
\label{sec:comp2previous}

As noted in Section \ref{sec:intro}, \citetalias{Baruteau2011} and \citetalias{LiYaPing2021} obtained conflicting results on whether prograde equal-mass circular BBHs in AGN discs would contract or expand.  Our simulations show that such stellar BBHs are contracting.  We now compare to these previous works in more details.  Admittedly, the local disc models employed in our simulations are geometrically different than these former global disc models.  That said, both previous studies concluded that the gas inside the Hill radius is largely responsible for the orbital evolution of the binary.

\citetalias{Baruteau2011} simulated a non-accreting binary (see their Section 6) with $q = 10^{-3}$ in a \rlr{locally isothermal} disc with $h=0.05$ and $\alpha=4\times 10^{-3}$ (the Shakura-Sunyaev parameter), where the binary separation $a_{\rm b}$ was resolved by about $\rlr{15}$ and $\rlr{10}$ cells in the radial and azimuthal directions, respectively.  The gravitational potential of the binary was modelled with a softening length $\epsilon=\rlr{0.5} a_{\rm b}$.  \citetalias{Baruteau2011} found that the binary would contract rapidly, where the hardening timescale is 1 to 2 orders of magnitude shorter than the migration timescale.  Though their modeling of the vicinity of the binary is limited by the resolution, the large $\epsilon$, and the lack of accretion, they properly resolved the spiral tails that harden the binary.  This finding is consistent with our analyses in Section \ref{subsubsec:runI_Tgrav_map}, where the two trailing small half bow shocks ($0.8a_{\rm b} \lesssim r \lesssim R_{\rm H}$; see Fig. \ref{fig:runI_Tgrav_map}) persistently lag and provide negative torques.

Since the binary used by \citetalias{Baruteau2011} is $3$ orders of magnitude more massive than ours and does not accrete gas, quantitative comparison is not straightforward.  \citetalias{Baruteau2011} argued that the disc parameters were chosen such that the binary-disc interactions can be rescaled and applied to a binary with $q \sim 10^{-5}$, where the gas gap profile and how the binary would migrate ``remain essentially unchanged''.  Such rescaling, however, affects the orbital evolution of the binary.  A rescaled experiment with $q=10^{-5}$ by \citetalias{Baruteau2011} showed that the hardening timescale shortens by one order of magnitude due to the different relative weight of the binary to the surrounding gas.  Thus, caution is needed in extrapolating the results obtained for larger $q$ to smaller $q$. 

\citetalias{LiYaPing2021} simulated an accreting binary (see their Model A) with $q = 2\times 10^{-3}$ in a \rlr{locally isothermal} disc with $h=0.08$ and $\alpha=6.35\times 10^{-3}$, where $a_{\rm b}$ was resolved by about $\rlr{25}$ cells and $\epsilon = \rlr{0.08} a_{\rm b}$.  In addition, the accretion was modelled by gradually removing gas within a distance of $r_{\rm acc} = \epsilon$ to each accretor.  \citetalias{LiYaPing2021} found that the binary would expand, rather than contract.  They argued that, by adopting a much smaller $\epsilon$, their simulations adequately resolved the CSD regions that soften the binary.  

In this work, we simulate an accreting binary with $q = 10^{-6}$ in a disc with $h=0.01$ \rlr{and with the $\gamma$-law EOS with $\gamma=1.6$}, where $a_{\rm b}$ is resolved by about $246$ cells, much higher than that in previous works.  Moreover, we adopt an absorbing accretion prescription (see Section \ref{subsec:acc_T}) with a sink radius $r_{\rm s}$ that is resolved by about $10$ cells.  This treatment is numerically robust (see Section \ref{subsubsec:deps_numParas}) and allows us to set $\epsilon=10^{-8} a_{\rm b}$ such that the accretion flows are more accurately modelled under the least modified binary potential (see also Fig. \ref{fig:torque_vs_rs} and Section \ref{subsubsec:r_sink_deps}).  Furthermore, our accretion prescription enables us to take into account the non-negligible torques resulted from accretion and pressure, which were ignored in previous works.

\rlr{Our simulations also sufficiently resolve the CSD regions that soften the binary, but we find that the trailing small half bow shocks harden the binary even faster (see Section \ref{subsubsec:runI_Tgrav_map}).  In our \citetalias{Li2022b}, we demonstrate that the main reason for the discrepancy between \citetalias{LiYaPing2021} and this work is the EOS.  The CSDs in isothermal simulations are much more massive, cooler, and less turbulent (than the CSDs seen in this work) and their positive gravitational torques dominate over those from other structures, leading to expanding binaries (see our \citetalias{Li2022b} for more details).}

Finally, we note that \citetalias{Baruteau2011} used the \texttt{FARGO} code \citep{Masset2000} and \citetalias{LiYaPing2021} used the \texttt{LA-COMPASS} code \citep{LiHui2005, LiHui2009}, while we use \texttt{ATHENA}.  All three codes have been extensively tested, though a detailed code comparison would be needed to attribute any differences to the algorithms (which is not our goal).

\section{Summary and Discussions}
\label{sec:summary}

\subsection{Key Results and Implications}
\label{subsec:res_impl}

We have studied the evolution of binary black holes (BBHs) embedded in AGN discs using a suite of 2D inviscid hydrodynamical simulations in local shearing boxes. We use the $\gamma$-law equation of state, and consider a range of values for the binary semi-major axis $a_b$ (relative to the Hill radius $R_H$), eccentricity $e_b$ and mass ratio $q_b=m_2/m_1$.  We adopt the BBH to SMBH mass ratio $q=m_b/M$ and the disc aspect ratio $h$ (measuring the gas sound speed) such that $hq^{-3}\simeq 1$, as appropriate for BBHs in AGN discs (see Eqs. \ref{eq:q}--\ref{eq:lambda} and Eqs. \ref{eq:c_s_v_b}--\ref{eq:Delta_V_K_v_b} for the relevant dimensionless parameters).  We use multi-level mesh refinements and an absorbing boundary condition to mimic BH accretion, resolving the accretion flow down to a few percent of the binary separation.  Our prescription for accretion onto the binary components and our on-the-fly post-processing treatment (described in Section \ref{subsec:acc_T}) robustly evaluate the accretion rate, the angular momentum and energy transfer rates onto the binary, taking account of the gravitational force (``dynamical friction'') and the hydrodynamical forces associated with gas accretion and pressure. These allow us to determine the long-term secular evolution of the binary for various parameters (see Tables \ref{tab:run_I} and \ref{tab:run_II}).

Our key findings are as follows:
\begin{enumerate}
  \item
  In all of our numerical runs, the accretion flow around the binary settles into a variable, but quasi-steady state after an initial transient phase, with almost identical large-scale flow structures   outside the Hill radius (see Figures \ref{fig:runI_snapshot} and \ref{fig:runII_snapshot}).  For prograde equal-mass circular binaries, the dominant accretion variability (see Figure \ref{fig:runI+II_breakdown}) has a frequency $2\Omega_{\rm b}'$, where $\Omega_{\rm b}'=\Omega_{\rm b} + \Omega_{\rm pre}$ is the apparent orbital frequency of the binary in the corotating (shearing-box) frame (see Eqs. \ref{eq:Omega_pre} and\ref{eq:Omega_bprime} and Figure \ref{fig:trends_eccs_PSD}).  As the binary eccentricity $e_b$ increases, the dominant variability frequency gradually switches to the radial epicyclic frequency $\Omega_{\rm b}=\sqrt{Gm_b/a_b^3}$.
  \item
  For all the cases studied in this paper except \texttt{Run II-q$_{\mathtt{b}}$} with $q_{\rm b} \leqslant 0.3$ (which have $e_b=0$ and $\lambda=R_H/a_b=5$), we find that the binary contracts with a time-averaged orbital decay rate $\langle\dot{a}_{\rm b} \rangle / a_{\rm b}$ of the order of a few times $\langle\dot{m}_{\rm b}\rangle / m_{\rm b}$ (see Tables \ref{tab:run_I} and \ref{tab:run_II}), where the mass accretion rate $\langle\dot{m}_{\rm b}\rangle$ scales as $\Sigma_\infty v_b a_b$ (where $\Sigma_\infty$ is the background disc surface density and $v_b=\sqrt{Gm_b/a_b}$).  The numerical values of $\langle\dot{a}_{\rm b}\rangle$ and $\langle\dot{m}_{\rm b}\rangle$ depend on $\lambda$, $e_b$, $q_b$, pro/retrograde rotation of the binary, and the size of the accretor.  We note that even when the binaries are contracting, those with small $q_{\rm b}$ or large $e_{\rm b}$ may gain angular momentum from accretion (i.e. $l_0>0$).
  \item 
  The three torque components, $\langle\dot{{L}}_{\rm b,acc}\rangle$, $\langle\dot{{L}}_{\rm b,pres}\rangle$ and $\langle\dot{{L}}_{\rm b,grav}\rangle$ (see Eq. \ref{eq:dotL_breakdown}), associated with accretion, pressure and gravity, respectively, can have comparable contributions to the total torque on the binary and thus to the orbital evolution (see Fig. \ref{fig:runI+II_breakdown}).  For prograde equal-mass circular binaries, the gravitational torque $\langle\dot{{L}}_{\rm b,grav}\rangle$ is largely determined by the persistent non-axisymmetric flow structures within $R_{\rm H}$ (see Fig. \ref{fig:runI_Tgrav_map}).
  \item 
  The physical size of the accretor (i.e. the sink radius $r_s$) affects the morphology of circum-single discs (CSDs) and the accretion rate.  For prograde equal-mass circular binaries, $\langle\dot{m}_{\rm b}\rangle$ roughly scales linearly with $r_{\rm s}$ (for $r_s$ between $0.02a_b$ and $0.08a_b$) (see Figs. \ref{fig:runI_snapshot_rs} and \ref{fig:RunI_trends_r_s}); a smaller $r_s$ results in a less truncated CSD, a more negative total torque (see Fig. \ref{fig:torque_vs_rs}), and a faster orbital decay rate.
  \item 
  Prograde equal-mass \textit{eccentric} binaries experience significant eccentricity damping when $e_b\gtrsim 0.2$, with the damping rate (in units of $\langle{\dot m}_b\rangle/m_b$) increasing with $e_b$.  A higher eccentricity generally leads to somewhat faster accretion and slower orbital decay rate (see Fig. \ref{fig:trends_eccs}).
  \item 
  For \textit{retrograde} equal-mass circular binaries, the CSDs cannot form due to severe ram pressure stripping; the accrete rate becomes much higher, leading to a faster orbital decay (see Fig. \ref{fig:RunI-ret}).
  \item 
  For prograde \textit{unequal-mass} circular binaries, the orbital decay rate (in units of $\langle{\dot m}_b\rangle/m_b$) generally decreases with decreasing $q_b$, and the binary may switch to orbital expansion for sufficiently small $q_b$ (as in Run II-$q_b$) (see Fig. \ref{fig:trends_qbs}).  The low-mass component ($m_2$) of the binary may accrete less mass ($\eta < 0.5$) at intermediate $q_{\rm b}$ due to the disrupted accretion flows; only at  sufficiently low $q_{\rm b}$ does $m_2$ accrete significantly more than $m_1$.
\end{enumerate}

It is of interest to compare the hardening timescale of the contracting binary with its migration timescale through the disc.  Our simulation results can be summarized as
\begin{equation}
  \frac{\langle\dot{a}_b\rangle}{a_b} = - \mathcal{F}{\Sigma_\infty a_b^2\Omega_b\over m_b},
\end{equation}
with $\mathcal{F}\sim 1$ for prograde binaries ($\mathcal{F}\simeq 0.4-1.3$ over a wide range of binary mass ratios and eccentricities) and $\sim 8-10$ for retrograde binaries (see Tables \ref{tab:run_I} and \ref{tab:run_II}).  On the other hand, the Type I migration rate of the binary in the disc is given by \citep[e.g.,][]{Ward1997,Cresswell2008}
\begin{equation}
\frac{\dot{R}}{R}\simeq - \frac{m_b\Sigma_{\infty} R^2}{M^2} h^{-2}\Omega_{\rm K},
\end{equation}
where $\Omega_K=\sqrt{GM/R^3}$.  The ratio between the binary contraction rate and the migration rate is then
\begin{equation}
\frac{\langle\dot{a}_{\rm b}\rangle / a_{\rm b}}{\dot{R}/R} \simeq \mathcal{F} h^2 q^{-4/3} \lambda^{-1/2}=10^4 \mathcal{F}\lambda^{-1/2}\left({h\over 10^{-2}}\right)^2 \left({q\over 10^{-6}}\right)^{-4/3},
\end{equation}
where $\lambda\equiv R_H/a_b\equiv Rq^{1/3}/a_b \gtrsim$ a few.  It is clear that the rate ratio is $\gg 1$ for all reasonable binary and disc parameters.

Our findings demonstrate that the hydrodynamical evolution of binaries embedded in accretion discs can be quite different from that of isolated binaries in their own circumbinary discs. The latter may experience orbital expansion instead of contraction \citep[][etc.]{Munoz2019, Moody2019, Munoz2020}.  Also, the accretion in a circumbinary disc is is regulated by viscosity, and the dominant variability frequency can be either $\simeq 0.2\Omega_{\rm b}$ or $\Omega_{\rm b}$, depending on the binary eccentricity and mass ratio \citep{Munoz2016, Munoz2020}.  We thus urge caution when predicting the orbital evolution of binaries in accretion discs based on the knowledge of isolated binaries.

\subsection{Possible Limitations and Further Works}
\label{subsec:limit} 

We note that our results are subject to several possible limitations.
\begin{itemize}
  \item[$\bullet$] \textbf{Parameters space:} Although our simulations have covered a range of binary eccentricities and mass ratios, a broader parameter survey would be desirable. Eqs. \ref{eq:c_s_v_b}--\ref{eq:V_s_v_b} show that the most important dimensionless parameters that determine the flow dynamics are $\lambda\equiv R_H/a_b$ and $h/q^{1/3}$.  Note that the ratio of the Bondi radius $r_B=G m_b/c_{\mathrm{s},\infty}^2$ to $a_b$ is $r_B/a_b=(v_b/c_{\mathrm{s},\infty})^2=(q^{1/3}/h)^2\lambda$.  In this paper, we have only covered $h/q^{1/3}=1$, and $\lambda=2.5,5$; it will be important to consider different values for these two parameters.  \rlr{In particular, binaries with $\lambda=2.5$ may experience long-term dynamical instability due to the perturbations of the SMBH.  We neglect this instability in this work since we want to explore the smallest $\lambda$ possible, which is also easier to model numerically.}  In addition, we have adopted the $\gamma$-law equation of state with $\gamma=1.6$; different values of $\gamma$ and more sophisticated equation of state may affect our results.
  \rlr{In our \citetalias{Li2022b}, we systematically study the dependence of our results on $h/q^{1/3}$, $\lambda$, and $\gamma$ with an extensive coverage of the parameter space.  We show that all three parameters play a significant role in determining the flow structure, the total torque on the binary, and its orbital evolution.}
  \item[$\bullet$] \textbf{Shearing box approximation:} This work uses the 2D local shearing box approximation in order to properly resolve the accretion flow around the binary.  This approximation comes with a natural geometric limitation: strictly speaking, our results are valid only for the cases where the Hill radius $R_{\rm H}=Rq^{1/3}$ is comparable to $H_{\rm g}$, the scale height of the disc, i.e. $q^{1/3} \sim h$.  For $R_{\rm H}\ll H_{\rm g}$, 3D simulations may be needed to better model the disc structure and take account of the meridional flow.  \rlr{\citet{Dempsey2022} conducted 3D shearing-box simulations and found that 3D simulations tend to have more negative torques than their 2D counterparts.}  For $R_{\rm H} \gg H_{\rm g}$, the binary is expected to open a deep gap in the disc. To study the resulting binary orbital evolution, a global disc model may be needed to self-consistently capture the gap profile (e.g., depth, width, etc.).
  \item[$\bullet$] \textbf{Sink accretion and viscosity:} 
  Our simulations adopt a sink prescription to mimic gas accretion onto the individual binary component.  However, we find that the mass accretion rate and the orbital evolution rate are influenced by the choice of the sink radius $r_s$ (see Section \ref{subsubsec:r_sink_deps}), implying the need of linking $r_{\rm s}$ to a physically motivated size of the accretor.  For instance, different components of unequal-mass binaries may require different sink radii, especially when $q_{\rm b} \ll 1$, which may change the accretion flow structure and alter the secular orbital evolution results.  \rlr{In addition, one may propose an alternative sink prescription where some angular momentum is kept outside the sink radius while gas is being accreted, mimicking the physical transport of angular momentum in discs.  This prescription would be possible with viscosity, but we only simulate inviscid hydrodynamics in this work.}  Including viscosity may regulate the accretion rate throughout the CSD and weaken the dependence of the binary orbital evolution rate on the sink radius.  We defer such studies to a future work \rlr{(Li \& Lai, in prep)}.
\end{itemize}

\section*{Acknowledgements}
This work has been supported in part by the NSF grant AST-2107796 and the NASA grant 80NSSC19K0444.  Resources supporting this work were provided by the NASA High-End Computing (HEC) Program through the NASA Center for Climate Simulation (NCCS) at Goddard Space Flight Center.

We thank the anonymous referee for useful suggestions.  RL thanks Kaitlin Kratter, Hui Li, Diego Muñoz, Ya-Ping Li, Adam Dempsey, Zoltan Haiman, Yan-Fei Jiang, Paul Duffell, and Barry McKernan for inspiring discussions and useful conversations. This research was supported in part by the National Science Foundation under Grant No. NSF PHY-1748958.

\section*{Data Availability}
The simulation data underlying this paper will be shared on reasonable request to the corresponding author.

\bibliographystyle{mnras}
\bibliography{refs}



\appendix

\begingroup 
\setlength{\medmuskip}{0mu} 
\setlength\tabcolsep{4pt} 
\setcellgapes{3pt} 
\begin{table*}
  \nomakegapedcells
  \caption{Simulation Results for Fiducial Runs without $\dot{\varpi}$}\label{tab:runs_nopomega}
  \makegapedcells 
  \linespread{1.025} 
  \begin{tabular}{lcccccccc|rrrrrr}
    \hline
    \texttt{Run}
    & $q_{\rm b}$
    & $e_{\rm b}$
    & $\displaystyle \frac{r_{\rm s}}{a_{\rm b}}$
    & $\displaystyle \frac{r_{\rm e}}{a_{\rm b}}$
    & $L_X\times L_Y$
    & \footnotesize $N_{\rm SMR}$
    & $\displaystyle \frac{a_{\rm b}}{\delta_{\rm fl}}$
    & Remarks
    & \makecell[c]{$\langle\dot{m}_{\rm b}\rangle$}
    & \makecell[c]{$\langle\dot{\mathcal{E}}_{\rm b}\rangle$}
    & \makecell[c]{$\ell_0$}
    & \makecell[c]{$\displaystyle \frac{\langle\dot{a}_{\rm b}\rangle}{a_{\rm b}}$}
    & \makecell[c]{$\langle\dot{e_{\rm b}^2}\rangle$}
    & \makecell[c]{$\eta$} \\

    &
    &
    & $[\%]$
    & $[\%]$
    & $[a_{\rm b}^2]$
    &
    &
    &
    & \makecell[c]{\footnotesize $[\Sigma_{\infty} v_{\rm b} a_{\rm b}]$}
    & \makecell[c]{\footnotesize $\displaystyle \left[\frac{\Sigma_\infty v_{\rm b}^3 a_{\rm b}}{m_{\rm b}}\right]$}
    & \makecell[c]{$[v_{\rm b} a_{\rm b}]$}
    & \makecell[c]{\small $\displaystyle \left[\frac{\langle\dot{m}_{\rm b}\rangle}{m_{\rm b}}\right]$}
    & \makecell[c]{\small $\displaystyle \left[\frac{\langle\dot{m}_{\rm b}\rangle}{m_{\rm b}}\right]$}
    & \\ 
    \hline\hline
    \texttt{I-no$\dot{\varpi}$}
    & $1.0$ & $0.0$ & $4$ & $4.75$ & $25\times25$ & $6$ & $245.76$
    & 
    & $0.17$ & $-0.68$ & $-0.47$ & $-6.77$ & $0.00$ & $0.50$ \\
    \texttt{II-no$\dot{\varpi}$}
    & $1.0$ & $0.0$ & $4$ & $4.75$ & $50\times50$ & $7$ & $245.76$
    & 
    & $0.17$ & $-0.46$ & $-0.17$ & $-4.40$ & $0.00$ & $0.50$ \\
    \hline
  \end{tabular} \\
  \begin{flushleft}
    {\large N}OTE --- Similar to Tables \ref{tab:run_I} and \ref{tab:run_II}, \texttt{Run I-no$\dot{\varpi}$} adopts $\lambda=2.5$ and \texttt{Run II-no$\dot{\varpi}$} adopts $\lambda=5$.  Both runs use $q=1$e-$6$ and $h=0.01$ and last $500\Omega_{\rm b}^{-1}$.  The results are time-averaged over the last $300\Omega_{\rm b}^{-1}$ for \texttt{Run I-no$\dot{\varpi}$} and over the last $240\Omega_{\rm b}^{-1}$ for \texttt{Run II-no$\dot{\varpi}$}.
  \end{flushleft}
\end{table*}
\endgroup

\section{Orbital Evolution without $\dot{\varpi}$}
\label{appsec:no_pomega}

We present the orbital evolution results of \texttt{Run I-no$\dot{\varpi}$} and \texttt{Run II-no$\dot{\varpi}$}, which are similar to our fiducial runs but without the apsidal precession (i.e., neglecting $\dot{\varpi}$ in Eq. \ref{eq:Omega_pre}).  Table \ref{tab:runs_nopomega} summarizes the key parameters and shows the time-averaged measurements for binary evolution.  We find that the binaries in both runs are contracting and their orbital decay rates are only somewhat slower than those in our fiducial runs.  This finding is expected since the lack of the apsidal precession only reduces the apparent binary orbital velocity and hence the torques acting on the binary. 

\section{Validation of Accretion Prescription and Force Computations}
\label{appsec:vali_BHL}

We test our prescription for accretion and our methods to evaluate the accretion rate and forces acting on an accretor (see Section \ref{subsec:acc_T}) through a suite of 2D Bondi-Hoyle-Lyttleton (BHL) accretion simulations \citep{Edgar2004}.  In a well established state of BHL accretion, we expect that the accretion rate to agree well with the analytical result.  Furthermore, we expect that the accretion rate and the total force acting on the accretor has little dependency on the evaluation radius $r_{\rm e}$ due to mass conservation and momentum balance \citep{Thun2016}.

We consider an accretor with a given mass $m$, a sink radius $r_{\rm s}$, and a velocity $v_\infty$, moves through an initially uniform, static background gas with surface density $\Sigma_{\infty}$, sound speed $c_{\mathrm{s},\infty}$, and a gamma-law equation of state with $\gamma = 3$ (the 2D equivalent of $5/3$ in 3D, above which the accretion does not have a sonic point; see Eq. \ref{eq:mdot_Bondi} below).  Our computational domain has a reference frame co-moving with the accretor such that gas is uniformly initialized with $\bm{u} = v_\infty \hat{\bm{x}}$.

The BHL formalism provides natural code units and analytical estimates for accretion rates \citep{Antoni2019}.  When $\mathcal{M}a \equiv v_\infty / c_{\mathrm{s},\infty} \gg 1$ (i.e., highly supersonic), the system eventually reaches a steady-state with a persistent bow shock in front the accretor \citep{Xu2019}.  The characteristic scale is the Hoyle-Lyttleton radius
\begin{equation}
  r_{\rm HL} = \frac{2 G m}{v_\infty^2},
\end{equation}
and the estimated accretion rates are
\begin{align}
  \dot{m}_{\rm HL} &= \frac{4 G m \Sigma_\infty}{v_\infty}, \label{eq:mdot_HL} \\
  \dot{m}_{\rm BHL} &= \dot{m}_{\rm HL} \left(\frac{\mathcal{M}a^2}{1 + \mathcal{M}a^2} \right)^{1/2} = \frac{4 G m \Sigma_\infty}{\sqrt{v_\infty^2 + c_{\mathrm{s},\infty}^2} }. \label{eq:mdot_BHL}
\end{align}
In our simulations, we use $r_{\rm HL}$, $r_{\rm HL}/c_{\mathrm{s},\infty}$, $G^{-1}$, and $\Sigma_\infty r_{\rm HL}^2$ as the code units for length, time, mass for accretor, and mass for background gas, respectively.

For $\mathcal{M}a \ll 1$, the flow is spherical and the Bondi accretion rate is
\footnote{See Li \& Lai, in preparation for derivation.}
\begin{equation}
  \dot{m}_{\rm Bondi} = q_{\rm s,2D}(\gamma) \frac{2\pi G m \Sigma_\infty}{c_{\mathrm{s},\infty}}, \label{eq:mdot_Bondi}
\end{equation}
where $q_{\rm s,2D}(\gamma) = \left[2/(3 - \gamma)\right]^{(3 - \gamma)/(2\gamma - 2)}$ and $q_{\rm s,2D}(\gamma = 3) = 1$. 

Similar to the binary modelled in the main text, we model the single accretor at the frame origin as an absorbing sphere (circular boundary) with $r_{\rm s} = 0.01 r_{\rm HL}$.  The root domain extends to $[-2, 6] r_{\rm HL} \times [-16, 16] r_{\rm HL}$ in $x$ and $y$, respectively.  We refine the mesh towards the accretor through 7 refinement levels such that the accretor is resolved with $r_{\rm s}/\delta_{\rm fl} = 10.24$ cells at the finest level.  The boundary condition along $x=-2 r_{\rm HL}$ is a uniform inflow of wind.  We choose the boundary conditions along $x=6 r_{\rm HL}$ to be outflow and along $y=\pm 16 r_{\rm HL}$ to be periodic.  We adopt a domain with a much larger $y$-extent so that the bow shocks only touch the $+x$ boundary, minimizing the influence of the boundary conditions on the flow structures. 

\begin{figure*}
    \centering
    \includegraphics[width=\linewidth]{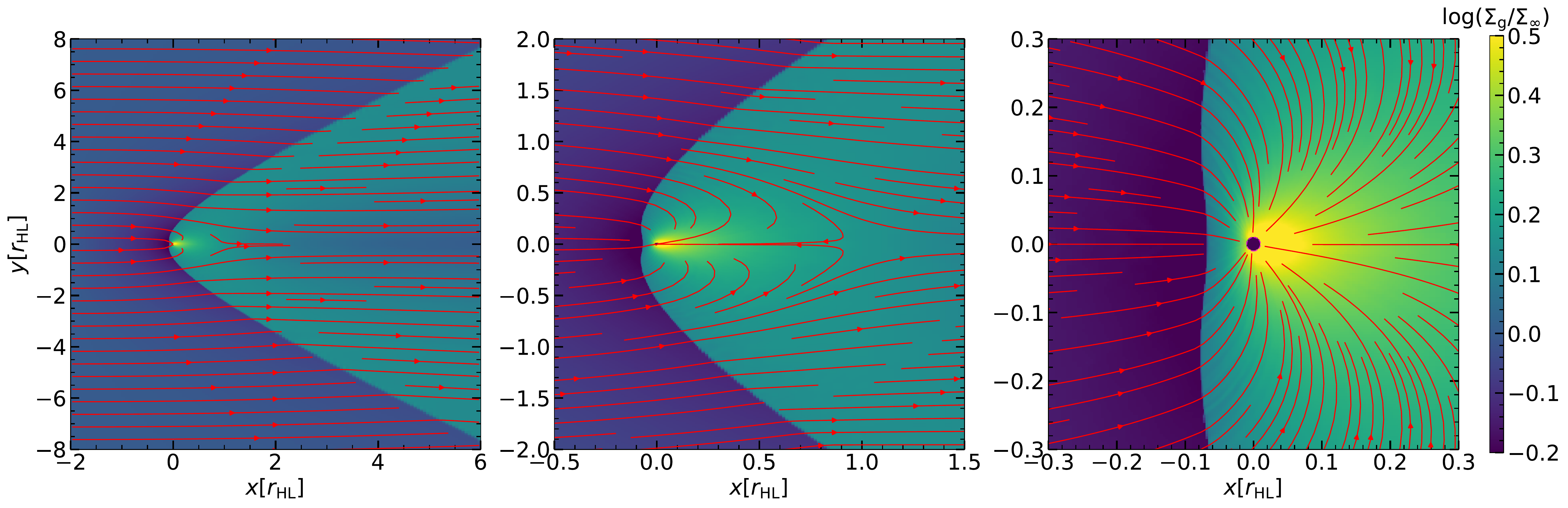}
    \caption{The final snapshot of our 2D BHL simulation with $\gamma=3$ and $\mathcal{M}a = 2$, where the mesh is refined progressively towards the accretor and the red streamlines show the detailed flow structures.  The \textit{magenta dashed} circle with a radius of $r_{\rm s}=0.01 r_{\rm HL}$ represents the sink radius.  \label{fig:BHL_snapshot}}
\end{figure*}

We simulate seven values of $\mathcal{M}a$ ($0.5$, $1$, $1.4$, $2$, $3$, $4$, $5$) and run each simulation for $60 r_{\rm HL}/c_{\mathrm{s},\infty}$.  Fig. \ref{fig:BHL_snapshot} shows the final snapshot of our run with $\mathcal{M}a=2$, where a vertically symmetric bow shock has formed on the upstream side and the gas close to the accretor flows almost radially onto the accretor, similar to what have been seen in previous studies \citep[e.g.,][]{Xu2019} .

\begin{figure}
  \centering
  \includegraphics[width=0.9\linewidth]{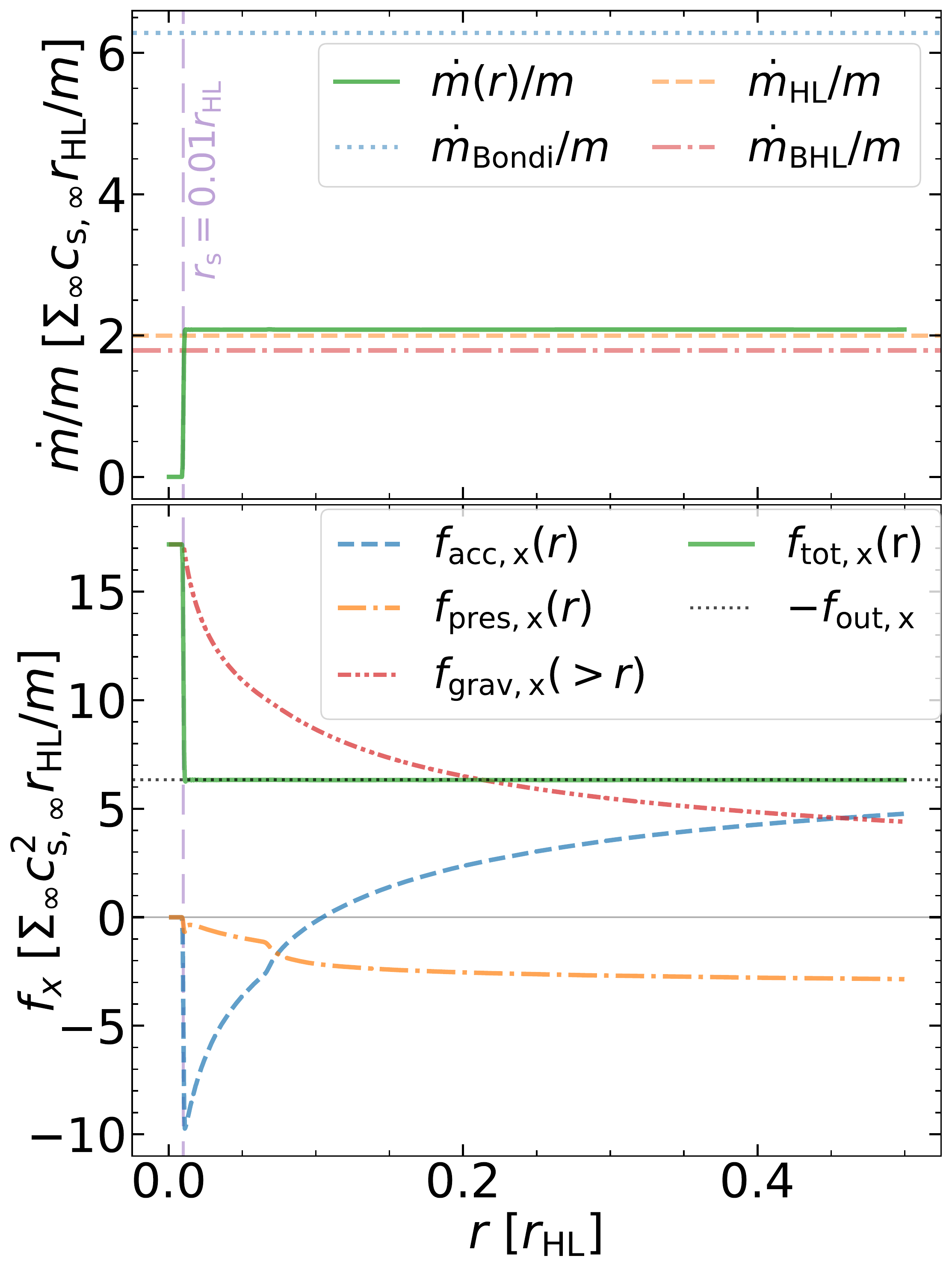}
  \caption{Measured specific accretion rate (\textit{green solid}) and specific forces ($\hat{\bm{x}}$-components) as a function of the evaluation radius, analyzed from the snapshot shown in Fig. \ref{fig:BHL_snapshot}.  The \textit{upper} panel also shows the analytical estimates for $\dot{m}_{\rm Bondi}$ (\textit{blue dotted}), $\dot{m}_{\rm HL}$ (\textit{orange dashed}), and $\dot{m}_{\rm BHL}$ (\textit{red dash-dotted}), all with a factor of $m^{-1}$.  The \textit{lower} panel presents the total specific force acting on the accretor $f_{\mathrm{tot},x}$ (\textit{green solid}), summed from the accretion force $f_{\mathrm{acc},x}$ (\textit{blue dashed}), the pressure force $f_{\mathrm{pres},x}$ (\textit{orange dash-dotted}), and the gravitational force $f_{\mathrm{grav},x}$ (\textit{red dash-dot-dotted}).  The \textit{black dotted} line denotes the specific force integrated from the outer boundary $f_{\mathrm{out},x}$ (see text for more details).  The \textit{violet dashed} vertical line across the two panels indicate the sink radius.  \label{fig:mdot_fx_BHL}}
\end{figure}

Based on the final snapshot, we linearly interpolate gas surface density, momentum density, and energy density onto a structured polar grid centered at the accretor that extends to $[0, 0.5] r_{\rm HL} \times [-\pi, \pi]$ in $r$ and $\theta$, respectively.  At each grid circle along the radial axis, or, in other words, at each evaluation radius, we calculate the accretion rate $\dot{m}$ based on Eq. \ref{eq:m_dot_i}.  Fig. \ref{fig:mdot_fx_BHL} presents all the measurements as a function of $r$.  We find that $\dot{m}$ is almost a constant outside the sink radius and is somewhat higher than $\dot{m}_{\rm BHL}$ and $\dot{m}_{\rm HL}$.  The slight offset towards $\dot{m}_{\rm Bondi}$ is because $\mathcal{M}a=2$ is not highly supersonic enough.

\begin{figure}
  \centering
  \includegraphics[width=0.8\linewidth]{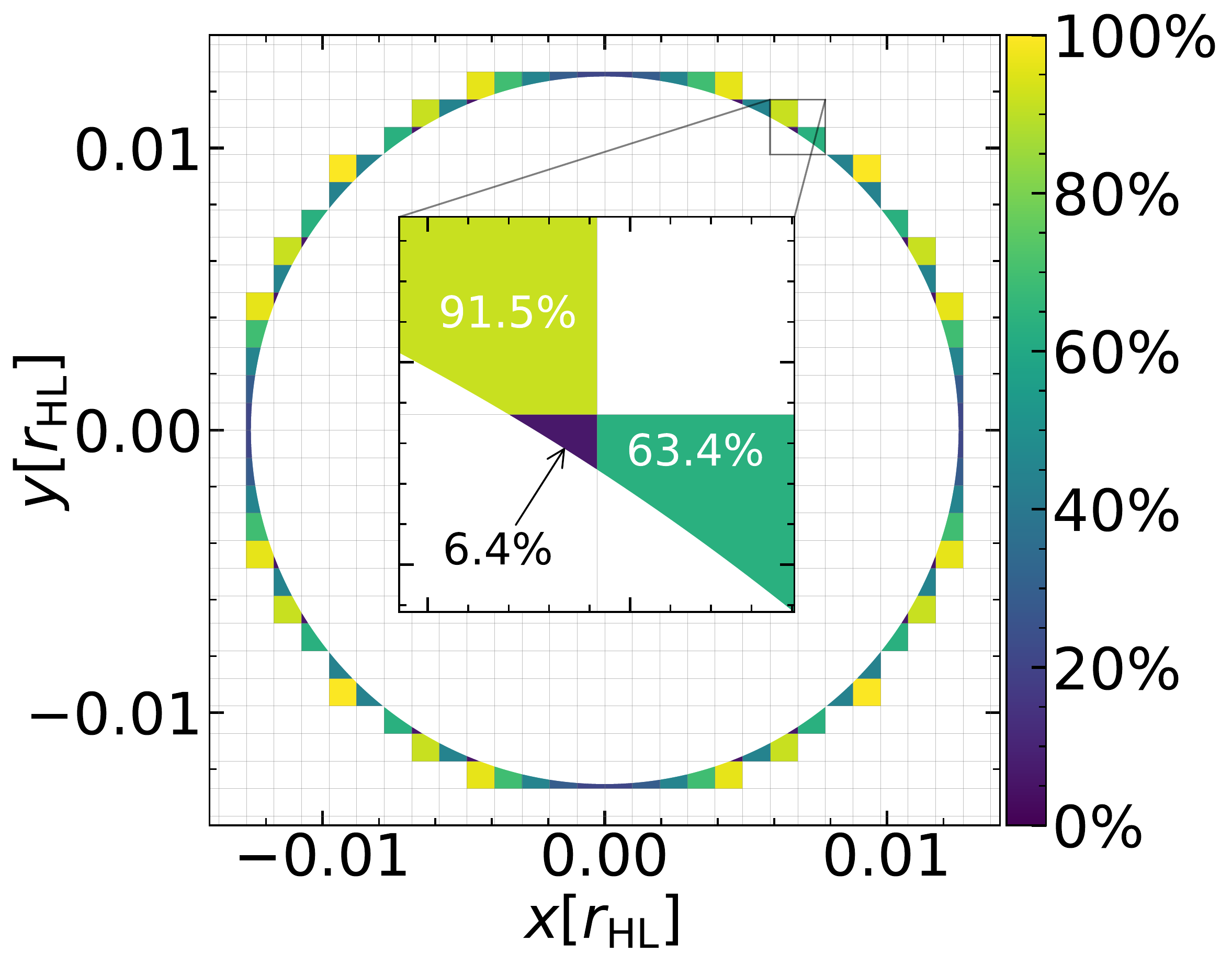}
  \caption{Demonstration on calculating the sub-cell contributions to $f_{\mathrm{grav},x}(>r)$ from cells intersecting a certain evaluation radius.  These cells are color-coded to show their area percentages that are outside the evaluation radius, which are multiplied by their specific gravitational forces on the accretor.  \label{fig:geos}}
\end{figure}

We also evaluate the $\hat{\bm{x}}$-components of all the specific forces $f_{\mathrm{acc},x}$, $f_{\mathrm{pres},x}$, and $f_{\mathrm{grav},x}(>r)$ acting on the accretor based on Eqs. \ref{eq:f_hydro}, \ref{eq:f_pres}, and \ref{eq:f_grav}.  The former two forces only rely on the interpolated quantities, while $f_{\rm grav,x}(>r)$ requires integrating specific gravitational forces from gas in cells at the Cartesian grid outside $r$.  To properly account for cells that intersect with a certain evaluation radius, we use a geometry engine \texttt{GEOS} \citep{GEOS} to compute their sub-cell contributions based on the area percentage that is outside $r$ (see Fig. \ref{fig:geos} for an example).

At each radius, the sum of these specific forces must be balanced by the corresponding momentum transport and pressure forces at the outer boundary of the area covered by the integral of $f_{\mathrm{grav},x}(>r)$ \citep[e.g., domain boundaries; see Section 3.5 in ][]{Thun2016}
\begin{equation}
  f_{\mathrm{acc},x} + f_{\mathrm{pres},x} + f_{\mathrm{grav},x}(>r) + f_{\mathrm{out},x} = 0, \label{eq:fx_balance}
\end{equation}
where $f_{\mathrm{out},x} = m^{-1} \oint (\Sigma_{\rm g} u_x \bm{u}  + P\hat{\bm{x}}) \cdot \textnormal{d}\bm{A}$ is an integral over the aforementioned outer boundary.  Fig. \ref{fig:mdot_fx_BHL} takes the entire computational domain as an example and demonstrates that the sum of the first three terms in Eq. \ref{eq:fx_balance} exactly cancel $f_{\mathrm{out},x}$.

\begin{figure}
  \centering
  \includegraphics[width=0.9\linewidth]{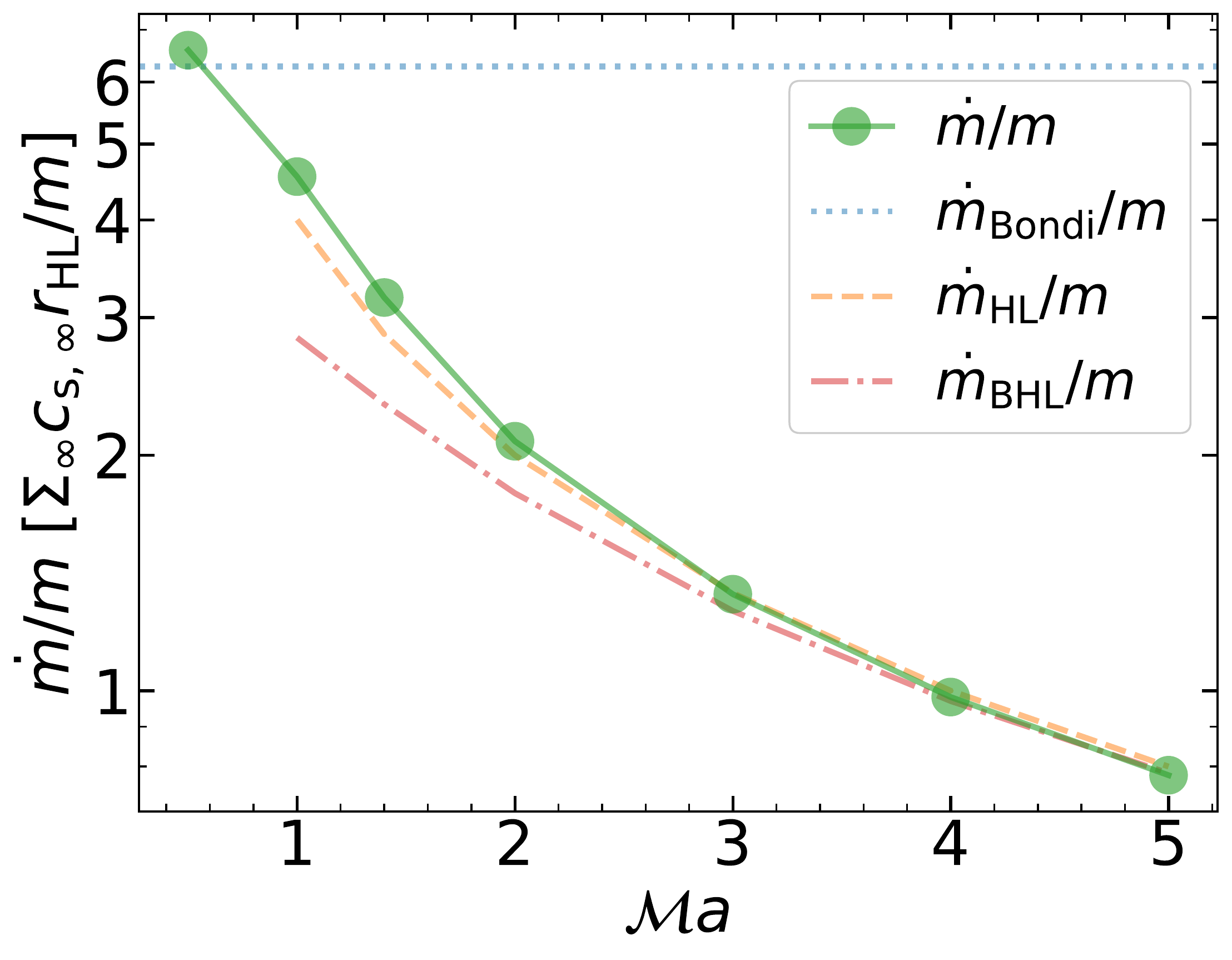}
  \caption{Specific accretion rate as a function of Mach number, measured from all the final snapshots of our 2D BHL simulations with $\gamma=3$.  Also shown are analytical estimates for $\dot{m}_{\rm Bondi}$ (\textit{blue dotted}), $\dot{m}_{\rm HL}$ (\textit{orange dashed}), and $\dot{m}_{\rm BHL}$ (\textit{red dash-dotted}), all with a factor of $m^{-1}$.  \label{fig:mdot_Ma_BHL}}
\end{figure}

We perform similar analysis on the final snapshots of runs with other values of $\mathcal{M}a$.  Fig. \ref{fig:mdot_Ma_BHL} shows that the measured $\dot{m}$ is in good agreement with $\dot{m}_{\rm BHL}$ for the highly supersonic cases (i.e., $\mathcal{M}a\gtrsim 4$).  As $\mathcal{M}a$ decreases, $\dot{m}$ gradually offsets towards a higher value, surpasses $\dot{m}_{\rm HL}$ for $\mathcal{M}a < 3$, and approaches $\dot{m}_{\rm Bondi}$ for $\mathcal{M}a\lesssim 1$.  In addition, we find that all cases exhibit force balance that satisfies Eq. \ref{eq:fx_balance}.

In summary, the findings in this section validate the robustness of our prescriptions for accretion and our methods to measure accretion rate and forces.  Our tests also confirm that, for a given sink radius $r_{\rm s}$, the measured $\dot{m}$ and $f_{\rm tot}$ (and thus the torques acting on the accretors in the main text) have little dependence on the choice of evaluation radius.

\bsp    
\label{lastpage}
\end{document}